%% file: CFT-09-013_temp.tex
\pdfoutput=1
\documentclass[11pt,twoside,a4paper,pdftex,cmspaper,final,collab]{cms-tdr}

\begin{document}\cmsNoteHeader{CFT-09-013}
%
%
%

%
%
\hyphenation{env-iron-men-tal}
\hyphenation{had-ron-i-za-tion}
\hyphenation{cal-or-i-me-ter}
\hyphenation{de-vices}
%
%
\RCS$Revision: 1.25 $
\RCS$Date: 2010/03/01 11:27:41 $
\RCS$Name:  $
\input{ptdr-definitions}
\cmsNoteHeader{09-013}
\title{Performance of the CMS Level-1 Trigger during Commissioning with Cosmic Ray Muons and LHC Beams}

\address[cern]{CERN}
\author[cern]{The CMS Collaboration}


\date{\today}

\abstract{The CMS Level-1 trigger was used to select cosmic ray muons and LHC beam events during data-taking runs in 2008, and to estimate the level of detector noise.  This paper describes the trigger components used, the algorithms that were executed, and the trigger synchronisation.  Using data from extended cosmic ray runs, the muon, electron/photon, and jet triggers have been validated, and their performance evaluated.  Efficiencies were found to be high, resolutions were found to be good, and rates as expected.}

\hypersetup{%
pdfauthor={},%
pdftitle={Performance of the CMS Level-1 Trigger during Commissioning with Cosmic Ray Muons and LHC Beams},%
pdfsubject={CMS},%
pdfkeywords={CMS, Trigger}}

\maketitle 

\input{Introduction.tex}

\input{L1Trigger.tex}

\input{CRAFTTriggers.tex}

\input{Synchronisation.tex}

\input{EmulatorValidation.tex}

\input{dt/DTPerformance.tex}

\input{csc/CSCPerformance.tex}

\input{rpc/RPCPerformance.tex}

\input{egamma/EGPerformance.tex}

\input{jet/JetPerformance.tex}

\section{Summary and Outlook}
\label{sect-summary}

The Level-1 trigger was operated stably during the period of LHC single beam operation, and later during the CRAFT cosmic ray data-taking period.  All muon sub-detector triggers have been synchronised and shown to provide good trigger efficiency.  The CSC muon candidate $\eta$, $\phi$, and $\pt$ assignments have been shown to work well, along with DT and RPC assignments.  The e/$\gamma$ trigger has been shown to be fully efficient.  The jet trigger was commissioned during CRAFT and shown to be efficient across the detector.  Together, these triggers have provided high quality cosmic ray event samples, as well as instrumental noise events necessary for understanding the detectors.  The LHC beam monitoring technical triggers and the CSC beam halo trigger were operated and synchronised during LHC single-beam operations.

Since the CRAFT data-taking in 2008, CMS has taken weekly cosmic ray runs, which have allowed a development and testing of additional functionality.  Optical links to the DT Track Finder for the $\theta$ view trigger primitives have been installed, improving the $\eta$ assignment.  Long runs have been taken with ``LHC-like'' look up tables in the DT track finder, allowing the $\pt$ assignment to be validated.  The links that share trigger primitives between DT and CSC systems in the barrel-endcap overlap region have been commissioned, and the DT trigger primitives have been synchronised at the CSC track finder.  The RPC and ECAL endcap triggers have been commissioned.  The remaining calorimeter energy sum triggers have been commissioned.  A six week long cosmic ray run in the summer of 2009 allowed a large cosmic ray and detector noise dataset to be acquired using the new functionality.  At the time of writing, analysis of this data is ongoing.  The Level-1 trigger system is expected to stably and efficiently trigger on LHC collisions for CMS in the forthcoming run.

\section*{Acknowledgements}

We thank the technical and administrative staff at CERN and other CMS Institutes, and acknowledge support from: FMSR (Austria); FNRS and FWO (Belgium); CNPq, CAPES, FAPERJ, and FAPESP (Brazil); MES (Bulgaria); CERN; CAS, MoST, and NSFC (China); COLCIENCIAS (Colombia); MSES (Croatia); RPF (Cyprus); Academy of Sciences and NICPB (Estonia); Academy of Finland, ME, and HIP (Finland); CEA and CNRS/IN2P3 (France); BMBF, DFG, and HGF (Germany); GSRT (Greece); OTKA and NKTH (Hungary); DAE and DST (India); IPM (Iran); SFI (Ireland); INFN (Italy); NRF (Korea); LAS (Lithuania); CINVESTAV, CONACYT, SEP, and UASLP-FAI (Mexico); PAEC (Pakistan); SCSR (Poland); FCT (Portugal); JINR (Armenia, Belarus, Georgia, Ukraine, Uzbekistan); MST and MAE (Russia); MSTDS (Serbia); MICINN and CPAN (Spain); Swiss Funding Agencies (Switzerland); NSC (Taipei); TUBITAK and TAEK (Turkey); STFC (United Kingdom); DOE and NSF (USA). Individuals have received support from the Marie-Curie IEF program (European Union); the Leventis Foundation; the A. P. Sloan Foundation; and the Alexander von Humboldt Foundation.

\bibliography{auto_generated}   
\clearpage

\cleardoublepage\appendix\section{The CMS Collaboration \label{app:collab}}\begin{sloppypar}\hyphenpenalty=5000\widowpenalty=500\clubpenalty=5000\input{CFT-09-013-authorlist.tex}\end{sloppypar}
\end{document}

%% file: ptdr-definitions.tex
%
%
%

\newcommand {\etal}{\mbox{et al.}\xspace} 
\newcommand {\ie}{\mbox{i.e.}\xspace}     
\newcommand {\eg}{\mbox{e.g.}\xspace}     
\newcommand {\etc}{\mbox{etc.}\xspace}     
\newcommand {\vs}{\mbox{\sl vs.}\xspace}      
\newcommand {\mdash}{\ensuremath{\mathrm{-}}} 

\newcommand {\Lone}{Level-1\xspace} 
\newcommand {\Ltwo}{Level-2\xspace}
\newcommand {\Lthree}{Level-3\xspace}

\providecommand{\ACERMC} {\textsc{AcerMC}\xspace}
\providecommand{\ALPGEN} {{\textsc{alpgen}}\xspace}
\providecommand{\CHARYBDIS} {{\textsc{charybdis}}\xspace}
\providecommand{\CMKIN} {\textsc{cmkin}\xspace}
\providecommand{\CMSIM} {{\textsc{cmsim}}\xspace}
\providecommand{\CMSSW} {{\textsc{cmssw}}\xspace}
\providecommand{\COBRA} {{\textsc{cobra}}\xspace}
\providecommand{\COCOA} {{\textsc{cocoa}}\xspace}
\providecommand{\COMPHEP} {\textsc{CompHEP}\xspace}
\providecommand{\EVTGEN} {{\textsc{evtgen}}\xspace}
\providecommand{\FAMOS} {{\textsc{famos}}\xspace}
\providecommand{\GARCON} {\textsc{garcon}\xspace}
\providecommand{\GARFIELD} {{\textsc{garfield}}\xspace}
\providecommand{\GEANE} {{\textsc{geane}}\xspace}
\providecommand{\GEANTfour} {{\textsc{geant4}}\xspace}
\providecommand{\GEANTthree} {{\textsc{geant3}}\xspace}
\providecommand{\GEANT} {{\textsc{geant}}\xspace}
\providecommand{\HDECAY} {\textsc{hdecay}\xspace}
\providecommand{\HERWIG} {{\textsc{herwig}}\xspace}
\providecommand{\HIGLU} {{\textsc{higlu}}\xspace}
\providecommand{\HIJING} {{\textsc{hijing}}\xspace}
\providecommand{\IGUANA} {\textsc{iguana}\xspace}
\providecommand{\ISAJET} {{\textsc{isajet}}\xspace}
\providecommand{\ISAPYTHIA} {{\textsc{isapythia}}\xspace}
\providecommand{\ISASUGRA} {{\textsc{isasugra}}\xspace}
\providecommand{\ISASUSY} {{\textsc{isasusy}}\xspace}
\providecommand{\ISAWIG} {{\textsc{isawig}}\xspace}
\providecommand{\MADGRAPH} {\textsc{MadGraph}\xspace}
\providecommand{\MCATNLO} {\textsc{mc@nlo}\xspace}
\providecommand{\MCFM} {\textsc{mcfm}\xspace}
\providecommand{\MILLEPEDE} {{\textsc{millepede}}\xspace}
\providecommand{\ORCA} {{\textsc{orca}}\xspace}
\providecommand{\OSCAR} {{\textsc{oscar}}\xspace}
\providecommand{\PHOTOS} {\textsc{photos}\xspace}
\providecommand{\PROSPINO} {\textsc{prospino}\xspace}
\providecommand{\PYTHIA} {{\textsc{pythia}}\xspace}
\providecommand{\SHERPA} {{\textsc{sherpa}}\xspace}
\providecommand{\TAUOLA} {\textsc{tauola}\xspace}
\providecommand{\TOPREX} {\textsc{TopReX}\xspace}
\providecommand{\XDAQ} {{\textsc{xdaq}}\xspace}

\newcommand {\DZERO}{D\O\xspace}     


\newcommand{\de}{\ensuremath{^\circ}}
\newcommand{\ten}[1]{\ensuremath{\times \text{10}^\text{#1}}}
\newcommand{\unit}[1]{\ensuremath{\text{\,#1}}\xspace}
\newcommand{\mum}{\ensuremath{\,\mu\text{m}}\xspace}
\newcommand{\micron}{\ensuremath{\,\mu\text{m}}\xspace}
\newcommand{\cm}{\ensuremath{\,\text{cm}}\xspace}
\newcommand{\mm}{\ensuremath{\,\text{mm}}\xspace}
\newcommand{\mus}{\ensuremath{\,\mu\text{s}}\xspace}
\newcommand{\keV}{\ensuremath{\,\text{ke\hspace{-.08em}V}}\xspace}
\newcommand{\MeV}{\ensuremath{\,\text{Me\hspace{-.08em}V}}\xspace}
\newcommand{\GeV}{\ensuremath{\,\text{Ge\hspace{-.08em}V}}\xspace}
\newcommand{\TeV}{\ensuremath{\,\text{Te\hspace{-.08em}V}}\xspace}
\newcommand{\PeV}{\ensuremath{\,\text{Pe\hspace{-.08em}V}}\xspace}
\newcommand{\keVc}{\ensuremath{{\,\text{ke\hspace{-.08em}V\hspace{-0.16em}/\hspace{-0.08em}c}}}\xspace}
\newcommand{\MeVc}{\ensuremath{{\,\text{Me\hspace{-.08em}V\hspace{-0.16em}/\hspace{-0.08em}c}}}\xspace}
\newcommand{\GeVc}{\ensuremath{{\,\text{Ge\hspace{-.08em}V\hspace{-0.16em}/\hspace{-0.08em}c}}}\xspace}
\newcommand{\TeVc}{\ensuremath{{\,\text{Te\hspace{-.08em}V\hspace{-0.16em}/\hspace{-0.08em}c}}}\xspace}
\newcommand{\keVcc}{\ensuremath{{\,\text{ke\hspace{-.08em}V\hspace{-0.16em}/\hspace{-0.08em}c}^\text{2}}}\xspace}
\newcommand{\MeVcc}{\ensuremath{{\,\text{Me\hspace{-.08em}V\hspace{-0.16em}/\hspace{-0.08em}c}^\text{2}}}\xspace}
\newcommand{\GeVcc}{\ensuremath{{\,\text{Ge\hspace{-.08em}V\hspace{-0.16em}/\hspace{-0.08em}c}^\text{2}}}\xspace}
\newcommand{\TeVcc}{\ensuremath{{\,\text{Te\hspace{-.08em}V\hspace{-0.16em}/\hspace{-0.08em}c}^\text{2}}}\xspace}

\newcommand{\pbinv} {\mbox{\ensuremath{\,\text{pb}^\text{$-$1}}}\xspace}
\newcommand{\fbinv} {\mbox{\ensuremath{\,\text{fb}^\text{$-$1}}}\xspace}
\newcommand{\nbinv} {\mbox{\ensuremath{\,\text{nb}^\text{$-$1}}}\xspace}
\newcommand{\percms}{\ensuremath{\,\text{cm}^\text{$-$2}\,\text{s}^\text{$-$1}}\xspace}
\newcommand{\lumi}{\ensuremath{\mathcal{L}}\xspace}
\newcommand{\Lumi}{\ensuremath{\mathcal{L}}\xspace}
%
\newcommand{\LvLow}  {\ensuremath{\mathcal{L}=\text{10}^\text{32}\,\text{cm}^\text{$-$2}\,\text{s}^\text{$-$1}}\xspace}
\newcommand{\LLow}   {\ensuremath{\mathcal{L}=\text{10}^\text{33}\,\text{cm}^\text{$-$2}\,\text{s}^\text{$-$1}}\xspace}
\newcommand{\lowlumi}{\ensuremath{\mathcal{L}=\text{2}\times \text{10}^\text{33}\,\text{cm}^\text{$-$2}\,\text{s}^\text{$-$1}}\xspace}
\newcommand{\LMed}   {\ensuremath{\mathcal{L}=\text{2}\times \text{10}^\text{33}\,\text{cm}^\text{$-$2}\,\text{s}^\text{$-$1}}\xspace}
\newcommand{\LHigh}  {\ensuremath{\mathcal{L}=\text{10}^\text{34}\,\text{cm}^\text{$-$2}\,\text{s}^\text{$-$1}}\xspace}
\newcommand{\hilumi} {\ensuremath{\mathcal{L}=\text{10}^\text{34}\,\text{cm}^\text{$-$2}\,\text{s}^\text{$-$1}}\xspace}


\newcommand{\zp}{\ensuremath{\mathrm{Z}^\prime}\xspace}


\newcommand{\kt}{\ensuremath{k_{\mathrm{T}}}\xspace}
\newcommand{\BC}{\ensuremath{{B_{\mathrm{c}}}}\xspace}
\newcommand{\bbarc}{\ensuremath{{\overline{b}c}}\xspace}
\newcommand{\bbbar}{\ensuremath{{b\overline{b}}}\xspace}
\newcommand{\ccbar}{\ensuremath{{c\overline{c}}}\xspace}
\newcommand{\JPsi}{\ensuremath{{J}/\psi}\xspace}
\newcommand{\bspsiphi}{\ensuremath{B_s \to \JPsi\, \phi}\xspace}
\newcommand{\AFB}{\ensuremath{A_\mathrm{FB}}\xspace}
\newcommand{\EE}{\ensuremath{e^+e^-}\xspace}
\newcommand{\MM}{\ensuremath{\mu^+\mu^-}\xspace}
\newcommand{\TT}{\ensuremath{\tau^+\tau^-}\xspace}
\newcommand{\wangle}{\ensuremath{\sin^{2}\theta_{\mathrm{eff}}^\mathrm{lept}(M^2_\mathrm{Z})}\xspace}
\newcommand{\ttbar}{\ensuremath{{t\overline{t}}}\xspace}
\newcommand{\stat}{\ensuremath{\,\text{(stat.)}}\xspace}
\newcommand{\syst}{\ensuremath{\,\text{(syst.)}}\xspace}

\newcommand{\HGG}{\ensuremath{\mathrm{H}\to\gamma\gamma}}
\newcommand{\gev}{\GeV}
\newcommand{\GAMJET}{\ensuremath{\gamma + \mathrm{jet}}}
\newcommand{\PPTOJETS}{\ensuremath{\mathrm{pp}\to\mathrm{jets}}}
\newcommand{\PPTOGG}{\ensuremath{\mathrm{pp}\to\gamma\gamma}}
\newcommand{\PPTOGAMJET}{\ensuremath{\mathrm{pp}\to\gamma +
\mathrm{jet}
}}
\newcommand{\MH}{\ensuremath{\mathrm{M_{\mathrm{H}}}}}
\newcommand{\RNINE}{\ensuremath{\mathrm{R}_\mathrm{9}}}
\newcommand{\DR}{\ensuremath{\Delta\mathrm{R}}}


\newcommand{\PT}{\ensuremath{p_{\mathrm{T}}}\xspace}
\newcommand{\pt}{\ensuremath{p_{\mathrm{T}}}\xspace}
\newcommand{\ET}{\ensuremath{E_{\mathrm{T}}}\xspace}
\newcommand{\HT}{\ensuremath{H_{\mathrm{T}}}\xspace}
\newcommand{\et}{\ensuremath{E_{\mathrm{T}}}\xspace}
\newcommand{\Em}{\ensuremath{E\!\!\!/}\xspace}
\newcommand{\Pm}{\ensuremath{p\!\!\!/}\xspace}
\newcommand{\PTm}{\ensuremath{{p\!\!\!/}_{\mathrm{T}}}\xspace}
\newcommand{\ETm}{\ensuremath{E_{\mathrm{T}}^{\mathrm{miss}}}\xspace}
\newcommand{\MET}{\ensuremath{E_{\mathrm{T}}^{\mathrm{miss}}}\xspace}
\newcommand{\ETmiss}{\ensuremath{E_{\mathrm{T}}^{\mathrm{miss}}}\xspace}
\newcommand{\VEtmiss}{\ensuremath{{\vec E}_{\mathrm{T}}^{\mathrm{miss}}}\xspace}

%

\newcommand{\ga}{\ensuremath{\gtrsim}}
\newcommand{\la}{\ensuremath{\lesssim}}
\newcommand{\swsq}{\ensuremath{\sin^2\theta_W}\xspace}
\newcommand{\cwsq}{\ensuremath{\cos^2\theta_W}\xspace}
\newcommand{\tanb}{\ensuremath{\tan\beta}\xspace}
\newcommand{\tanbsq}{\ensuremath{\tan^{2}\beta}\xspace}
\newcommand{\sidb}{\ensuremath{\sin 2\beta}\xspace}
\newcommand{\alpS}{\ensuremath{\alpha_S}\xspace}
\newcommand{\alpt}{\ensuremath{\tilde{\alpha}}\xspace}

\newcommand{\QL}{\ensuremath{Q_L}\xspace}
\newcommand{\sQ}{\ensuremath{\tilde{Q}}\xspace}
\newcommand{\sQL}{\ensuremath{\tilde{Q}_L}\xspace}
\newcommand{\ULC}{\ensuremath{U_L^C}\xspace}
\newcommand{\sUC}{\ensuremath{\tilde{U}^C}\xspace}
\newcommand{\sULC}{\ensuremath{\tilde{U}_L^C}\xspace}
\newcommand{\DLC}{\ensuremath{D_L^C}\xspace}
\newcommand{\sDC}{\ensuremath{\tilde{D}^C}\xspace}
\newcommand{\sDLC}{\ensuremath{\tilde{D}_L^C}\xspace}
\newcommand{\LL}{\ensuremath{L_L}\xspace}
\newcommand{\sL}{\ensuremath{\tilde{L}}\xspace}
\newcommand{\sLL}{\ensuremath{\tilde{L}_L}\xspace}
\newcommand{\ELC}{\ensuremath{E_L^C}\xspace}
\newcommand{\sEC}{\ensuremath{\tilde{E}^C}\xspace}
\newcommand{\sELC}{\ensuremath{\tilde{E}_L^C}\xspace}
\newcommand{\sEL}{\ensuremath{\tilde{E}_L}\xspace}
\newcommand{\sER}{\ensuremath{\tilde{E}_R}\xspace}
\newcommand{\sFer}{\ensuremath{\tilde{f}}\xspace}
\newcommand{\sQua}{\ensuremath{\tilde{q}}\xspace}
\newcommand{\sUp}{\ensuremath{\tilde{u}}\xspace}
\newcommand{\suL}{\ensuremath{\tilde{u}_L}\xspace}
\newcommand{\suR}{\ensuremath{\tilde{u}_R}\xspace}
\newcommand{\sDw}{\ensuremath{\tilde{d}}\xspace}
\newcommand{\sdL}{\ensuremath{\tilde{d}_L}\xspace}
\newcommand{\sdR}{\ensuremath{\tilde{d}_R}\xspace}
\newcommand{\sTop}{\ensuremath{\tilde{t}}\xspace}
\newcommand{\stL}{\ensuremath{\tilde{t}_L}\xspace}
\newcommand{\stR}{\ensuremath{\tilde{t}_R}\xspace}
\newcommand{\stone}{\ensuremath{\tilde{t}_1}\xspace}
\newcommand{\sttwo}{\ensuremath{\tilde{t}_2}\xspace}
\newcommand{\sBot}{\ensuremath{\tilde{b}}\xspace}
\newcommand{\sbL}{\ensuremath{\tilde{b}_L}\xspace}
\newcommand{\sbR}{\ensuremath{\tilde{b}_R}\xspace}
\newcommand{\sbone}{\ensuremath{\tilde{b}_1}\xspace}
\newcommand{\sbtwo}{\ensuremath{\tilde{b}_2}\xspace}
\newcommand{\sLep}{\ensuremath{\tilde{l}}\xspace}
\newcommand{\sLepC}{\ensuremath{\tilde{l}^C}\xspace}
\newcommand{\sEl}{\ensuremath{\tilde{e}}\xspace}
\newcommand{\sElC}{\ensuremath{\tilde{e}^C}\xspace}
\newcommand{\seL}{\ensuremath{\tilde{e}_L}\xspace}
\newcommand{\seR}{\ensuremath{\tilde{e}_R}\xspace}
\newcommand{\snL}{\ensuremath{\tilde{\nu}_L}\xspace}
\newcommand{\sMu}{\ensuremath{\tilde{\mu}}\xspace}
\newcommand{\sNu}{\ensuremath{\tilde{\nu}}\xspace}
\newcommand{\sTau}{\ensuremath{\tilde{\tau}}\xspace}
\newcommand{\Glu}{\ensuremath{g}\xspace}
\newcommand{\sGlu}{\ensuremath{\tilde{g}}\xspace}
\newcommand{\Wpm}{\ensuremath{W^{\pm}}\xspace}
\newcommand{\sWpm}{\ensuremath{\tilde{W}^{\pm}}\xspace}
\newcommand{\Wz}{\ensuremath{W^{0}}\xspace}
\newcommand{\sWz}{\ensuremath{\tilde{W}^{0}}\xspace}
\newcommand{\sWino}{\ensuremath{\tilde{W}}\xspace}
\newcommand{\Bz}{\ensuremath{B^{0}}\xspace}
\newcommand{\sBz}{\ensuremath{\tilde{B}^{0}}\xspace}
\newcommand{\sBino}{\ensuremath{\tilde{B}}\xspace}
\newcommand{\Zz}{\ensuremath{Z^{0}}\xspace}
\newcommand{\sZino}{\ensuremath{\tilde{Z}^{0}}\xspace}
\newcommand{\sGam}{\ensuremath{\tilde{\gamma}}\xspace}
\newcommand{\chiz}{\ensuremath{\tilde{\chi}^{0}}\xspace}
\newcommand{\chip}{\ensuremath{\tilde{\chi}^{+}}\xspace}
\newcommand{\chim}{\ensuremath{\tilde{\chi}^{-}}\xspace}
\newcommand{\chipm}{\ensuremath{\tilde{\chi}^{\pm}}\xspace}
\newcommand{\Hone}{\ensuremath{H_{d}}\xspace}
\newcommand{\sHone}{\ensuremath{\tilde{H}_{d}}\xspace}
\newcommand{\Htwo}{\ensuremath{H_{u}}\xspace}
\newcommand{\sHtwo}{\ensuremath{\tilde{H}_{u}}\xspace}
\newcommand{\sHig}{\ensuremath{\tilde{H}}\xspace}
\newcommand{\sHa}{\ensuremath{\tilde{H}_{a}}\xspace}
\newcommand{\sHb}{\ensuremath{\tilde{H}_{b}}\xspace}
\newcommand{\sHpm}{\ensuremath{\tilde{H}^{\pm}}\xspace}
\newcommand{\hz}{\ensuremath{h^{0}}\xspace}
\newcommand{\Hz}{\ensuremath{H^{0}}\xspace}
\newcommand{\Az}{\ensuremath{A^{0}}\xspace}
\newcommand{\Hpm}{\ensuremath{H^{\pm}}\xspace}
\newcommand{\sGra}{\ensuremath{\tilde{G}}\xspace}
\newcommand{\mtil}{\ensuremath{\tilde{m}}\xspace}
\newcommand{\rpv}{\ensuremath{\rlap{\kern.2em/}R}\xspace}
\newcommand{\LLE}{\ensuremath{LL\bar{E}}\xspace}
\newcommand{\LQD}{\ensuremath{LQ\bar{D}}\xspace}
\newcommand{\UDD}{\ensuremath{\overline{UDD}}\xspace}
\newcommand{\Lam}{\ensuremath{\lambda}\xspace}
\newcommand{\Lamp}{\ensuremath{\lambda'}\xspace}
\newcommand{\Lampp}{\ensuremath{\lambda''}\xspace}
\newcommand{\spinbd}[2]{\ensuremath{\bar{#1}_{\dot{#2}}}\xspace}

\newcommand{\MD}{\ensuremath{{M_\mathrm{D}}}\xspace}
\newcommand{\Mpl}{\ensuremath{{M_\mathrm{Pl}}}\xspace}
\newcommand{\Rinv} {\ensuremath{{R}^{-1}}\xspace}

%
%
\hyphenation{en-viron-men-tal}

%% file: Introduction.tex
\section{Introduction}

The primary goal of the Compact Muon Solenoid (CMS)\footnote{A glossary of acronyms used in this paper may be found in Ref.~\cite{CMS:2008zzk} p309} experiment~\cite{CMS:2008zzk} is  
to explore particle physics at the TeV energy scale exploiting the proton-proton  
collisions delivered by the Large Hadron Collider (LHC)~\cite{Evans:2008zzb}.   During 
October-November 2008 the CMS  collaboration conducted a month-long data taking exercise, known as  
the Cosmic Run At Four Tesla (CRAFT), with the goal of commissioning the  
experiment for extended operation \cite{CRAFTGeneral}.
With all installed detector systems participating, CMS recorded 270 Million
cosmic ray triggered events with the solenoid at its nominal axial field
strength of 3.8~T.  Prior to CRAFT, in September 2008, CMS observed the muon halo from 
single circulating beams and received several single shot ``beam splash'' events.  In a beam splash event, the beam is steered onto closed collimators upstream 
of CMS, releasing $\mathcal{O}(10^5)$ muons that produce signals in most channels of the detector. 

\begin{figure}[!hbt]
  \centering
 \includegraphics[width=0.98\textwidth]{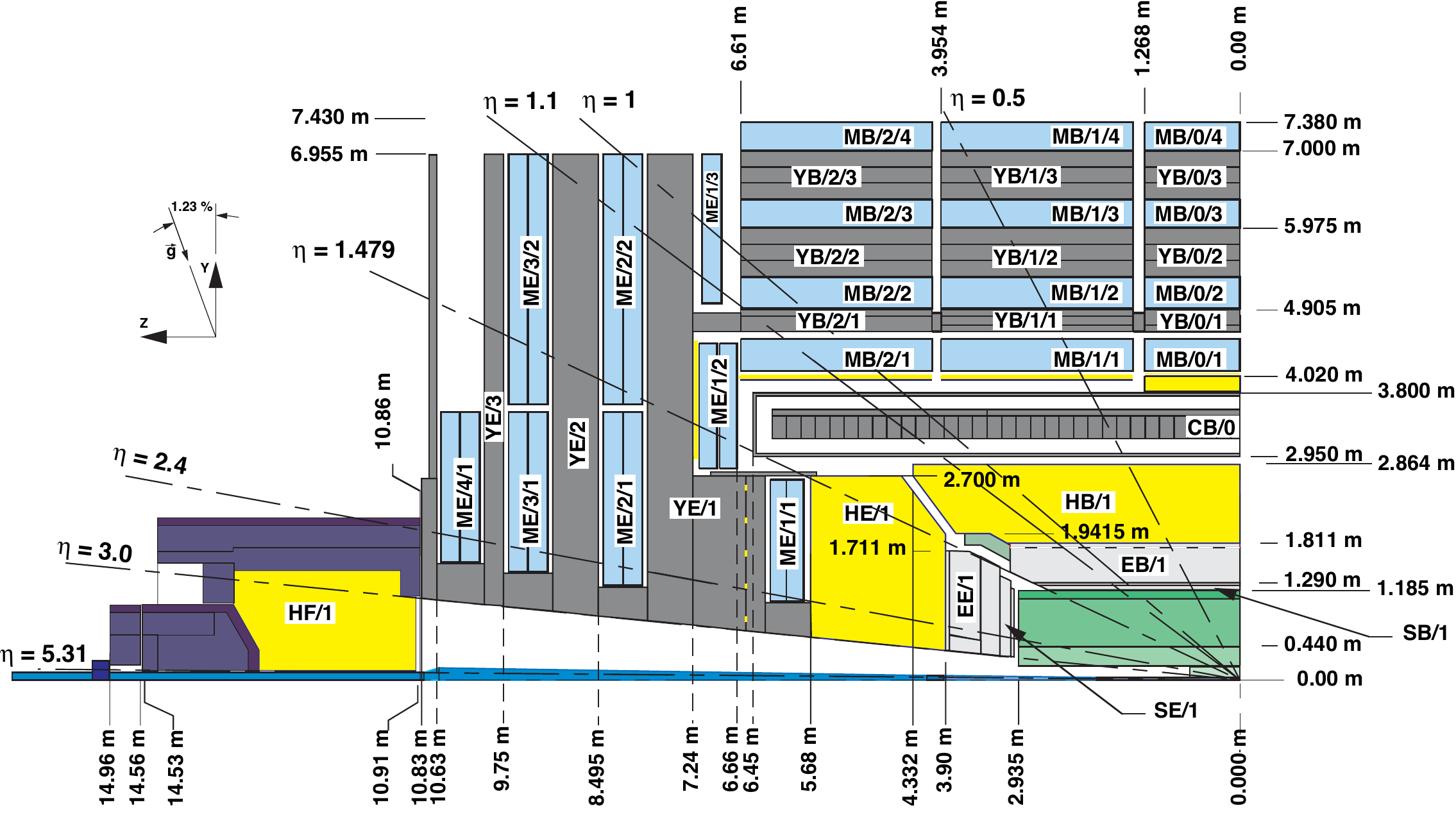}  
  \caption{Cross-section through the CMS detector in the $y$-$z$ plane.}
  \label{fig:cms}
\end{figure}

A detailed description of the CMS detector can be found in Ref.~\cite{CMS:2008zzk}.  
Figure~\ref{fig:cms} shows a cross-section through the detector.  The central feature of the apparatus is a superconducting
solenoid, of 6~m internal diameter.  Within the field volume are the
silicon pixel and strip trackers, the crystal electromagnetic
calorimeter (ECAL) and the brass-scintillator hadron calorimeter
(HCAL).  Muons are measured in drift tube chambers (DT), resistive plate chambers (RPC), and cathode strip chambers (CSC), embedded in the steel
return yoke.  Beyond the magnet yoke endcaps are iron-quartz forward hadron calorimeters (HF).  The first level (L1) of the CMS trigger system, composed of custom hardware
processors, is designed to select one potentially interesting event from
every thousand, in less than 1~$\mu$s processing time, using information from
the calorimeters and muon detectors.  The High Level Trigger
(HLT) processor farm further decreases the event rate to the order of
100~Hz, before data storage. CMS uses a right-handed coordinate system, with the origin at the nominal interaction point, the $x$-axis pointing to the centre of the LHC, the $y$-axis pointing up (perpendicular to the LHC plane), and the $z$-axis along the anticlockwise-beam direction. The polar angle, $\theta$, is measured from the positive $z$-axis and the azimuthal angle, $\phi$, is measured in the $x$-$y$ plane.

The CMS L1 trigger was commissioned before and during CRAFT and the single beam operation.  
Overall, the L1 trigger performed well during these periods; efficiencies were high and resolutions were good. 
Subsequent analysis of the data has allowed the performance to be quantified, and presented in this paper.  The task of commissioning such a large and complex system was challenging.  The trigger hardware is described in Section \ref{sec:l1trigger}, and its use in CRAFT and LHC single beam operation are described in Section \ref{sec:crafttriggers}.  The results of trigger synchronisation are described in Section \ref{sec:synchronisation}.   Section \ref{sec:emulator} describes the comparison of the trigger data with a software emulation of the system.  The performance measurements are described in Sections \ref{sec:dtperformance}--\ref{sec:jetperformance}, covering muon triggers from three subdetectors, e/$\gamma$ and jet triggers.  In general, these analyses use only fractions of the CRAFT dataset,
because the performance is evaluated using events similar to those of LHC collisions for which the L1 trigger design was optimised.

%% file: L1Trigger.tex
\section{CMS Level-1 Trigger}
\label{sec:l1trigger}

The CMS L1 trigger is described in detail in Ref.~\cite{Dasu:2000ge}.  In brief, the calorimeters and the muon
subdetectors provide trigger primitives in the form of local energy
deposits in calorimeter trigger towers and track segments or hits in
muon chambers.  Regional and global processors identify trigger objects: electron, jet, and muon candidates, and energy sums. 
A full set of trigger primitives are produced every 25~ns, a period known as a ``bunch-crossing'' (BX).  For LHC collisions, this identifies the trigger primitives and resulting trigger objects with a particular proton-proton collision.  Objects are ranked and sorted. They form the basis for trigger decisions taken by
the final L1 stage, the Global Trigger (GT), according to programmable
algorithms. The Trigger Control System (TCS) determines if the
subdetectors are ready to read out the event, and if the data acquisition (DAQ) system is ready to receive it.  Data from trigger primitives, regional energy sums, muon candidates from each sub-detector, and final trigger objects are sent in parallel to the DAQ for each accepted event. Control and monitoring of the L1 trigger operation are performed centrally by dedicated software.

\subsection{Muon Triggers}
Drift tube chambers in the barrel of the detector and cathode strip
chambers in the endcap regions provide tracking up to pseudorapidity $\vert \eta \vert = 2.4$ and 
trigger information up to $\vert \eta \vert = 2.1$. Resistive plate chambers cover up to $\vert \eta \vert = 1.6$ and
are used mainly for triggering purposes. 
The chambers are mounted in the return yoke of the solenoid that is composed
of five wheels in the barrel region and three disks in each of
the endcaps. 
Wheels and disks are subdivided into azimuthal sectors covering approximately
$30^{\circ}$ or $60^{\circ}$ (with some overlap).
The DT and CSC share trigger information in the overlap region, enabling each of the three muon subdetectors to 
deliver its own list of up to four muon candidates, 
ranked and sorted according to decreasing reconstruction
quality and transverse momentum, to the Global Muon Trigger (GMT)
\cite{Sakulin:2005qx}.  This then combines them and forwards up to four candidates to the GT.

\subsubsection{DT Trigger}
\label{sec:DTtrigger}

The DT system consists of 250 ``chambers'' arranged in four muon stations (MB1, MB2, MB3, and MB4) embedded in the steel yoke of CMS. Each DT chamber consists of staggered planes of drift cells. Four planes form a superlayer. The three innermost stations are made of chambers with three 
superlayers; the inner and outer superlayers measure $\phi$ while the central 
superlayer measures $\eta$. The fourth muon station 
has only $\phi$-superlayers.

The on-chamber electronics produce trigger primitives, consisting of information about identified track segments. They are generated separately for the
``$\phi$-view'' (in the $x-y$ plane) and the
``$\theta$-view'' (in the $r-z$ plane).   Up to
two primitives per chamber and per bunch-crossing (BX) are generated in the $\phi$-view, each primitive comprising the track segment position, direction and a quality code.  The latter encodes the number of drift cell hits that were found aligned by the trigger logic, labelling a primitive built with four or three hits from the four staggered layers of a superlayer, ``High'' (H) or ``Low'' (L) quality, respectively.  If a proper matching, or correlation, between the segments from the two superlayers in the $\phi$-view is found, the primitive is labelled HH, HL, or LL. If a proper correlation is not found, single H or L segments are output. In the $\theta$-view, segments are
accepted only if their direction is compatible with a track that originated
from the interaction point. Thus, no information about direction is
output, and the positions of valid segments are encoded into a bit pattern.

Trigger primitives from a given muon sector are sent to the Sector
Collector electronics, located outside of the detector. The signals from
each station are synchronised, coded, and forwarded through high-speed
optical links \cite{Guiducci:2007sc} to the Drift Tube Track Finder
(DTTF) located in the underground counting room adjacent to the
detector. The DTTF also receives trigger primitives from the CSCs for
the barrel-endcap overlap region. The DTTF system performs matching
between trigger primitives received from the DT stations and
assigns a quality code as well as $\phi$, $\eta$, charge and
transverse momentum ($\pt$) values to the reconstructed muon track.
The track matching is based on extrapolation. The standard algorithms used to identify muons from 
LHC collisions are described in detail
in Ref.~\cite{Ero:2008zz}.  Different algorithms are used for cosmic
ray triggers, and are described in Section~\ref{sec:crafttriggers}.
The Sector Collector and the DTTF also read out their input and output
data for several time samples around the triggered event for
diagnostics and monitoring.

\subsubsection{CSC Trigger}
The CMS endcap muon system includes 468 trapezoidal cathode strip
chambers with different $\phi$ coverage, arranged to form four disks at each endcap (stations ME1, ME2, ME3, and ME4). Each station is in turn subdivided into rings of chambers as follows: ME1 has three rings of chambers (ME1/n, n=1,2,3), ME2 and ME3 stations have two rings of chambers, and ME4 has one ring of chambers. Each chamber consists of six layers equipped with anode wires and cathode strips.

In each chamber, the track segment position, angle and bunch
crossing, are first determined separately in the nearly
orthogonal anode and cathode views.  The cathode readout is
optimised to measure the $\phi$-coordinate, while the anode readout is optimised to
identify the bunch crossing. The front-end
electronics boards reconstruct track segments using
pattern-recognition firmware based on pattern templates. These
templates require track segments in cathode as well as anode views to
point towards the interaction point, with an angular acceptance, of
order one Radian, depending on the station. The track segments from
the cathode and anode readout from each chamber are finally
combined into 3-dimensional local tracks, which are the CSC trigger
primitives.  The trigger primitives are collected by the Muon Port
Cards, which sort them and send up to three candidates to the CSC
Track Finder (CSCTF) via optical fibres.

The CSCTF matches trigger primitives to form complete tracks and
determine their $\pt$, $\eta$, $\phi$ and charge. The CSCTF
functionality is described in more detail
in Ref.~\cite{Acosta:2003is}. For the purpose of track finding, the
CSC detector is logically partitioned into six $60^{\circ}$ azimuthal
sectors per endcap.  The trigger primitives from each sector are
received and processed by single Sector Processor boards.  The CSCTF
also receives trigger primitives from the DT system for the
barrel-endcap overlap region.  The CSCTF is optimized to cope with the
non-axial magnetic field present in the endcap region.  Thus, the
algorithms of the CSCTF are inherently 3-dimensional to achieve
maximum background rejection, in particular for low momentum tracks.  
In addition to the $\pt$, $\eta$, $\phi$, and charge,
each track identified by the CSCTF carries a quality code.  
This quality code is used along with the $\pt$ to sort the candidates;
 the highest ranking four are sent to the Global Muon Trigger. 
The quality code is a two-bit word that is used to indicate the
expected coarse $\pt$ resolution.  Quality~3 (high $\pt$ resolution)
refers to a three- or four-segment track with one of the segments in
ME1. Quality~2 (medium $\pt$ resolution) refers to a 2-segment
coincidence with one of the segments in ME1. Quality~1 (low $\pt$
resolution) refers to any other 2-segment coincidence. Quality~3
candidates, with $5<\pt<35$ GeV/c, are expected to have about $20\%$
resolution in $\pt$, while quality~2 are expected to have about
$30\%$.  In addition to identifying muons originating from the
interaction point, the CSCTF identifies tracks from ``halo muons'',
coming from the interaction of the LHC beam with the gas particles in
the beam pipe or with the beam pipe itself. This set of muons,
parallel to the beam line, has proven to be very useful at the LHC
start-up to align the several endcap disks~\cite{CFT-09-016}.

\subsubsection{RPC Trigger}
In the barrel and endcap regions, the DT and CSC chambers are complemented by 
double-gap resistive plate chambers. The RPCs are arranged in six layers in the barrel region and three layers in the
forward regions. 
They have excellent timing
resolution, of the order of 1~ns. Their main purpose is to identify
the bunch-crossing in which the detected muon was emitted. They
also assign track parameters. 
The RPC trigger is based on the spatial and temporal coincidence of hits in several layers. The Pattern Comparator trigger logic \cite{Andlinger:1994ve} compares signals from all four muon stations to predefined hit patterns in order to find muon candidates. Muon $p_T$, charge, $\eta$, and $\phi$ are assigned according to the matched pattern.
The algorithm requires a minimum number of hit planes, which varies
with the $p_T$ and location of the muon. Either 4/6 (four out of six),
4/5, 3/4 or 3/3 hit layers are minimally required. A quality value,
encoded in two bits, reflects the number of hit layers.
Analog signals from the chambers are digitized by Front End Boards, then zero-suppressed and assigned to the proper bunch crossing by a system of Link Boards located in the vicinity of the detector.  They are then sent via optical links to
Trigger Boards located in the underground counting room.  
Each of the 84 Trigger Boards can produce up to four muon candidates for every bunch crossing.  A system of two 
Half-Sorter Boards followed by a Final Sorter Board sorts the candidates by quality and $\pt$, and sends up to eight muon candidates, four from the barrel and four
from the endcaps, to the GMT.

\subsubsection{Global Muon Trigger}

The Global Muon Trigger receives up to 4 candidates from each of the DTTF
and CSCTF and up to 8 candidates (4 in the barrel, and 4 in the endcap) from the RPC
trigger.  Look-up tables (LUTs) are used to combine candidates
identified by more than one sub-detector, and to assign a quality code
based on the number of subdetectors involved, as well as on the
quality of the track candidates, as assigned by the track-finders. The four highest quality muon candidates are forwarded to the GT.  The GMT also reads out its input and output data for 3 time samples around each triggered event, for diagnostics, monitoring, and to indicate regions of interest to HLT.

\subsection{Calorimeter Triggers}

For triggering purposes the barrel and endcap calorimeters are subdivided in trigger towers.  The pattern of energy deposited in those towers is analyzed to identify electron/photon and jet candidates, and the tower energies are summed to obtain the candidate transverse energy ($\et$).  A trigger primitive is generated for each trigger tower in the ECAL and HCAL, up to $\vert\eta\vert=3.0$.  The towers have the same segmentation in both the ECAL and HCAL. Their size is $\Delta\eta \times \Delta\phi = 0.087 \times 0.087$ in the barrel and in the endcaps up to $\vert\eta\vert=1.8$.  For $\vert\eta\vert > 1.8$, the tower segmentation in $\eta$ increases to $\Delta\eta = 0.1 - 0.35$.  Trigger primitives from the forward region, which covers the range $3.0 < \vert\eta\vert < 5.0$, are used for jet and energy sum triggers only.  A single trigger primitive is generated for each HF trigger region, which are equal to $3\eta\times2\phi$ readout towers, and are of constant size; $\Delta\eta \times \Delta\phi = 0.5 \times 0.349$.  The initial energy scale for calorimeter triggers was derived from test beam results, and in the case of HCAL was further fine-tuned using Monte-Carlo simulations.

\subsubsection{ECAL Trigger Primitives}
\label{ECALTPG}
The ECAL trigger primitive generation (TPG) starts in the on-detector front-end electronics after digitisation of the signal, by summing the energy from each $\mbox{PbWO}_4$ crystal in a strip of five in the $\phi$ direction and converting the result to $\et$, taking into account the electronics gains and calibration coefficients.   An amplitude filter is applied to the strip sum consisting of the weighted sum of five 25~ns time samples, taking into account the expected signal shape and residual pedestal to be dynamically subtracted.  Finally, a peak finder applied to three consecutive time samples in
a sliding window requires the amplitude of the central sample to be
maximum, keeping this value as a measure of the transverse energy
contained in the strip.  The $\et$ values from five adjacent strips in $\eta$ are then summed and the $\et$ estimate for the trigger tower is transferred to the Regional Calorimeter Trigger (RCT).  The $\et$ value is encoded in 8 bits. In addition, a fine-grain veto bit is set for each trigger tower if the highest two adjacent strips in the tower contain less than 90\% of the total $\et$.  This gives some indication of the lateral shower shape, and can be used to reject L1 electron/photon (e/$\gamma$) candidates that result from physical jets.

\subsubsection{HCAL Trigger Primitives}
\label{HCALTPG}
HCAL signals are digitised on-detector and the data transmitted to the HCAL trigger and readout boards via optical fibres.  The TPG processing for barrel and endcap is different from the forward calorimeter, and is described first.  The barrel and endcap trigger primitives are formed by first linearising the received signal, using LUTs that are programmed to account for individual channel gains and pedestals.  The trigger towers are the same size in $\eta\times\phi$ as the readout towers, but energies from separate longitudinal readout channels are summed.  The pulse energy is obtained by summing two adjacent 25~ns time
samples and the peak time is found by a peak finder applied to three
consecutive samples, as described for ECAL trigger data.  The resulting energy value for the trigger tower is compressed before being sent to the RCT, using an analytical compression function that has no loss of precision at low energies and matches the calorimeter resolution at high energies.
The forward calorimeter trigger primitives are generated by linearising signals from the front-end and converting to $\et$, again accounting for channel gains and pedestals.  These are then summed over $3\eta\times2\phi$ towers to give a trigger region of $0.5\times0.349$, which is not too large since the forward calorimeter is only used in jet and energy sum triggers.  The pulses are short, so no temporal sums or peak detection are required.  $\et$ values are sent to the RCT.  A fine-grain bit, used by dedicated minimum bias triggers, is set for each HF trigger region if one or more of the 6 readout towers entering the sum has $\et$ above a programmable threshold. 

The CMS barrel HCAL includes a ``tail catcher'' outside the magnet solenoid (HO).  Signals from this detector are not included in the HCAL trigger primitives, but a technical trigger is generated that requires a single readout segment to be above a threshold. 

\subsubsection{Regional Calorimeter Trigger}
The RCT receives the ECAL and HCAL trigger primitives in 18 electronics crates, each covering one half of the detector in $z$ and $40^{\circ}$ in $\phi$.  The RCT Receiver Cards use LUTs to decompress the HCAL values to $\et$.  The Electron Identification Cards then identify e/$\gamma$ candidates up to $\vert\eta\vert\sim2.5$, using a sliding window algorithm based on $3\times3$ trigger towers, with the central tower of the $3\times3$ window required to have greater $\et$ than its neighbours.  The resulting candidates are classified as isolated or non-isolated, according to the ECAL trigger primitive fine-grain veto information, and the ratio of HCAL to ECAL $\et$, calculated in the RCT \cite{CMS:2008zzk}. 
The $\et$ of the e/$\gamma$ candidate is taken as the sum of that in the central tower and its highest $\et$ neighbour, and a coarse position is assigned as the centre of the $4\times4$ tower region in which the candidate is contained. Each RCT crate transmits up to four isolated and four non-isolated e/$\gamma$ candidates to the Global Calorimeter Trigger (GCT).  

In addition, the Receiver Cards sum the ECAL and HCAL tower $\et$ values over non-overlapping $4\times4$ towers (for barrel and endcaps) and forward these region sums via the Jet Summary card to the GCT.  For each region, the RCT sends a $\tau$-veto bit to the GCT, which indicates that the tower energy is spread out over multiple towers, rather than contained in a small number of contiguous towers, and hence is not consistent with a $\tau$-lepton decay.  The HF trigger regions are forwarded directly to the GCT without processing.

\subsubsection{Global Calorimeter Trigger}
The GCT hardware \cite{Stettler:2006zz} has been completely redesigned since the L1 Trigger Technical Design Report \cite{Dasu:2000ge}, to take advantage of new technologies and improve the robustness of this complex system.  The e/$\gamma$ candidates and region sums are received from the RCT crates by 63 Source Cards, which serialize the data and transmit them to the main GCT crate via optical fibres.  The e/$\gamma$ candidates are received by two Electron Leaf cards, which sort them based on $\et$, and forward the highest four isolated and highest four non-isolated candidates to the GT.  Six Jet Leaf cards process the region sums, finding jets and summingx $\et$.  Two Wheel cards, each covering a half-detector in $z$, then sort and select the jets, and calculate energy sums.  Finally, a single Concentrator card performs final jet sorting and calculates full detector energy sums.  Jet candidates are identified using a $3 \times 3$ sliding window of trigger regions (equivalent to $12 \times 12$ trigger towers, or $1.05 \times 1.05$ in $\eta \times\phi$).  The jet-finder algorithm is described in detail in Ref.~\cite{Iles:2006zz}.  After jets are found, LUTs are used to apply a programmable $\eta$-dependent jet energy scale correction.  Jets found with $|\eta| > 3.0$ are classified as forward jets.  Those found with $|\eta| < 3.0$ are classified as central or $\tau$, depending on the OR of the nine $\tau$-veto bits associated with the 9 regions in the $3 \times 3$ window.  The GCT also calculates total and missing $\et$ from the trigger regions, and total and missing $H_T$.  The total $H_T$ is the scalar sum of $\et$ identified in jets, and missing $\et$ is the corresponding vector sum in the $x-y$ plane.  Finally, minimum-bias trigger quantities are formed by summing  $\et$ in rings around the beampipe in the HF calorimeter (for $4 < | \eta | < 4.5$ and $4.5 < | \eta | < 5$), and by counting fine-grain bits set by the HF TPG.  The four highest $\et$ jets in each of the central, $\tau$ and forward categories are sent to the GT, along with $\et^{total}$, $\et^{miss}$, $H_T^{total}$, $H_T^{miss}$ and the minimum-bias quantities.  The GCT transmits all input and output data to the DAQ for each triggered event, to be used for diagnostics, monitoring and HLT regions of interest.

\subsection{Global Trigger}
The main task of the Global Trigger is to 
reject or to accept events for readout and further processing by the
high-level trigger. Before performing trigger algorithm calculations,
it has to first receive and synchronise the muon and calorimeter input data. This task is achieved by several Pipelined Synchronizing Buffer (PSB) cards. The data are then transmitted to the Global Trigger Logic (GTL) board. This unit is programmed to provide a menu of up to 128 algorithms, which can transform logical combinations of L1 trigger objects (muons, jets, e/$\gamma$, calorimeter transverse energy sum, etc.) with selection criteria (energy/momentum thresholds, etc.) into decision bits. These bits can be enabled to contribute to a final OR of decisions which determines whether the data are read out. In addition, a special PSB receives up to 64 simple on/off signals, called technical triggers, that can be added to contribute to the final OR \cite{Jeitler:2006zz}. Random triggers can also be generated using a linear congruential random number generator. Input data for 3 time samples around the triggered event are read out by the GT Front-End Module (GTFE). 
The GT boards are housed in a single crate, which also contains the GMT and the TCS. 

Besides combining and propagating triggers from subdetectors, the GT provides a throttling mechanism to assure that all triggered events can be completely recorded by the DAQ system. Part of this mechanism is the application of programmable trigger rules, which prevent accumulation of triggers in short time intervals. The rules used are: no more than 1 trigger in 3 BX's, 2 in 25, 3 in 100, 4 in 240. Moreover, front-end buffers of subdetectors can signal to the GT that they are filling up, which results in the GT interrupting trigger activity until the buffers are emptied and the flag removed.  Counters in the GT record the overall trigger rate and the individual rate of each trigger algorithm and technical trigger, as well as dead time counters that record the amount of time during which triggers were inhibited.

\subsection{Trigger Software and Operation}
The trigger system is controlled and monitored centrally, using the Trigger Supervisor \cite{MagransdeAbril:2005dk} and XDAQ \cite{Gutleber:2003cd} software frameworks. The configuration of the trigger electronics is managed by the Trigger Supervisor, using predefined configuration conditions stored in a database. At run start, the shift personnel are able to choose the configuration of each subsystem from a list of keys provided by subsystem experts, allowing flexibility in the trigger configuration whilst also ensuring reproducibility and reducing possibility for human error.  The shift personnel are also able to enable, mask, or prescale individual trigger algorithms, providing robustness against unexpected detector conditions. The configuration data and trigger masks used for each run are recorded in online databases and stored offline for use in offline analysis, such as validation of the trigger operation using emulators.  During the run, shift personnel monitor the trigger system through direct monitoring of system status, via hardware registers and the GT counters described above, and through data quality monitoring histograms of the actual data recorded.

%% file: CRAFTTriggers.tex
\section{L1 Trigger during CRAFT and LHC Single Beam Operations}  %
\label{sec:crafttriggers}

In this section, the trigger operation during CRAFT and LHC single beam periods is described, including the hardware that was operational and how the trigger was configured for cosmic rays, beam splash events, and single circulating beams.

\subsection{Muon Triggers}

The muon trigger systems are designed to identify muons originating from
the interaction point (IP) with high acceptance and efficiency. However, cosmic ray muons
arrive from many directions and the vast majority that traverse the detector do not come close to the IP.  As a result, the detector acceptance and/or the track segment matching
efficiency is not optimal for cosmic rays. During CRAFT, the muon trigger configuration
was adjusted to give the highest possible rate of cosmic ray muons.

Unlike muons from bunched beam interactions, cosmic ray muons arrive
uniformly distributed in time.  The DT electronics devoted to the
bunch crossing identification requires a fine synchronisation to the
phase of LHC collisions. Cosmic ray muons arriving at a marginal time with
respect to the optimal phase can be detected as lower quality,
or out of time trigger primitives.  Segments
reconstructed using a single superlayer, called uncorrelated triggers,
were allowed only if of H type (Section \ref{sec:DTtrigger}) and
confirmed by a coincidence with the trigger primitives from the
$\theta$-view. Details of the DT trigger primitives configuration
and performance can be found in \cite{CFT-09-022}. 

To improve the cosmic ray muon acceptance, the DTTF extrapolation mechanism was relaxed.  A track was generated if a muon in one sector, or crossing two neighbouring sectors, produced at least two trigger primitives at the same bunch-crossing (BX) in two different stations - with no requirements on their position or direction.  This configuration is referred to as ``open LUTs'', while the configuration used for muons originating from LHC collisions is known as ``closed LUTs''.  A consequence of this was that no $\pt$ assignment to DTTF track candidates was 
possible. 

Five DTTF modules were not operational and thus masked (7\% of
the system) and the internal connections to allow track finding across
sector boundaries were installed but not commissioned. Finally, the
link system connecting the $\theta$-view output of the DT  trigger primitive generators was not yet commissioned; as a consequence the DTTF system could only assign low-resolution $\eta$ values to track candidates.

Similarly, to improve the cosmic muon acceptance, the CSCTF was operated in a mode where a muon candidate is generated from a single trigger primitive, and assigned quality 0.  The CSC trigger primitives were formed using standard LHC collision pattern templates. This was performed in addition to the regular mode of operation, where muon candidates are generated from several matching trigger primitives, and assigned higher quality codes.  The halo muon algorithm was operational during CRAFT and first beam.  DT trigger primitives were not yet included in the CSCTF track finding in the region of the DT-CSC overlap.

For the RPC system, triggers were supplied only by the barrel.  The RPC Pattern Comparator trigger
electronics were fully installed and functional.  The hit patterns that can be identified by the RPC trigger are constrained by the connectivity between RPC strips and the Pattern Comparator.  The strips that can be compared to a given pattern template are arranged in cones radiating from the IP.  To achieve good acceptance for cosmic muons, the Pattern Comparator was programmed with patterns that produce a muon candidate from the logical OR of all strips in each cone.  Measurement of muon sign and $\pt$ is not possible with these pattern template. In addition to this, the coincidence requirements were loosened by allowing the coincidence of 3 out of 6 detector layers in the barrel. Muon candidates were assigned a quality value based solely on the number of planes that fired (0 to 3, corresponding to 3 to 6 firing planes).  The standard ``ghost busting'' algorithm \cite{CMS-2002/022} was applied to prevent single muons producing more than one candidate in logical cones that overlap in space.  A single candidate was selected on the basis of higher quality followed by higher $\phi$ value.

The main function of the Global Muon Trigger during CRAFT was to synchronise triggers from the three muon systems. Another important function was to record  L1 muon track candidates in a unified format in the event data such that the performance of individual muon trigger systems could be conveniently analyzed.  Other functions, like smart quality assignment and cancellation or merging of duplicate candidates (described in Ref.~\cite{Sakulin:2005qx}) were applied but not actively used.

\subsection{Calorimeter Triggers} 

During CRAFT, the calorimeter triggers were configured to trigger on instrumental noise and energy deposited by cosmic rays.  Only the ECAL barrel was used to provide e/$\gamma$ triggers, since the ECAL endcap trigger electronics were not installed yet.  The ECAL trigger primitive transverse energy was sent to the RCT on a linear scale, with a least significant bit (LSB) corresponding to 250 MeV, the maximum possible value being 63.75~GeV.  In order to minimize the contribution from noise, trigger primitives below 750~MeV were suppressed. This value corresponds to between three and four times the noise.  HCAL was operated in the standard way for cosmic and LHC runs.  For most of CRAFT, all three calorimeter parts (barrel, endcap and forward) were active. 
The barrel and endcap HCAL trigger primitives were sent to the RCT using an 8-bit non-linear scale in energy. The HF trigger primitives were sent on a linear transverse energy scale, with the LSB of 250 MeV.

The full Regional Calorimeter Trigger was used during CRAFT. The default RCT configuration used during cosmic ray runs produced e/$\gamma$ candidates from ECAL barrel trigger primitives, and region sums from the sum of ECAL (barrel) and HCAL (barrel, endcap and forward) trigger primitives. Noisy or absent ECAL and HCAL channels were masked in the RCT LUTs.  A total of 8.5\% of ECAL trigger towers were masked; the number has since been reduced through finer-granularity (crystal) masks.  Less than 1\% of HCAL channels were masked.  The RCT input LUTs were generated from the scales provided by ECAL and HCAL, to give linear $\et$ with a 250 MeV LSB.  The isolation, ``fine-grain'', and H/E criteria were ignored in the production of e/$\gamma$ candidates.  The e/$\gamma$ candidate $\et$, which is the sum of $\et$ in a pair of contiguous ECAL trigger towers, was transmitted to GCT (and thence to GT) on a linear scale with a LSB of 500~MeV.  This algorithm is referred to as an e/$\gamma$ trigger, although the requirement is simply an ECAL energy deposit above a configurable cut.


The GCT jet and e/$\gamma$ trigger algorithms were enabled during CRAFT.  The global energy sum and minimum-bias algorithms had not been commissioned at that time.  No jet energy corrections were applied to the jet $\et$.  The output jet $\et$ scale was chosen to be linear with 2~GeV steps from zero to 126~GeV.

\subsection{Global Triggers}
\label{sec:craft-gt}

During cosmics data taking, only the simplest single object algorithms were enabled in the GT, with no threshold for muons and the lowest energy threshold allowed by the noise rate for calorimeter objects.  The trigger algorithms enabled in the Global Trigger during CRAFT were :
\begin{itemize}
\item{L1\_SingleMuOpen : any muon candidate from any sub-detector with any $\pt$}
\item{L1\_SingleEG1 : single e/$\gamma$ candidate with $\et > 1 \GeV$}
\item{L1\_SingleJet10 : single jet candidate with $\et > 10 \GeV$}
\end{itemize}

Since the outer section of the hadron calorimeter (placed behind the magnet coil in the central part of CMS), called the HO, does not contribute to the jet trigger, a self-triggering HO technical trigger was introduced to provide insight into the noise behavior of this device, and to investigate timing.

The average rate of each of these triggers during CRAFT is given in Table~\ref{tab:rates}, along with the total overall rate.  The DT in coincidence with the RPC trigger gave 120~Hz rate, and the CSC with the RPC trigger yielded 20~Hz.  The rate of coincidence between calorimeter triggers and the single open muon trigger was: 0.5~Hz for the single e/$\gamma$ trigger, 0.9~Hz for the single jet trigger, and 0.15~Hz for the HO technical trigger.


\begin{table}[h]
  \begin{center}
    \caption{Nominal L1 trigger rates during CRAFT.}
    \label{tab:rates}
    \begin{tabular}{lcccccccc} \hline
      Trigger & Rate (Hz) \\ \hline
      L1\_SingleMuOpen & 300 \\    
      L1\_SingleMuOpen (DT only) & 240 \\    
      L1\_SingleMuOpen (RPC only) & 140 \\    
      L1\_SingleMuOpen (CSC only) & 60 \\    
      L1\_SingleEG1 & 23 \\      
      L1\_SingleJet10 & 140 \\       
      HO technical trigger & 14 \\ \hline
      Total & 475 \\ \hline
    \end{tabular}
  \end{center}
\end{table}

Complementary to these ``physics'' triggers, the GT system routinely provided calibration triggers with a rate of 100~Hz preceded by control signals used to fire calibration systems in various subdetectors. During such a calibration sequence all ``physics'' triggers were disabled for about 5 $\mu s$, introducing a dead-time of 3\%.  Note that such calibration triggers occur during the LHC abort gap when running with beam, so this does not imply a real deadtime for LHC physics triggers. Finally, low rate Poissonian distributed random triggers were added to the trigger mix.

During LHC single beam operations in 2008, the GT was programmed to trigger on CSC beam halo muon candidates, as well as technical triggers from LHC beam pickup monitors (BPTX) and the HF detector.  The BPTX system uses signals from electrostatic devices, located on the beam pipe 175~m from CMS in each direction, that produce signals synchronous with a passing proton bunch.  For single circulating beams, a technical trigger is generated from the BPTX upstream of CMS.  The HF technical trigger required a single tower with energy above a threshold of 15 GeV.

%% file: Synchronisation.tex
\section{Synchronisation}
\label{sec:synchronisation}

For efficient triggering, all parts of the CMS detector must produce trigger signals synchronously for the same event.  For LHC collisions, synchronisation to the same BX is relatively easily achieved.  Synchronisation for cosmic ray muon detection, however, was more challenging.  Cosmic ray muons arrive asynchronously, from all directions, and the time of flight for a relativistic particle to traverse the detector is much greater than the clock period.  Despite this, the detector was well synchronised, as described in this Section.

\subsection{DT Synchronisation}

\begin{figure}[htbp]
\begin{center}
\includegraphics[width=0.93\textwidth]{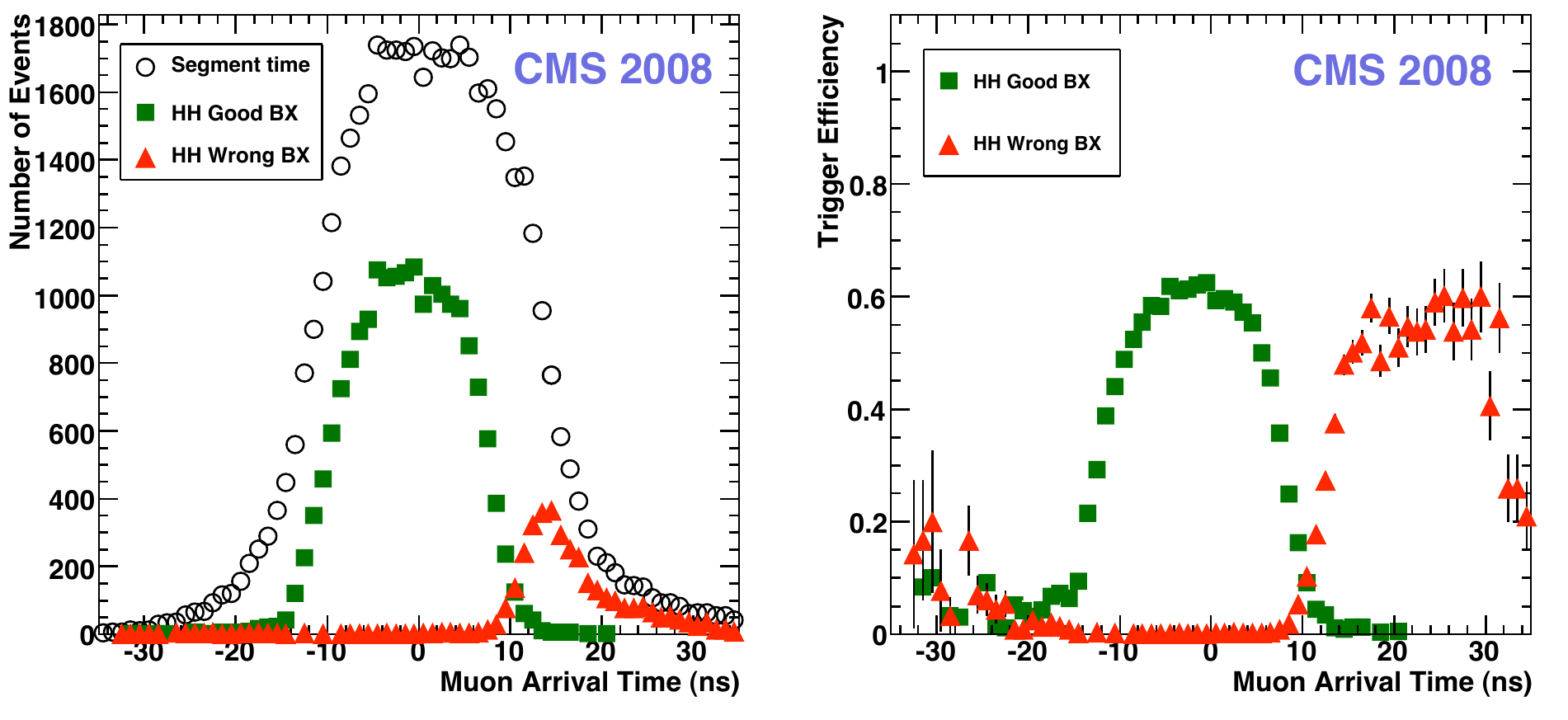} 
\caption{Distribution of the arrival time of muon
  track segments ({\em left, open circles}), and for segments which also have an associated HH
  local trigger in the station at the correct ({\em squares}) and
  neighbour BX ({\em triangles}). 
  Ratio between the HH-triggered and all muon track
  segments corresponding to the HH trigger efficiency ({\em right}),
  shown for the correct ({\em squares}) and neighbour BX ({\em triangles}).}
\label{fig:DTTPGEff}
\end{center}
\end{figure}

In order to provide a cosmic ray muon trigger, the behavior of the DT trigger primitives with respect to 
the arrival time of the muons was investigated.  Cosmic rays arrive with a flat timing distribution, 
while the DT trigger was designed to identify the bunch crossing of muons from 
beam collisions. The time at which the muon crosses the chamber is
computed as an additional parameter of the track segment object
delivered by the offline local reconstruction \cite{CFT-09-012}, with
a resolution of $\sim$~3~ns. 
In the left plot of Fig.~\ref{fig:DTTPGEff}, the time distribution of muons in a given
chamber is shown, as well as the tracks triggered with HH quality code, at 
the ``good'' and neighbour BX respectively. HH triggers can provide a precise 
BX assignment, and their efficiency as a function of time was shown to be a 
precise indicator of the trigger synchronisation in dedicated test-beams with 
bunched muons \cite{Aldaya:2006cp}. The ratio of the inner and outer curves 
from this plot corresponds to the efficiency for HH triggering, and is shown in 
Fig.~\ref{fig:DTTPGEff} ({\em right}). The HH efficiency  plateau in both BX
curves is about 60\%, as expected for a well timed-in trigger system, 
while the HH efficiency decreases in the region between the two 
consecutive BXs. In this region, the overall efficiency is recovered 
by lower quality triggers, whose BX identification power is lower. 
This analysis of the trigger performance with respect to the muon track 
time has been developed as a tool in the initial process of fine 
synchronisation of the DT trigger to the LHC bunch crossing time \cite{CFT-09-025}.

The local trigger synchronisation was specifically adjusted for
cosmic ray triggers. Cosmic ray muons generally cross the detector from top to bottom, so 
the system was synchronised to take this into account. The trigger latency of the chambers of the
top sectors was increased using configurable pipelines in
the Sector Collector modules, accounting for a maximum time of flight to
the bottom chambers of about 50~ns, or 2~BX. Thus, when a single muon crosses two sides of the
detector, two segments from different detector regions are sent to the DTTF in the
majority of cases for the same BX, so that the DTTF system sends two muon
track candidates to the GMT at the same BX. 

\begin{figure}[htbp]
\begin{center}
\includegraphics[angle=90, width=0.5\textwidth]{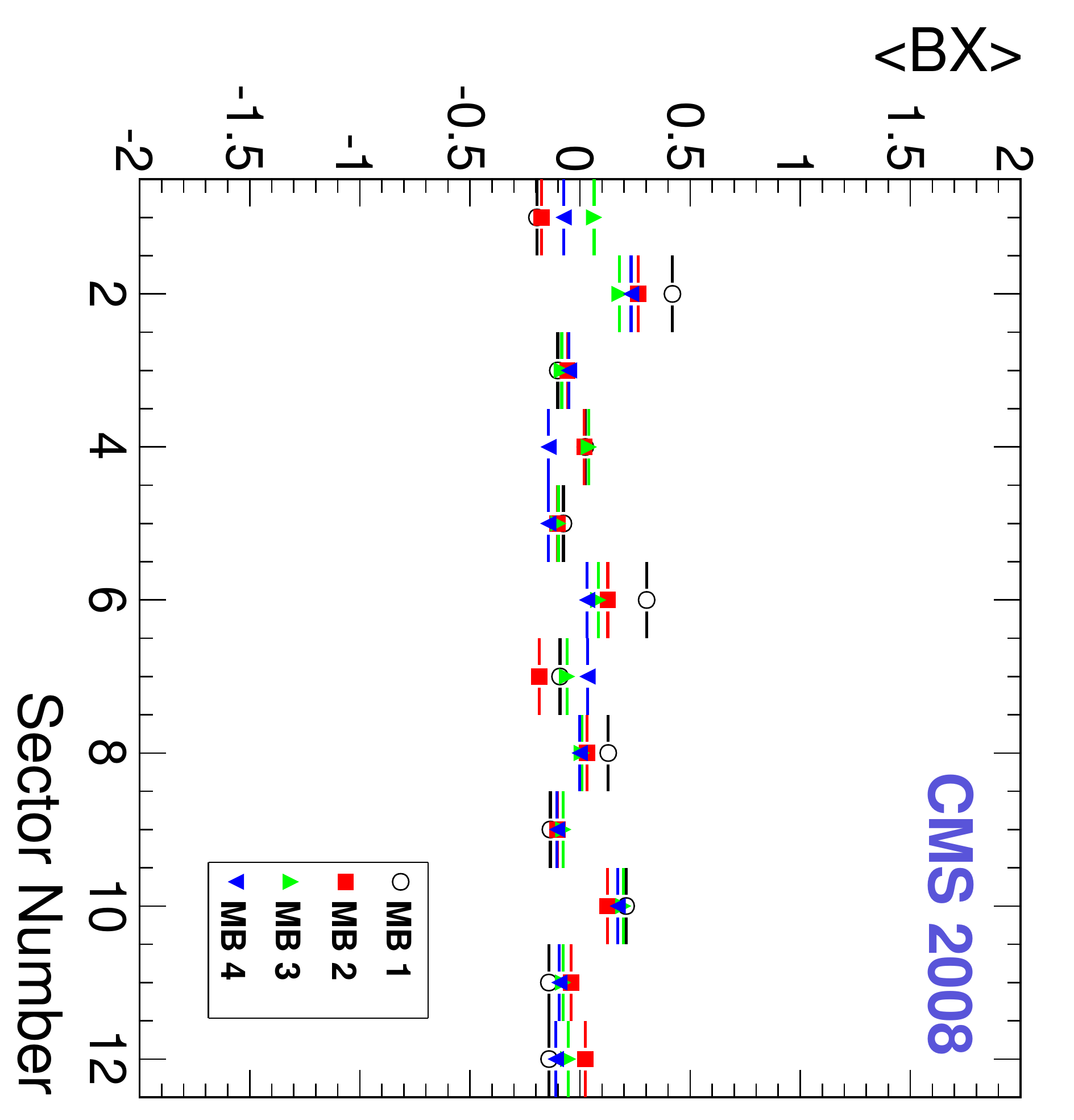}
\caption{Mean BX from DT chambers of wheel 0.}
\label{fig:DTTPGBX}
\end{center}
\end{figure}

The synchronisation parameters were obtained
with dedicated runs, by means of checking the bunch crossing
distribution of trigger segments from different detector regions with respect
to a reference DT sector. Fig.~\ref{fig:DTTPGBX} shows the resulting mean BX 
for each of the chambers in wheel 0, which are compatible to about one third of a BX.

\subsection{CSC Synchronisation}

\begin{figure}[!hbt]
  \centering
  \includegraphics[angle=90,width=0.47\textwidth]{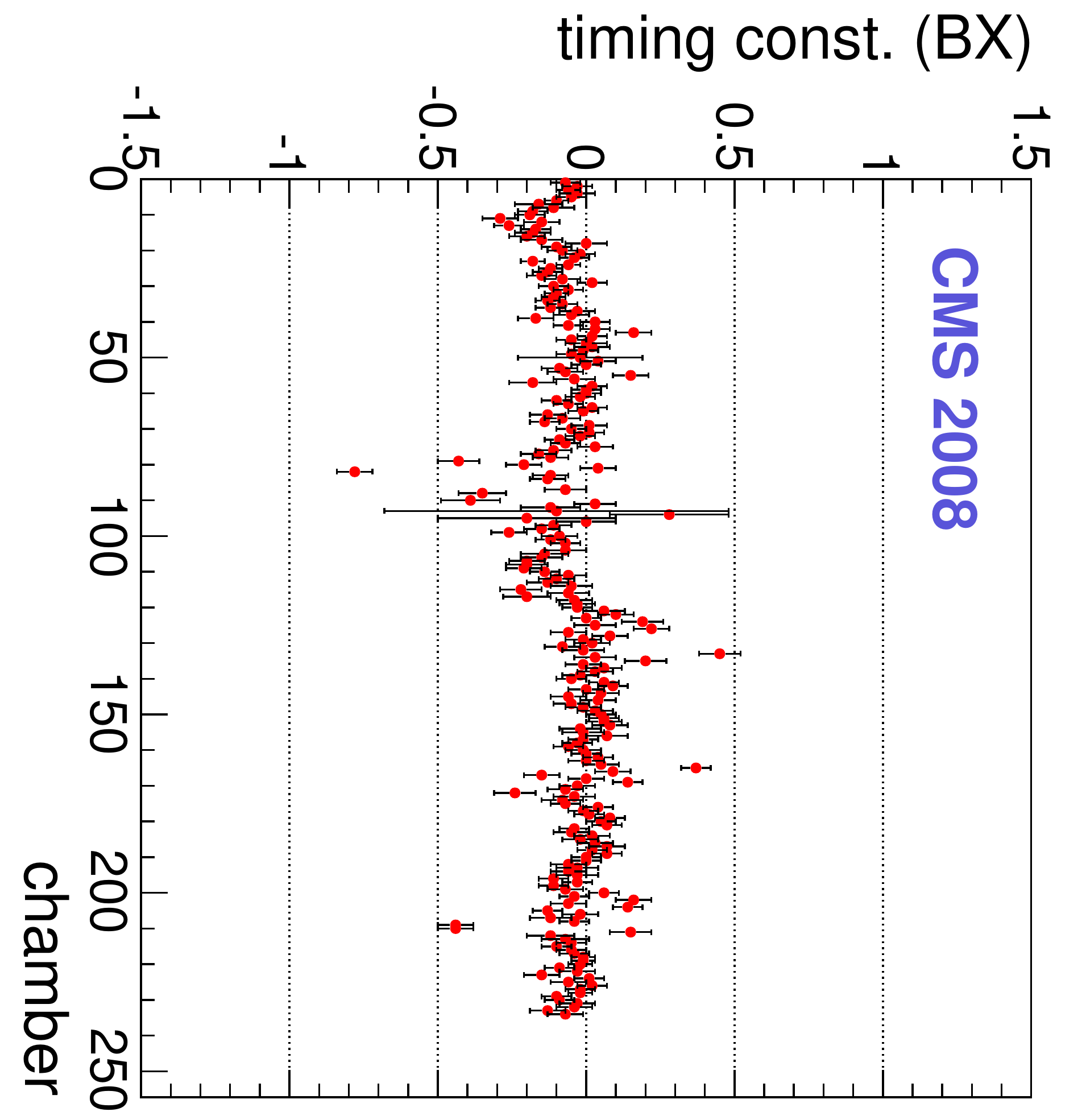}  
  \includegraphics[angle=90,width=0.47\textwidth]{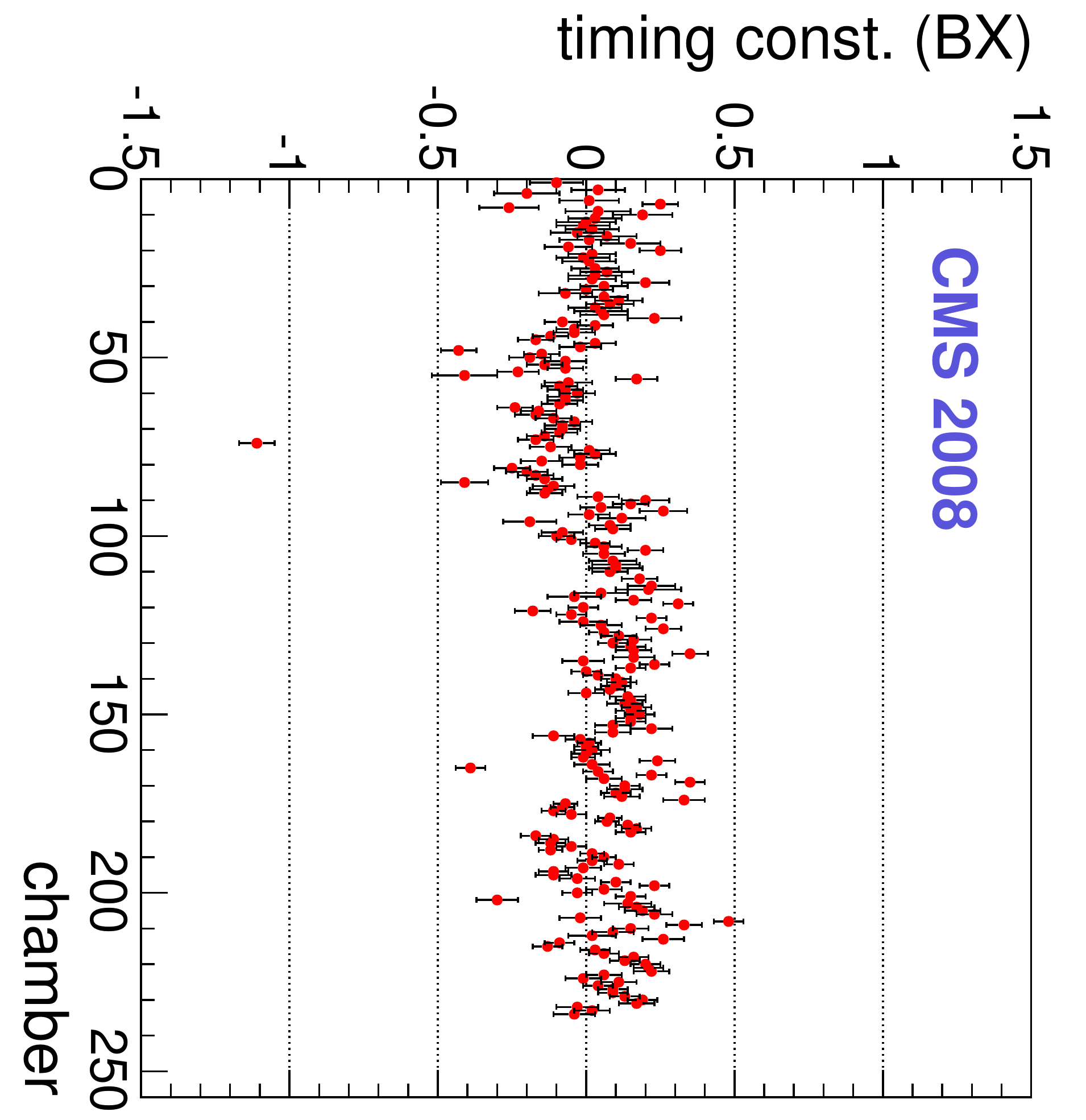}  
  \caption{CSC inter-chamber timing constants from CRAFT for the plus ({\em left}) and minus ({\em right}) endcaps
    after adjustment of delays based on CRAFT analysis. }
  \label{fig:craftconst}
\end{figure}

For CRAFT, a coarse delay of 1~BX was introduced for the upper chambers with respect to the bottom. Precise inter-chamber synchronisation of the CSC detector is achieved by measuring the 
arrival time at the CSCTF of trigger primitives from the same event, on a 
chamber by chamber basis.  For each endcap, the mean relative time of signal arrival in each chamber is measured relative to a single reference chamber.  These measurements are used to construct a 
global $\chi^2$, minimization of which can yield optimal timing constants for each 
chamber simultaneously.  The mean arrival time of signals from all CSC chambers, 
after adjustment using this method are shown in Fig. \ref{fig:craftconst}.  They 
indicate the precision of CSC inter-chamber synchronisation achieved during CRAFT to be around 0.15~BX.


\subsection{RPC Synchronisation}

The RPC Link Boards were synchronised such that the cosmic ray
muons crossing top and bottom parts of the detector produce triggers
in the same BX.
The initial settings were calculated from fibre and cable lengths and
assuming the time of flight to the chambers as the time to cover the
distance from the outermost layer of the top barrel sector for a
straight, vertical track. The distance between the chambers was
computed from the detector geometry. Adjustments to these settings
were obtained from dedicated data-taking runs, in which only the RPC
trigger was enabled. 

Distributions of the muon hit BX with respect to the BX of the trigger were produced for each Link Board and corrected timing constants were calculated. The different parts of the detector were synchronised in consecutive steps. First, the corrections for the Link Board of the bottom sector of wheel 0 were found, based on data in which triggers coming only from that sector were enabled. In the next run, again only triggers from the bottom sector of wheel 0 were enabled, and used as a reference for the top sectors of wheels -1, 0, 1.  In a similar way, the other parts of the detector were synchronised; the corrections for the bottom sectors of wheels -1, 0, 1 were calculated with respect to the muons triggered in the top sector of wheel 0, the top sectors of wheels -2 and 2 were synchronised with respect to the bottom sector of wheels -1 and 1, and so on up to the endcap.  More details can be found in Ref.~\cite{CFT-09-010}.

\subsection{Calorimeter Synchronisation}

The ECAL trigger primitives were synchronised by measuring the signal arrival time in DT-triggered events.  Ten time samples were read out, and the signal peak was required to be in the 6th 25~ns time sample.  Initial delays for each channel were set according to fibre length from the detector.  No additional channel by channel synchronisation was required.

The sampling phase of each HCAL channel was individually adjusted to compensate for differences in particle time of flight from the interaction point, and signal propagation delays in the scintillator tile fibres.  These delays were determined in the test beam, and validated using the beam splash events.  Laser test pulses distributed to most portions of the detector were also used to check the synchronisation.  These methods demonstrate that the HCAL sampling had an RMS variation of 2~ns during CRAFT.  In addition to the uniformity of sampling phase, it is possible that the digitized samples shift latency by 1~BX during their transfer from the front ends to the HCAL trigger/readout cards.  During CRAFT, the system of optical latency control was still imperfect, and about 0.2\% of the data was shifted by 1~BX.  Measures have been taken since CRAFT to correct this.

\subsection{Global Synchronisation}

\input{gtgmt/SynchronizationWithCosmics.tex}

\subsection{Synchronisation with LHC Beam}

\input{gtgmt/SynchronizationWithBeam.tex}

%% file: gtgmt/SynchronizationWithCosmics.tex
\begin{figure}[htbp]
\begin{center}
\includegraphics[angle=90, width=0.99\textwidth]{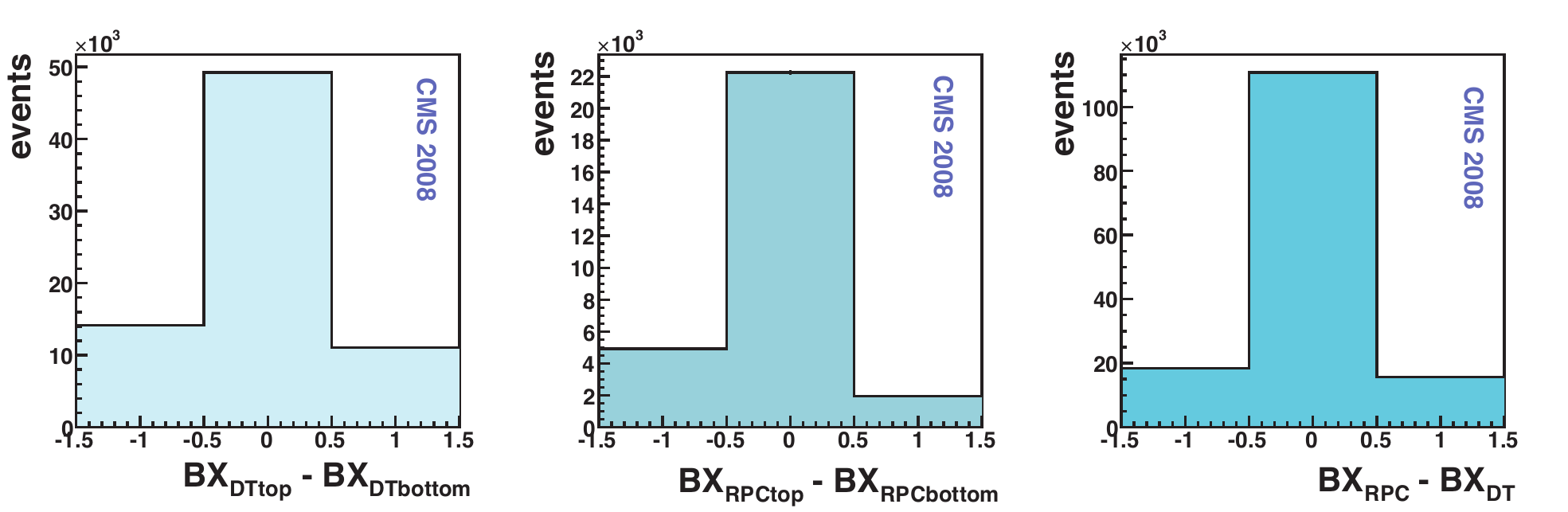}
\caption{Time differences at the GMT in terms of BX between L1 muon
candidates - created in most cases by the same cosmic ray muon - from the top half and the bottom half of the DT system ({\em left}) and the RPC ({\em middle}) and between the RPC and DT system ({\em right}). The majority of the signals are synchronised and the skew at the clock edges is balanced.}
\label{figSYNCDTRPC}
\end{center}
\end{figure}
\begin{figure}[htbp]
\begin{center}
\includegraphics[angle=90, width=0.66\textwidth]{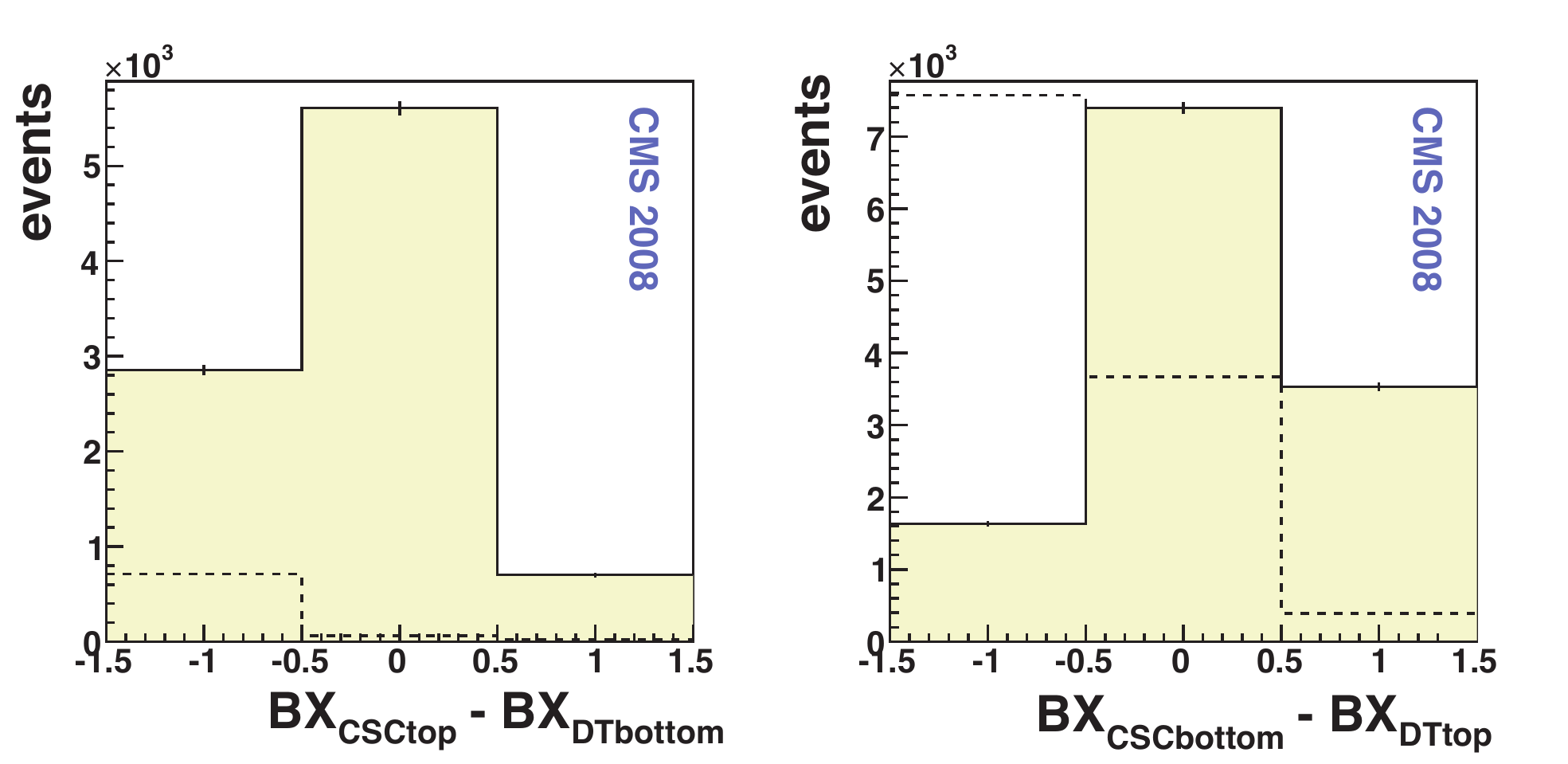}
\caption{Time differences at the GMT in terms of BX between L1 muon candidates -
created in most cases by the same cosmic ray muon - from top half of the CSC and bottom part of the DT system ({\em left}), and from bottom part of the CSC and top part of the DT system ({\em right}).  The dashed line shows the situation before the modification to CSC timing (top delayed by 2~BX and bottom delayed by 1~BX) mentioned in the text.}
\label{figSYNCDTCSC}
\end{center}
\end{figure}

After each of the three muon systems was internally synchronised, it
was necessary to make sure that the signal created by the same muon in
different detectors enters the GMT in the same clock cycle. With
cosmic rays, this is only possible to a limited extent, depending on
the level of internal subdetector synchronisation. Using latency
calculations of upstream trigger pipelines, it was possible to
determine rough delay settings at the GMT inputs. Fine tuning of these
delays was then performed using the cosmic ray data. One method is to
measure the signal arrival time from a particular subdetector with
respect to triggers from another.  A direct comparison is possible
using the readout of the GMT, which records all input muon candidates
and reads 3 consecutive clock cycles centered at the
trigger. Figs.~\ref{figSYNCDTRPC} and \ref{figSYNCDTCSC} show that in
most cases L1 muon candidates from different muon systems, induced by
the same cosmic ray muon, arrive at the same BX. Occasional difference by
1 BX is unavoidable due to the fact that cosmic rays are asynchronous
to the clock of the experiment and because the relative
synchronisation between different detector parts obtained with cosmic
rays has a finite precision of several ns.  The CSC timing was 
adjusted during CRAFT, to improve the synchronisation.  The top half of
CSC was delayed by 2~BX and the bottom half was delayed by 1~BX.  The 
dashed line in Fig.~\ref{figSYNCDTCSC} shows the situation before this adjustment, 
and the solid line shows the situation afterwards.

\begin{figure}[htbp]
\begin{center}
\includegraphics[angle=90, width=0.66\textwidth]{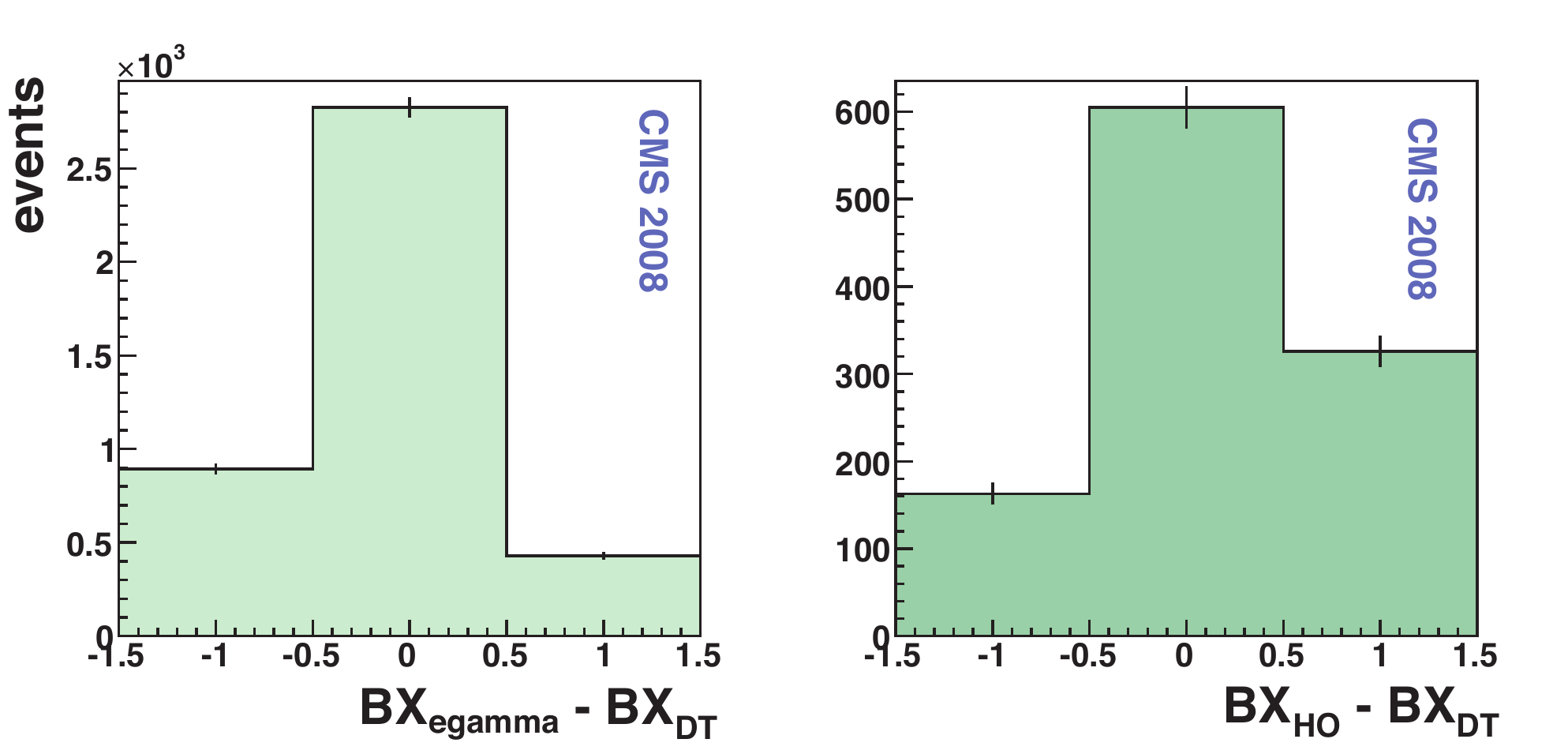}
\includegraphics[angle=90, width=0.33\textwidth]{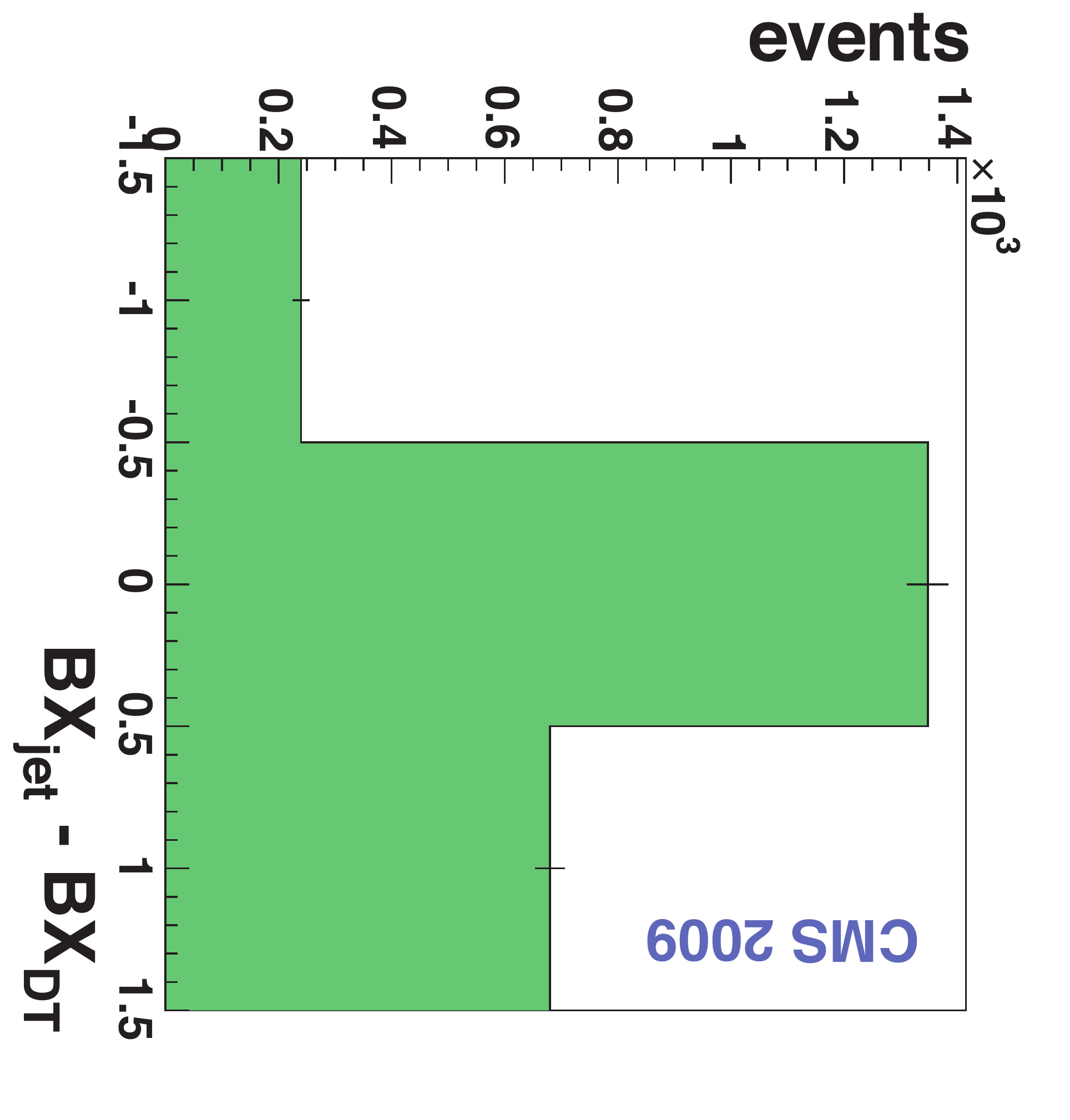}
\caption{Time differences at the GT input between e/$\gamma$ triggers
({\em left}), HO technical triggers ({\em middle}), and jet triggers
({\em right}, 2009 data), with respect to L1 muon candidates from the DT system. }
\label{figSYNCDTcalo}
\end{center}
\end{figure}

Similarly, delays of calorimeter trigger inputs to the Global Trigger
have been adjusted to provide the highest coincidence rate above the
noise continuum. In this way, the e/$\gamma$ trigger and technical
triggers from HCAL were adjusted.
The timing of the jet trigger, which was in the early stage of commissioning during CRAFT, was adjusted shortly after the CRAFT exercise. Relative timing of calorimeter
triggers with respect to DT muon triggers is shown in
Fig.~\ref{figSYNCDTcalo}.

%

%% file: gtgmt/SynchronizationWithBeam.tex
\begin{figure}[tp]
\begin{minipage}[b]{0.5\textwidth}
    \centering
    \includegraphics[width=0.95\textwidth]{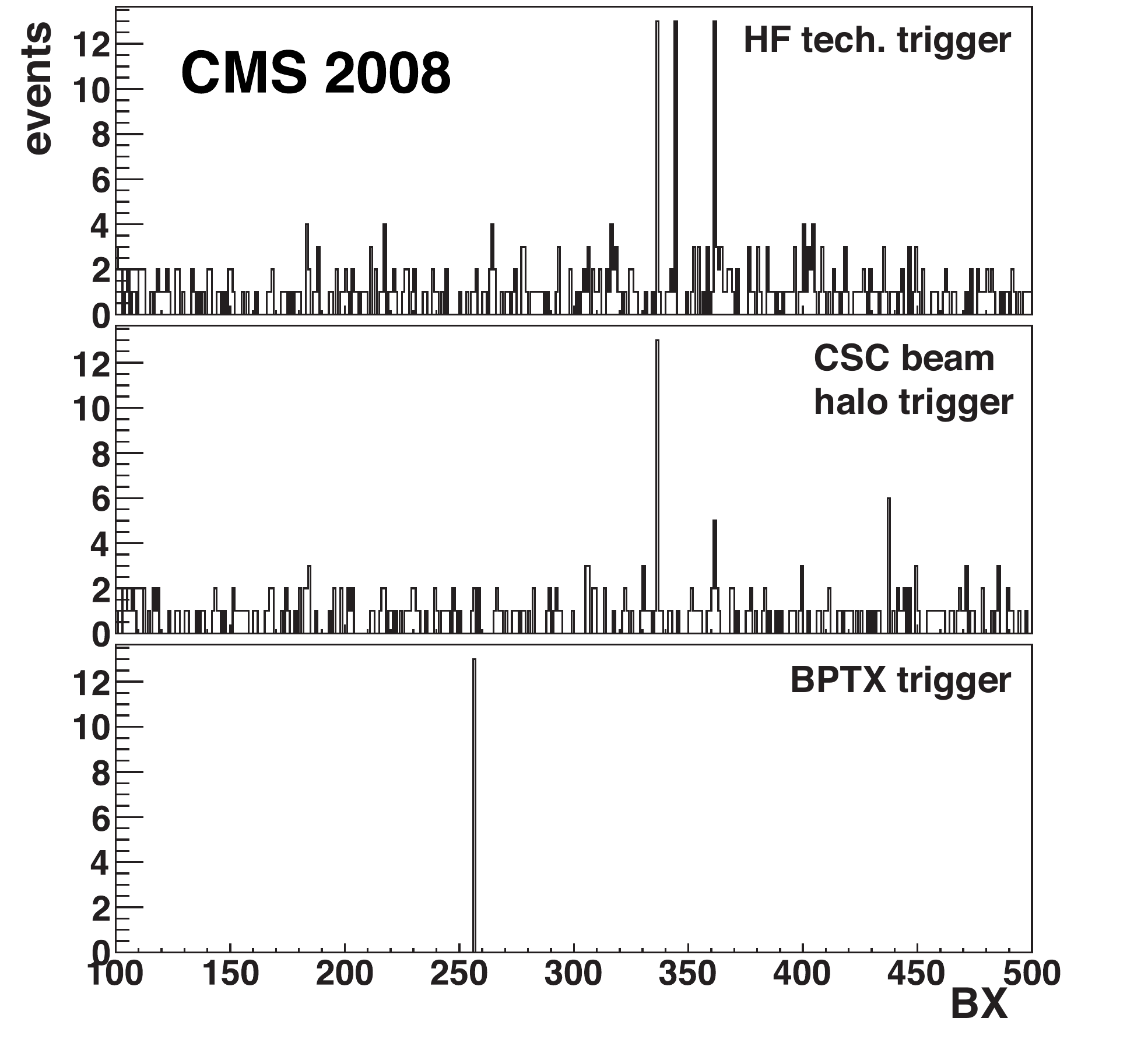}
    \caption{Measurement of the time delay in BX between the BPTX trigger
and previously synchronised CSC beam halo and HF triggers, using beam
splash events.}
    \label{fig:splashsync}
\end{minipage}
\hspace{0.5cm}
\begin{minipage}[b]{0.5\textwidth}
    \centering
    \includegraphics[width=0.95\textwidth]{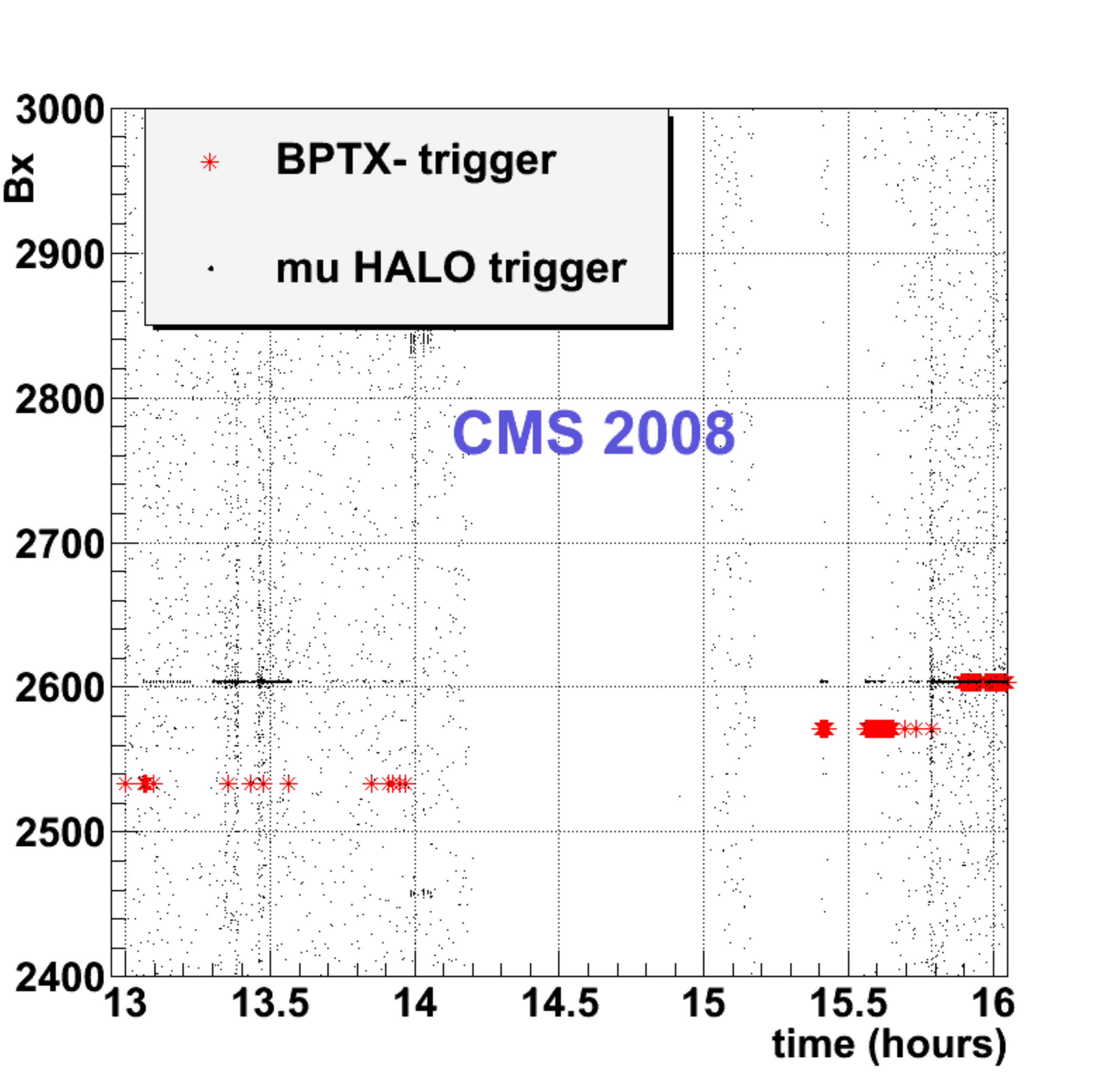}
    \caption{Synchronisation of the BPTX trigger with CSC beam halo trigger during circulating LHC beam.}
    \label{fig:beamsync}
\end{minipage}
\end{figure}

During LHC operations, the BPTX technical triggers were enabled. Their timing with respect to
muon and calorimeter triggers could be tested only in the presence of
the beam. Initial synchronisation was achieved using the ``beam splash'' events. The
beam producing these events was always injected at the same phase with respect to the orbit
signal. Using the muon beam halo
trigger provided by the CSC system and the HF technical trigger, which
both have very low background from cosmic rays and noise (total rate less than
3~Hz), it was possible to see a clear signal in the BX distribution
just after a few beam shots (Fig.~\ref{fig:splashsync}). The BX
distribution is produced by counting LHC clocks (40~MHz) and resetting
the counter by a signal derived from the LHC orbit signal. The
upstream CSC endcap was delayed by 2 BX to provide a trigger in
coincidence with the downstream endcap. This information was then used to delay the BPTX trigger signal
and align them with muon and calorimeter triggers. Satellite peaks in
the HF are due to afterpulses in HF phototubes filtered through
trigger rules. The procedure was repeated also with the circulating
beams.  Fig.~\ref{fig:beamsync} shows the distribution of
the CSC muon halo and BPTX triggers within the LHC orbit, 
as a function of time.  The rate of both triggers increases during
periods when LHC beam is circulating.  As can be seen, the delay associated with the
BPTX trigger was adjusted over the course of several LHC fills, and
brought into synchronisation with the halo trigger.

%% file: EmulatorValidation.tex
\section{Hardware Validation using Emulators}
\label{sec:emulator}

A full bit-level software emulation of the L1 trigger was developed alongside the hardware \cite{Ghete:2009}.  This serves two purposes: to simulate the trigger response in CMS simulation and to monitor the operation of the hardware.  In the latter role, the input to a particular trigger subsystem is read out and used as input to a bit-level software emulation of that subsystem.  The output of the emulator can be compared with the output of the subsystem for each event, to validate the trigger operation.  In this section, the results of such comparisons performed with the CRAFT data are discussed.  These comparisons were run in offline analysis, and in automated online and offline data quality monitoring processes.

One general problem encountered when comparing emulator and hardware processing, is the need to ensure that the emulator is configured in the same way as the hardware.  In future data-taking operations, automated database transfers synchronised with online run control will allow automatic configuration of the emulators running offline, and hence allow fully automated validation of every event stored.  Unfortunately, during CRAFT, the database transfer system had not been fully commissioned, and subsystem configurations changed from run to run.  Therefore, a subset of events has been used to validate each subsystem, rather than the entire CRAFT dataset.  The results, presented below for each subsystem, show that the L1 trigger processing is in good agreement with the software emulation.  Disagreements arise in some subsystems, at the few \% level at most.  Such disagreements generally indicate subtle differences between hardware and emulator algorithms, or hardware problems and have been followed up on since CRAFT.

\subsection{Muon Triggers}

Validation of the DT trigger is complicated by the fact that the DT
trigger primitives are digitised using a different technique and clock phase from that
used for the full granularity readout.  Precision time measurements are made using 
TDC data, whereas the trigger track segment time is calculated using a digital mean timer technique.
Different clock phases are used to eliminate
clock skew between trigger boards and achieve the best synchronisation
for muons coming from LHC collisions.  The difference in clock phase
must be taken into account when emulating the trigger response from
the full detector readout.  For muons with LHC timing, the trigger
primitives can be correctly emulated.  For cosmic ray muons, which have a
flat distribution in time, the emulator cannot be expected to
reproduce muon times that are close to the trigger clock edge. 
Neverthless, a data-emulator comparison was performed for $3 \times 10^6$
events, and 99\% agreement was found in the trigger efficiency as a function
of track position and impact angle. 
This result is consistent with what can be expected from the emulator given the timing issue outlined above.  

The remaining muon trigger subsystems, apart from CSCTF, were validated by emulating outputs from read out input data.  The DTTF validation was performed for 1 Million events in a run that used ``closed LUTs'' (the ``open LUTs'' used for cosmic ray data taking are not emulated) where the muon remains in a single DT sector.  100\% agreement was observed between emulator and data.  For a typical run, the emulated CSC trigger primitives agreed with those in the data in 99.5\% of events.  The remaining 0.5\% were due to a minor firmware error that has since been corrected. The RPC trigger validation showed disagreement between data and emulator in $\sim$2\%  of cases, again for a typical run, coming almost exclusively from muon candidates in particular detector regions.  The GMT validation was performed regularly, with a typical run showing 100\% agreement between data and emulator. 

\subsection{Calorimeter Triggers}

The calorimeter readout includes full granularity ECAL crystal and HCAL tower data, as well as the trigger primitives sent to RCT.  The trigger readout includes e/$\gamma$ candidates and region sums at the output of the RCT, and the e/$\gamma$ and jet candidates at the output of GCT.  Validation of the calorimeter trigger processing therefore involves emulation of :

\begin{itemize}
\item{ECAL trigger primitives from full granularity crystal data;}
\item{HCAL trigger primitives from full granularity HCAL towers;}
\item{RCT e/$\gamma$ candidates and region sums from ECAL trigger primitives in data;}
\item{GCT e/$\gamma$ and jet candidates from RCT e/$\gamma$ candidates and region sums in data.}
\end{itemize}

The ECAL validation was performed on 10 Million events, constituting the bulk of runs where ECAL crystal data was not zero-suppressed.  After accounting for masked channels, agreement was observed between the emulated $\et$ and fine-grain bit and the data in more than 99.9\% of trigger primitives. 

The HCAL validation was performed on over 50 Million events.  Both $\et$ sums and the HF fine-grain bit were compared between emulator and data. The level of disagreement observed between emulator and data was less than $1 \times 10^{-6}$.


The RCT validation takes ECAL and HCAL towers from ECAL/HCAL
readout data and produces emulated e/$\gamma$ candidates, which are then
compared with those read out by the GCT. Disagreements at the level of
a few percent were observed in 2008 due to masked channels that were not
emulated and latency instability in the HCAL data. 
During cosmic ray data-taking in 2009, the comparison is
performed on a regular basis and shows perfect agreement between data
and emulator.

The GCT emulator validation was performed on 20 Million CRAFT events.  Agreement was observed between hardware and emulator for 100\% of e/$\gamma$ candidates.  A small error in the implementation of the jet-finding algorithm was discovered in the comparison of the jet candidates.  In approximately 0.05\% of cases a jet was incorrectly labeled as passing the $\tau$-lepton veto.  This has no impact on the efficiency and resolution studies presented later in this paper, and the firmware has since been corrected.

%% file: dt/DTPerformance.tex
\section{Drift-Tube Trigger Performance}
\label{sec:dtperformance}

\begin{figure}[htbp]
\begin{center}
\includegraphics[width=0.47\textwidth]{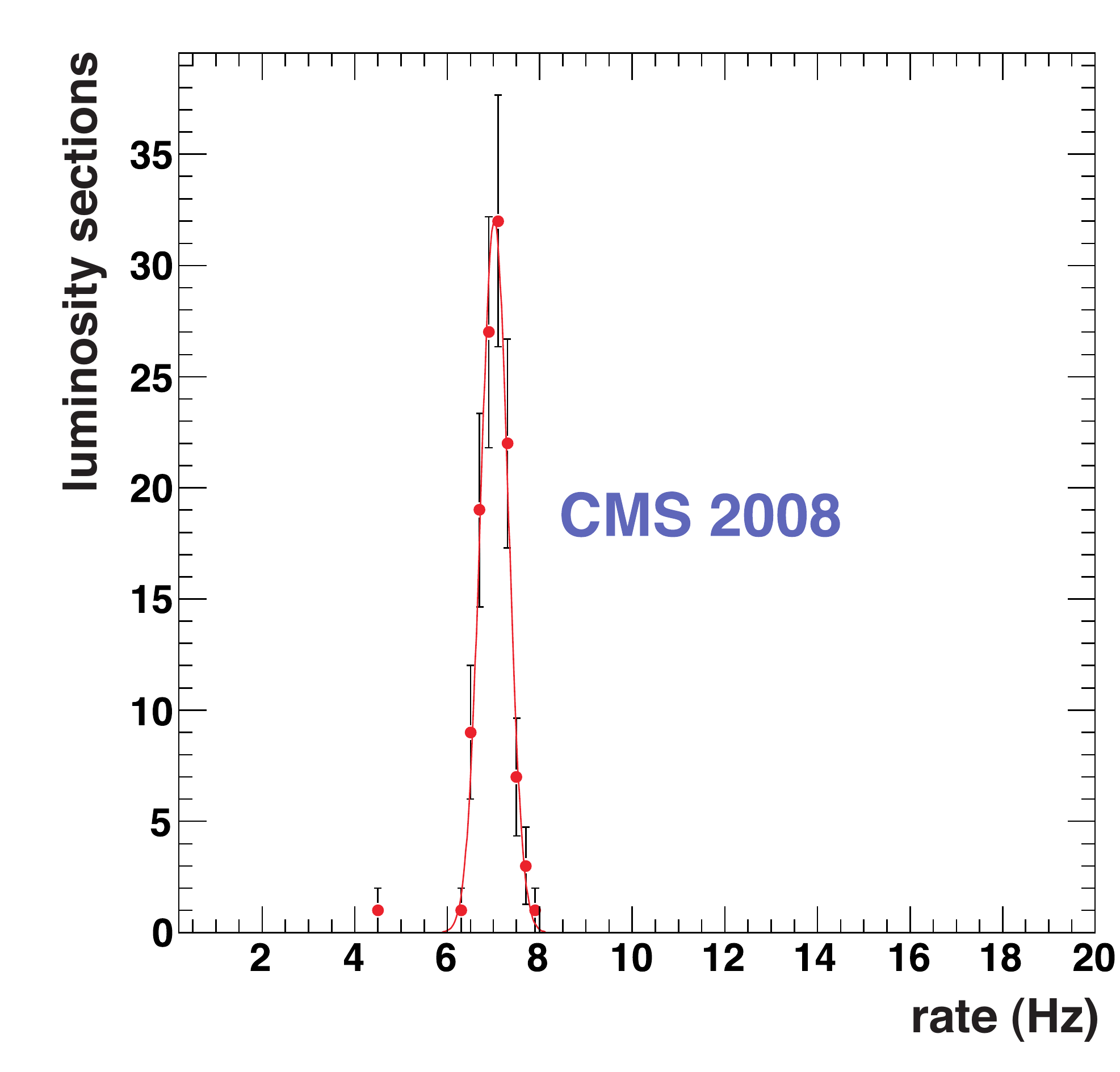}
\includegraphics[width=0.47\textwidth]{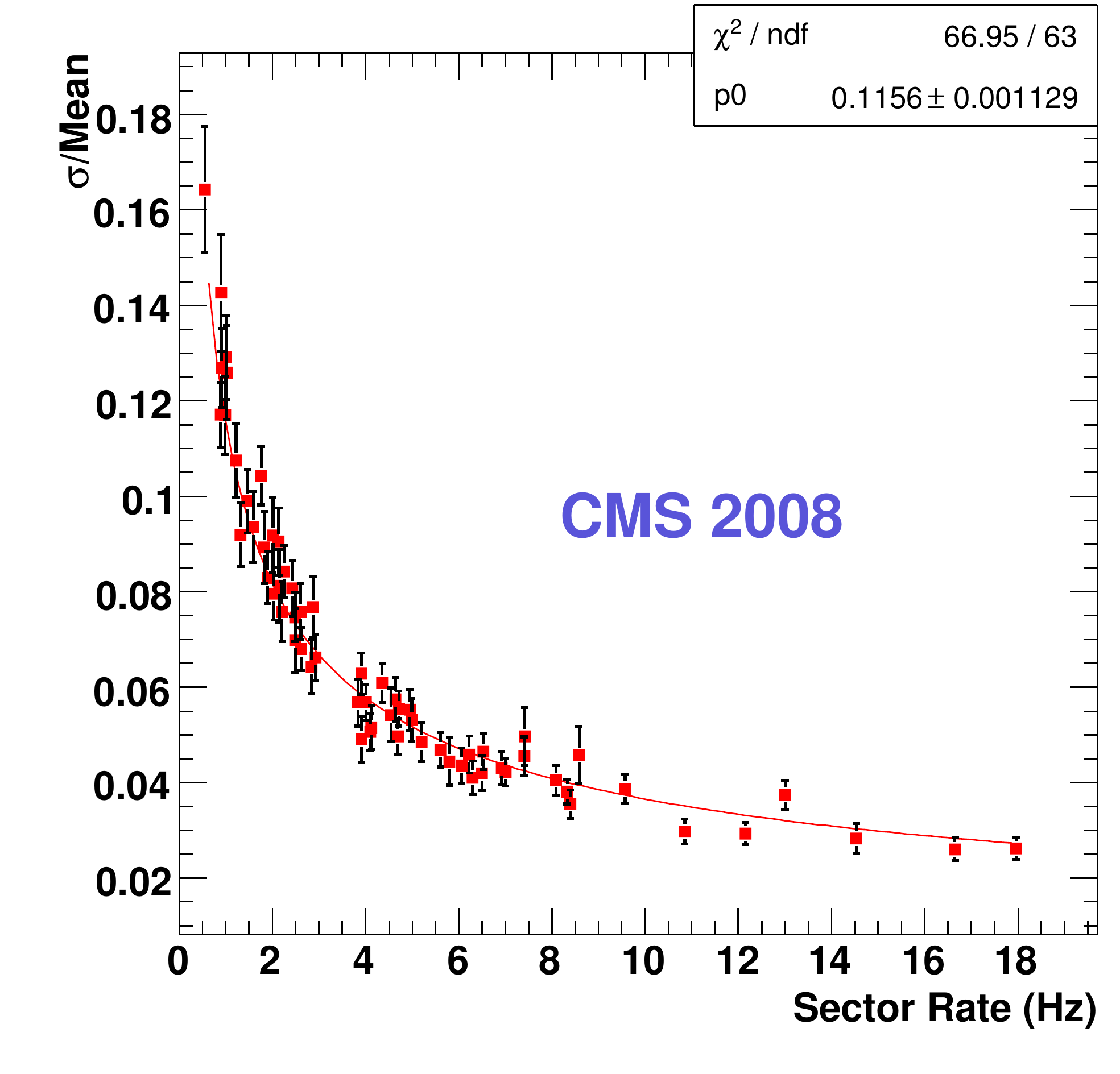}
\caption{The rate distribution of a particular sector of wheel 0 ({\em left}).  Fit to the $\sigma/mean$ of all measured sector rates ({\em right}).}
\label{fig:DTTFRates}
\end{center}
\end{figure}

Operating in ``open LUTs'' mode, the DTTF delivered about 240 Hz of muon candidates from the whole DT detector.
The rate stability was tested by counting the number of DTTF tracks
collected from each sector per ``luminosity section''; a period of time lasting $\sim$~93 seconds. For each run, a
sample of rate measurements was collected in a histogram, to which a single
Gaussian function was fit. An example is shown in the left plot of
Fig.~\ref{fig:DTTFRates} for sector 8 in wheel 0. The L1 trigger
system could start to deliver L1 accepts several seconds after the data
acquisition was started; thus the first luminosity section, which appears as an isolated point at the left of the Gaussian peak in the rate plot, has incomplete statistics and is not considered in the fit. The mean and sigma of the Gaussian were used to compute the $\sigma/mean$ for each active sector.
In the absence of biases the $\sigma/mean$ is expected to scale with the square root of the number of events collected, consistent with a Poissonian distribution. To test this assumption another fit was performed on the distribution of the $\sigma/mean$ for each active sector with the function:

$$
f(x)=\frac{p0}{\sqrt(x)}
$$

The results of the fit are shown in the right plot of
Fig.~\ref{fig:DTTFRates}. The measured~$p0$~is 10\%~higher than the
naive expectation~${1/\sqrt{L}}\sim0.103$, where $L$ is the luminosity section length,
due to trigger dead-time not taken into account in the rate
calculation (see Section~\ref{sec:craft-gt}) and other possible effects under
investigation. 


The coordinates assigned by the DTTF were compared to the coordinates
obtained from the offline reconstruction of muon tracks using the muon detectors only (``standalone muons''). 
The comparison was performed for $\phi$ and $\eta$ coordinates, but 
no $\pt$ assignment study is possible with the CRAFT data (Section 
\ref{sec:crafttriggers}). The L1 muon candidate position is reported at
the extrapolated intersection of the track with a cylinder at the radius of the MB2 station. 
The offline track was therefore propagated to the MB2 cylinder, and the position of the
intersection point compared to the trigger data.
Events with tracks in masked sectors or with known hardware problems
were rejected in the analysis. 
 
\begin{figure}[htbp]
\begin{center}
\includegraphics[width=0.43\textwidth,angle=90]{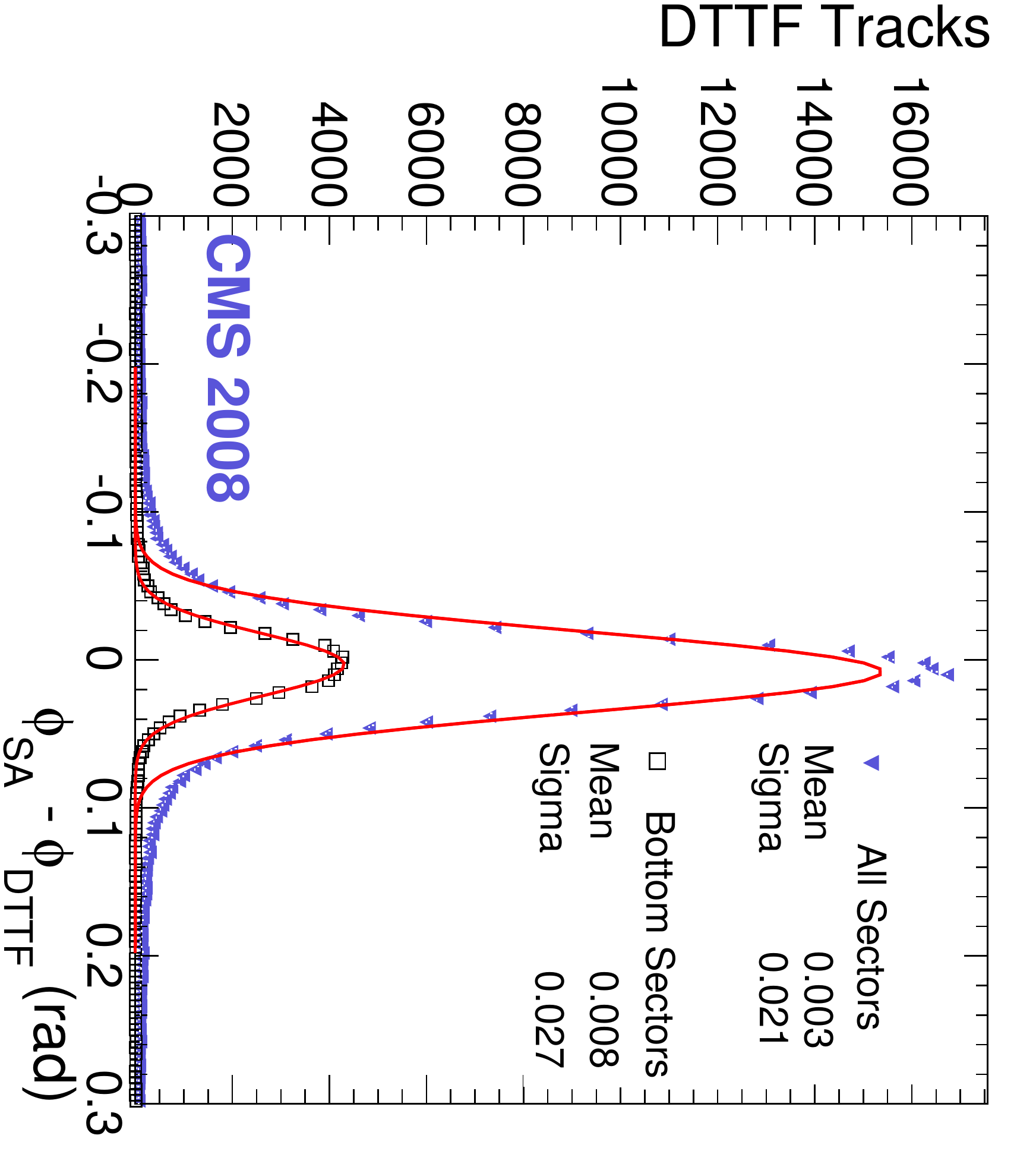}
\includegraphics[width=0.43\textwidth,angle=90]{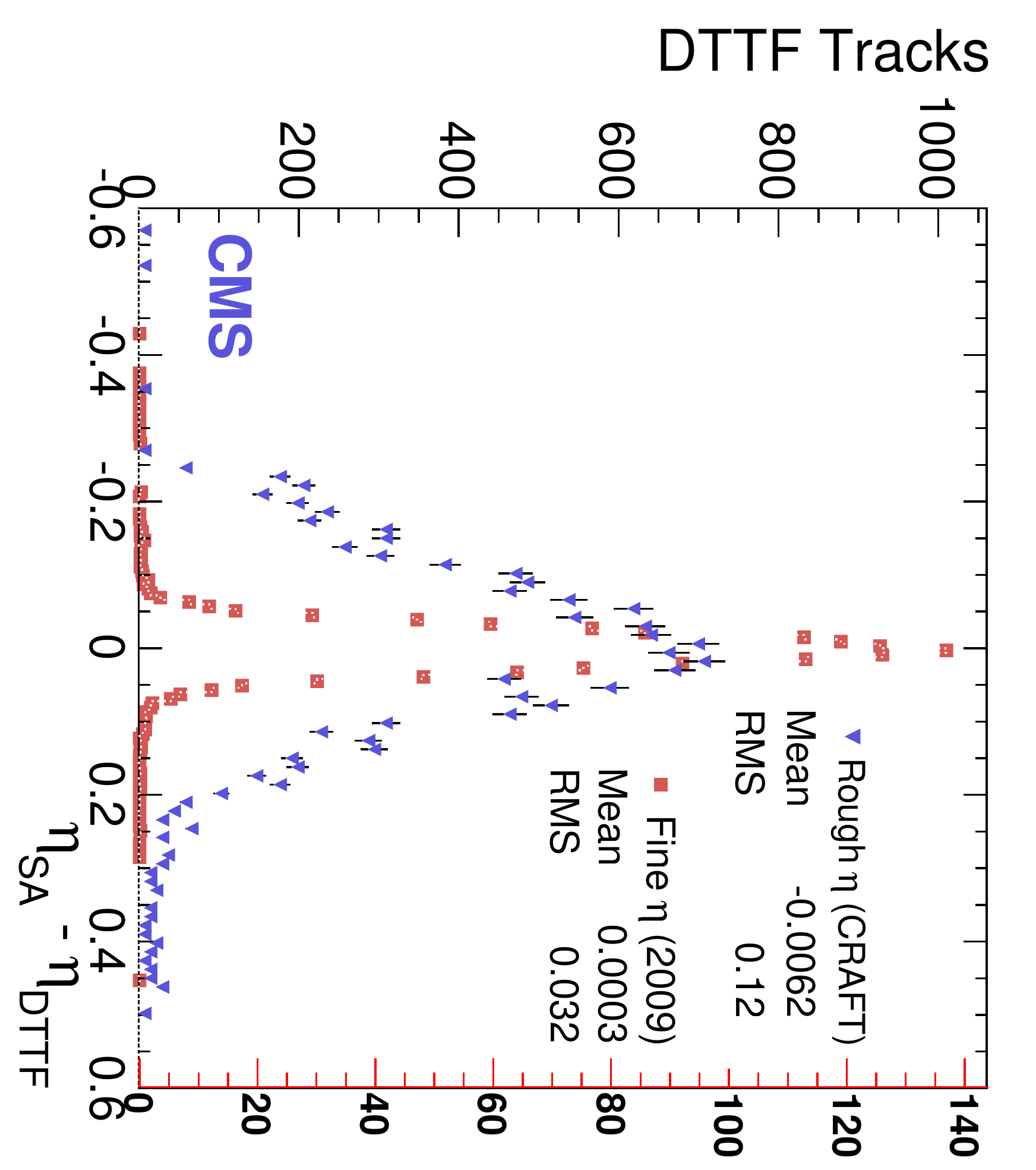}
\caption{DTTF $\phi$ resolution ({\em left}). The $\phi$ difference between DTTF and
  offline standalone muons ({\em SA}) is shown. Both the distribution obtained from all
  sectors ({\em triangles}) and the one obtained from bottom sectors only
  (sectors 9,10 and 11, {\em squares}) are shown.
  DTTF $\eta$ resolution ({\em right}). The $\eta$
  difference between DTTF and offline standalone muons ({\em SA}) is shown.
  The low-resolution $\eta$ assignment, the only possibility during
  CRAFT 2008, is shown with blue triangles. For comparison,
  high-resolution $\eta$ assignment is shown with red squares (from 2009 data).}

\label{fig:DTResolution}
\end{center}
\end{figure}

The difference between the coordinate from the reconstructed track and
the DTTF $\phi$ value is shown in the left plot of Fig.~\ref{fig:DTResolution}. Only positive wheels were used, as the negative wheel coordinate assignments were not implemented properly. This was a consequence
of misconfiguration of the hardware modules delivering
trigger primitives, which was corrected and validated after CRAFT.
Two histograms are shown, one including
all sectors and one for bottom sectors only (sectors 9, 10 and 11).
In the bottom sectors the muon direction, and hence multiple scattering and
energy loss effects, are LHC-like, so the resolution improves to $\sigma\sim0.021$~rad and the tails in the distribution are removed.

As mentioned in Section \ref{sec:crafttriggers}, the DTTF $\eta$
assignment was not yet commissioned in CRAFT, as trigger primitive
$\theta$-view information was not yet delivered to the track finder.
Nevertheless, a subset of the system
could assign low-resolution $\eta$ values using the
$\phi$-view primitives. The right plot of
Fig.~\ref{fig:DTResolution} shows the difference of
trigger and offline $\eta$ values assigned during CRAFT.  For comparison, the same quantity 
is shown for cosmic ray data taken in 2009, at a time when
fine-resolution $\eta$ assignment based on trigger primitive
$\theta$ hits was possible. The $\eta$ assignment between the DTTF and 
the reconstructed muons is in good agreement.

The DT trigger efficiency was evaluated using offline standalone reconstructed
muon tracks, that were required to intersect the CMS tracker volume.  
Only tracks traveling from the top to the bottom of the detector were kept, as low $\pt$ muons
can bend back and exit the detector from the top side.  A minimum track momentum
of 5~GeV/c was required, and only tracks with at least 20 hits in total from
DT and RPC detectors were kept, as this ensures the presence of local track segments in
at least two stations resulting in acceptable $\pt$ resolution.

Tracks reconstructed in the bottom half of the detector, with a matching trigger candidate, were used to
probe the efficiency of the top half of the detector in an unbiased way.
This was done by propagating the track to the second muon station in the
top and looking for a matching trigger candidate. If a DTTF track was
found, the trigger was considered efficient in this event.

\begin{figure}[htbp]
\begin{center}
\includegraphics[width=0.47\textwidth,angle=90]{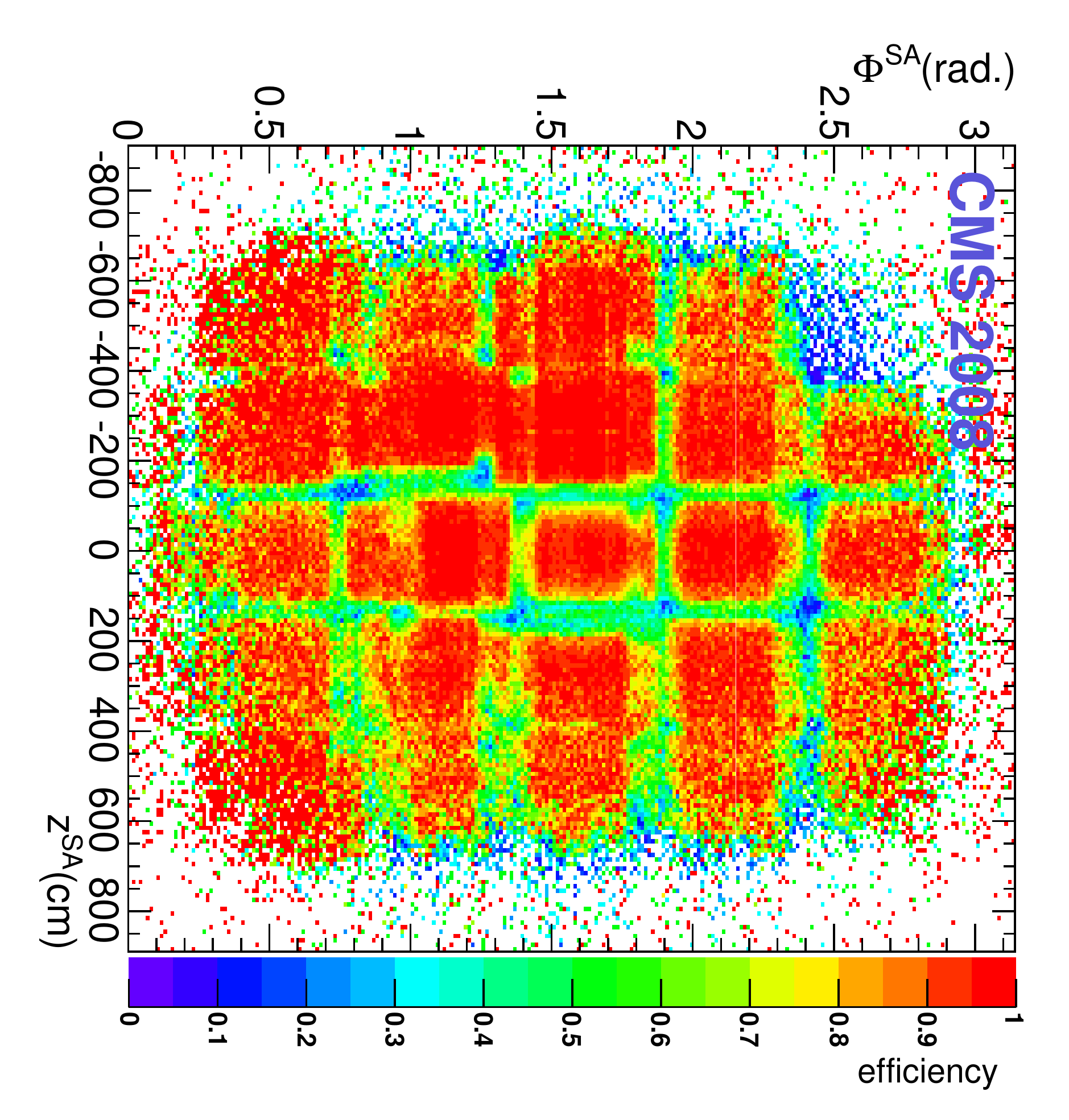}
\includegraphics[width=0.47\textwidth,angle=90]{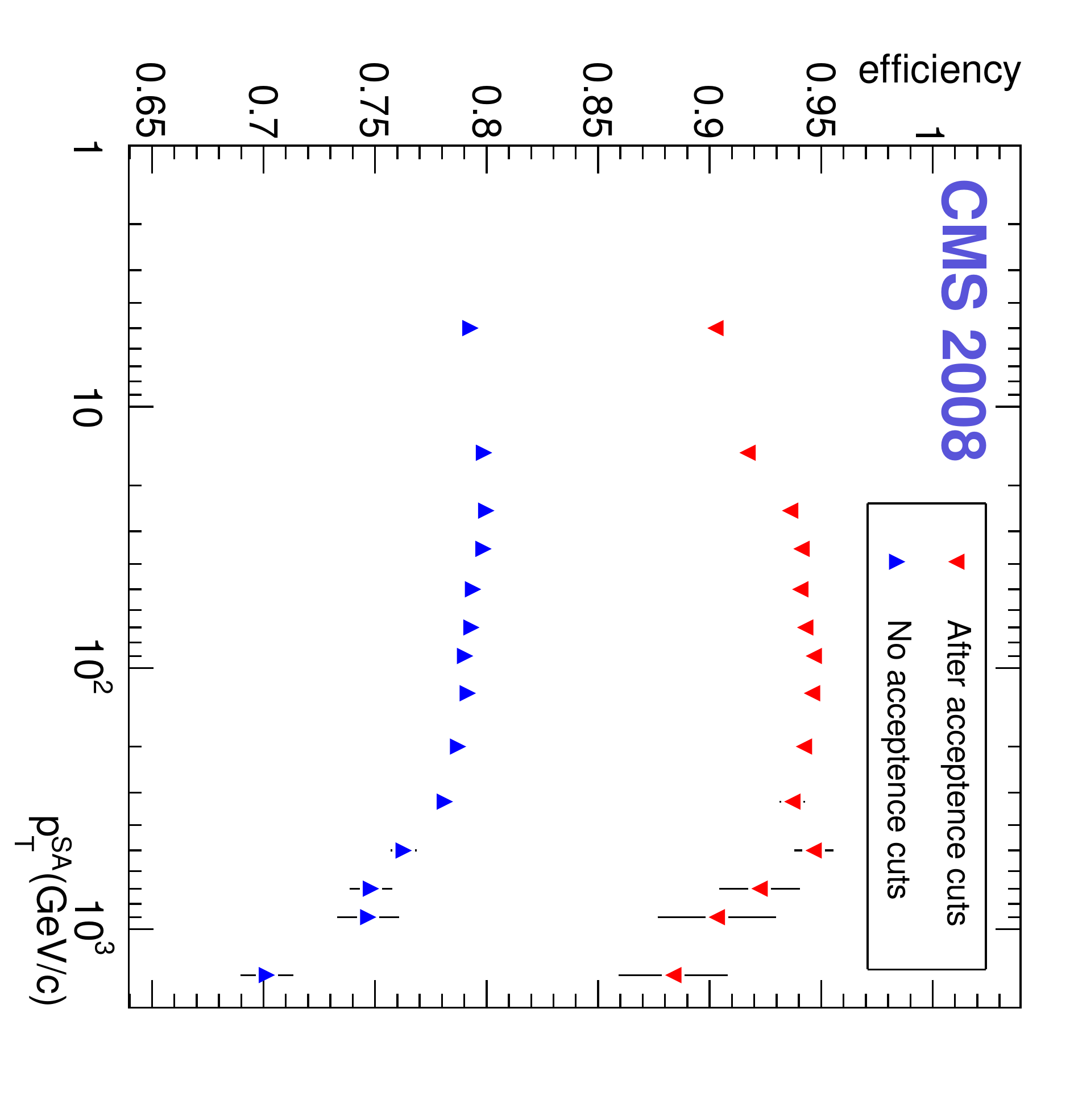}
\caption{Left: DT trigger efficiency for the upper half of the detector as a function of ($\phi,z$) of the
  standalone track (SA), computed at the position of the MB2
  station. Right: DT trigger efficiency as a function of the $\pt$ of the
  standalone track; the two data series correspond to efficiencies
  computed with and without acceptance cuts (see text).}
\label{fig:DTEff}
\end{center}
\end{figure}

In Fig.~\ref{fig:DTEff}, left, a ($\phi$,$z$) map of the efficiency
computed in this way is shown.  Besides the very low occupancy around
$\phi\sim0$ and $\phi\sim\pi$, due to the low rate of horizontal cosmic
rays, the main features visible are a lower
efficiency in cracks between detector sectors, and a whole sector
missing (wheel -2, sector 6, closest to $\phi=\pi$).  This was due to a malfunctioning hardware
module that had been masked. 

To check the intrinsic performance of the DT trigger system, tracks passing 
through the central portion of the top 3 sectors only were considered.  
Tracks passing within $5^{\circ}$ in $\phi$, or 50 cm in z, of a
sector boundary were ignored. The results are summarized in 
Fig.~\ref{fig:DTEff} ({\em right}), where the efficiency versus the $\pt$ of the muon track is
shown, before and after the removal of the crack regions. The
efficiency after the removal of the cracks reaches about 95\%, 
while it drops to about 80\% without the acceptance cut. The
acceptance losses between wheels are due to the loose pointing
requirements used to select the muons which allow a significant
fraction of vertical muons.

%% file: csc/CSCPerformance.tex
\section{Cathode Strip Chamber Trigger Performance}
\label{sec:cscperformance}

\begin{figure}[!htb]
  \centering
  \includegraphics[angle=90, width=0.47\textwidth]{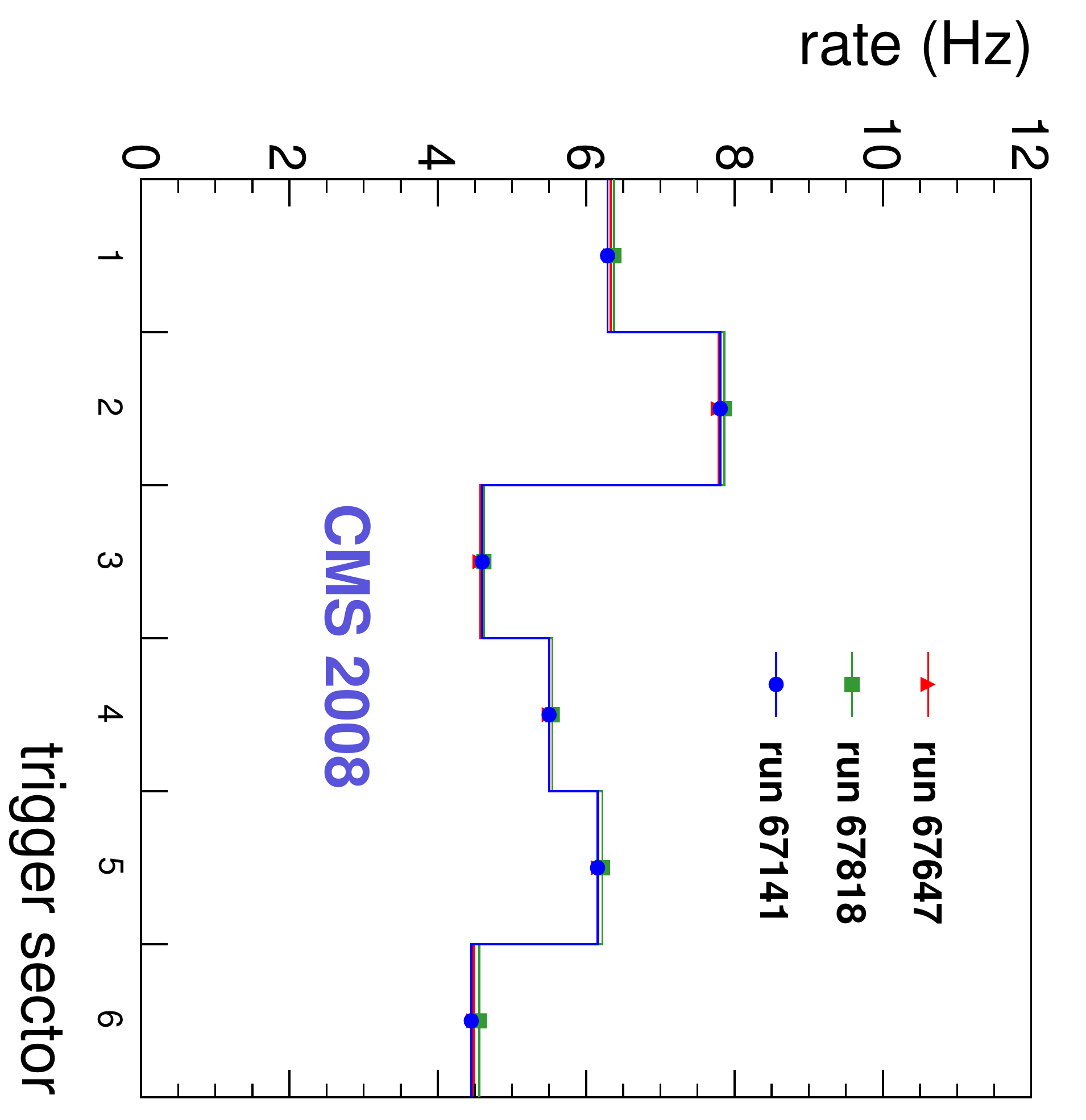}
  \includegraphics[angle=90, width=0.47\textwidth]{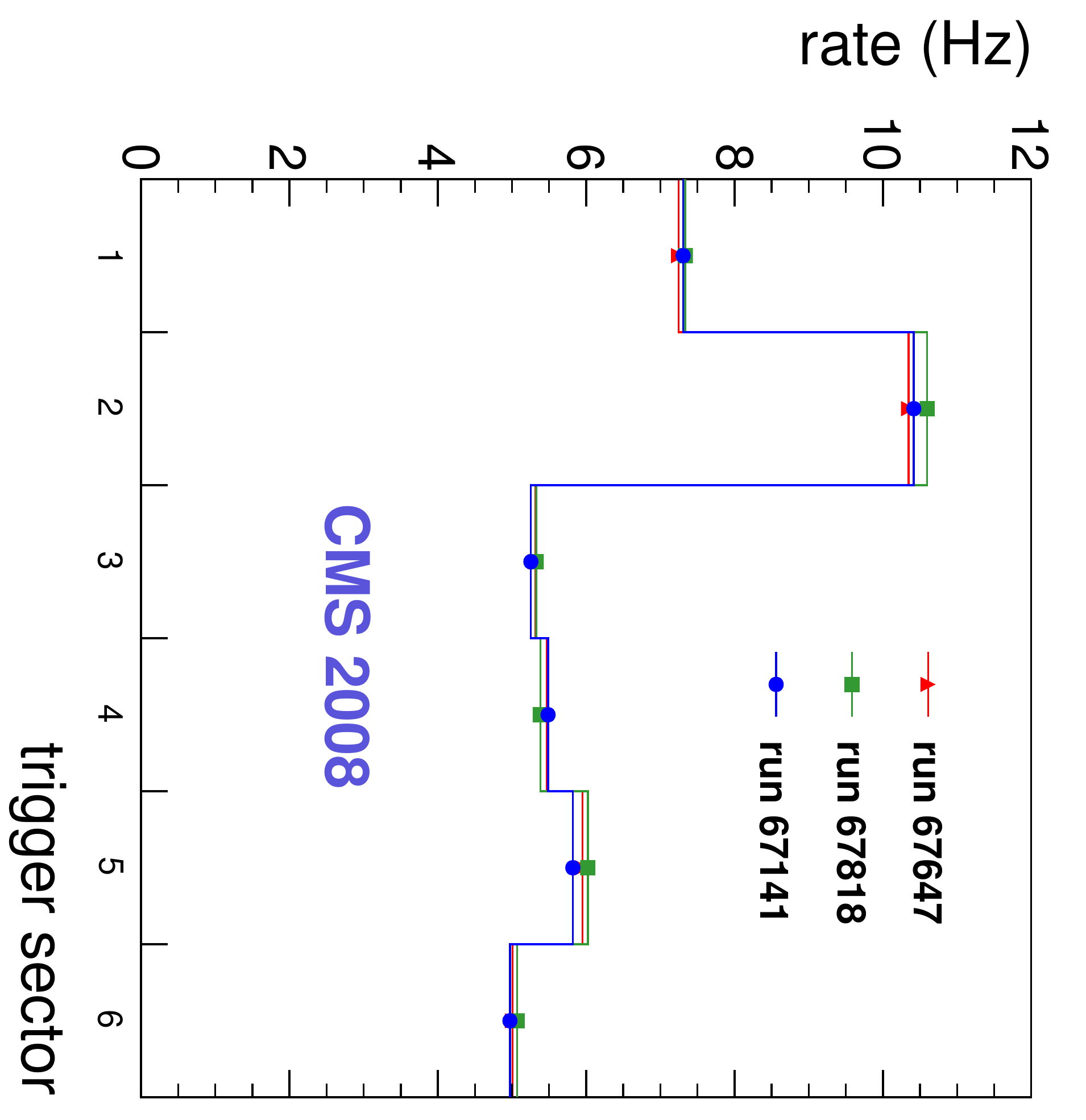}
  \caption{CSC trigger rates by sector during CRAFT.
    Several typical long runs were selected to calculate the trigger
    rates. Rates from these runs show good agreement.
    The left plot shows the trigger rate from the $z>0$ endcap; the
    right plot is for the $z<0$ endcap.}
  \label{fig:cscrates}
\end{figure}

The total output trigger rate from the CSCTF was $\sim$ 60~Hz. The distributions
of trigger rates in each $60^{\circ}$ CSC trigger sector are shown in
Fig.~\ref{fig:cscrates}. The trigger rate ranges from 4.5~Hz to 10.5~Hz for
different trigger sectors according to their different positions. Several
features of the trigger rate distributions are visible: the differences of top (1-3) and
bottom (4-6) sectors and the other $\phi$ dependencies reflect the spatial
distribution of cosmic rays penetrating CSC chambers, as well as the angular
acceptance of strip and wire trigger primitive pattern templates. There are also
asymmetries between endcaps, which are caused by the higher muon rate at the
negative side of the detector, which is below the CMS main access shaft.

The assignment of $\phi$, $\eta$, and $\pt$ by the CSCTF has been compared with
that of offline reconstructed muons.  While the aim of the CSCTF is to identify
collision muons, cosmic ray muons may arrive from all possible directions.  For this
reason, these studies use only muons whose direction points to the interaction
point, to resemble the expected behavior of collision muons. In addition, all candidates tagged 
by the CSCTF as halo muons were removed, along with those where only one segment
was found in the CSCTF, since the $\phi$ assignment of such candidates was not
properly implemented at the time of CRAFT.

\begin{figure}[!htb]
  \centering
  \includegraphics[angle=90, width=0.47\textwidth]{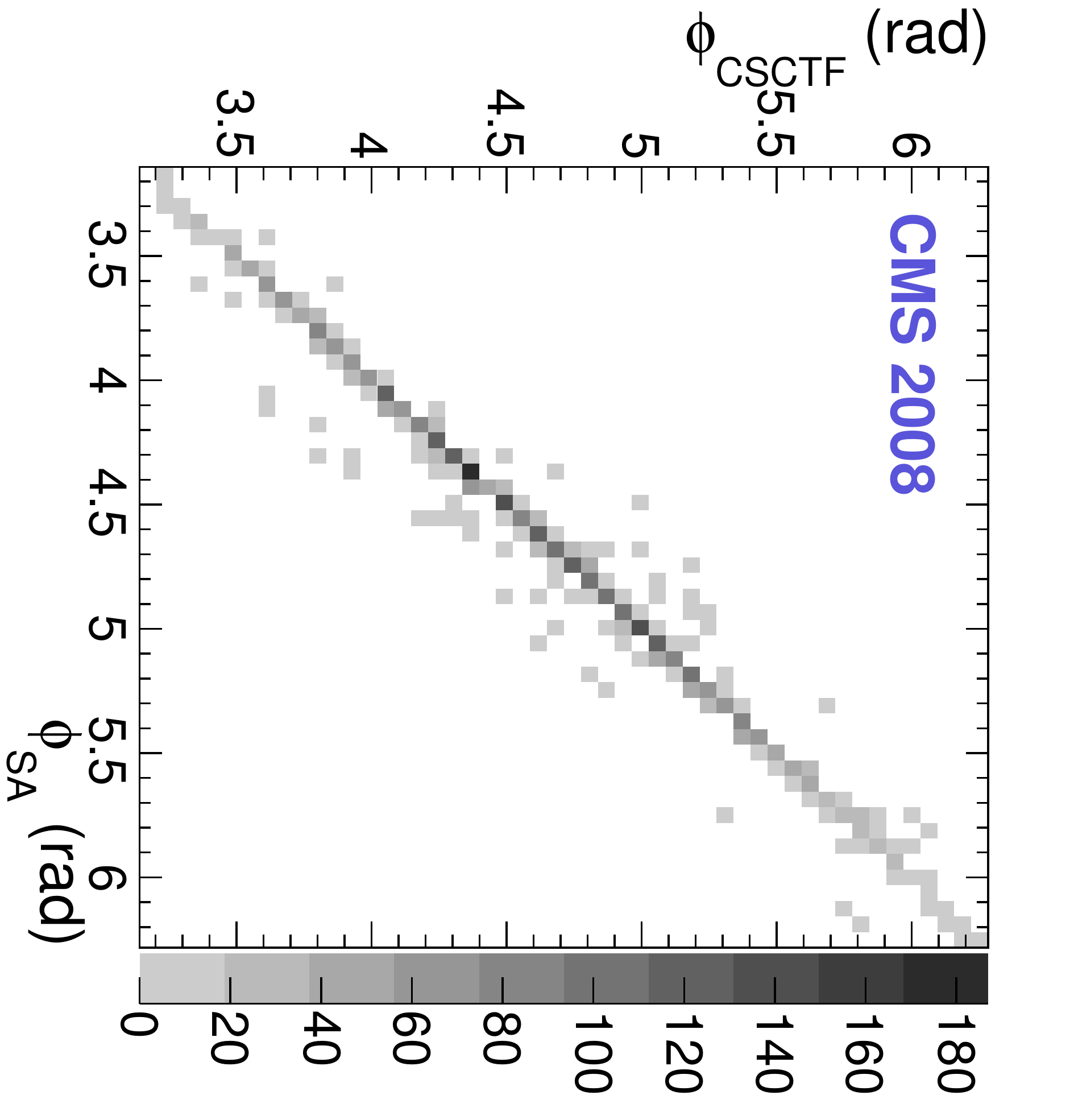}
  \includegraphics[angle=90, width=0.45\textwidth]{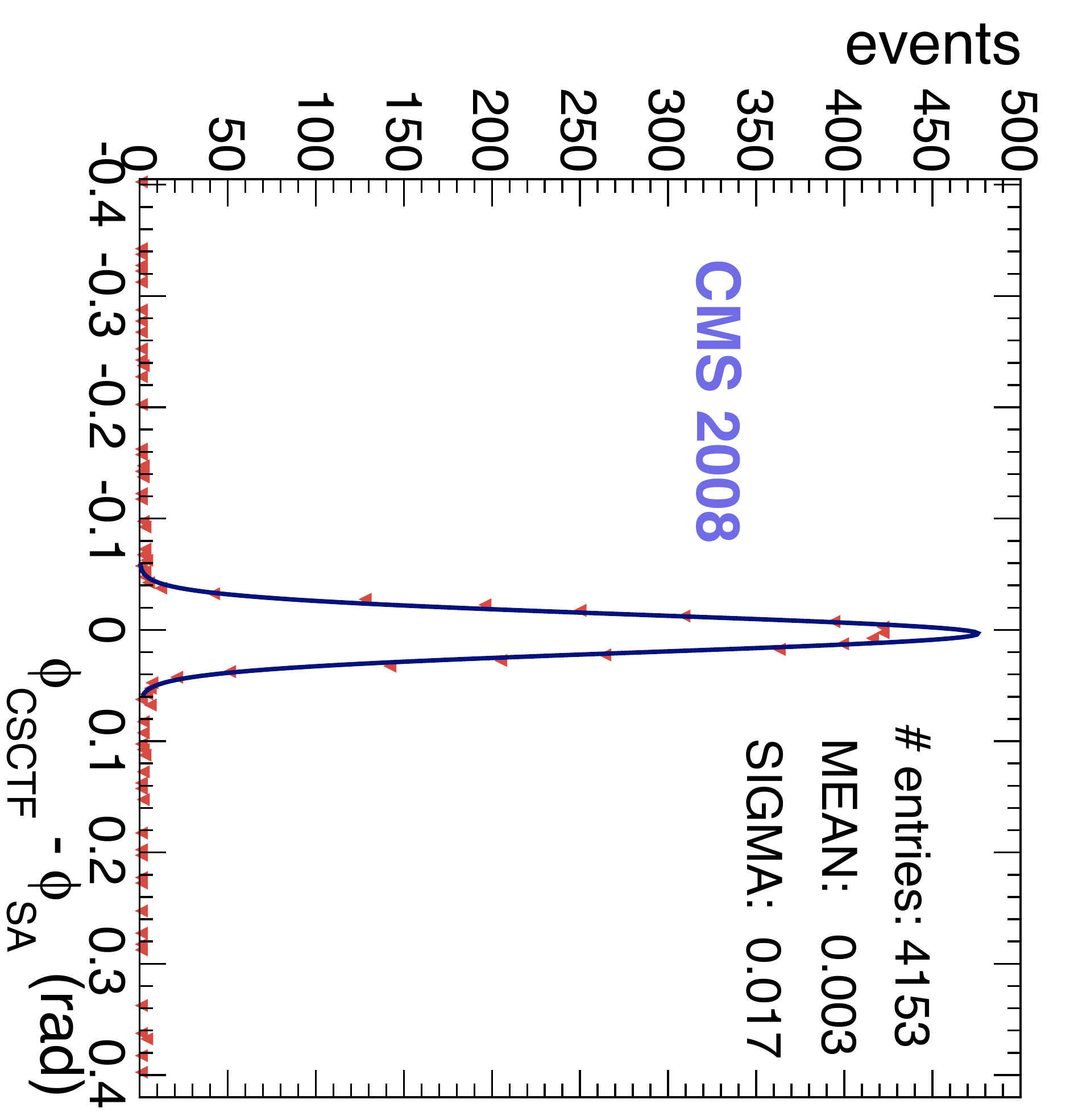}
  \caption{CSCTF $\phi$ resolution. Left: comparison between the $\phi$ measured
    by the CSCTF (``$\phi_{CSCTF}$'') and the $\phi$ estimated from the offline
    standalone reconstruction (``$\phi_{SA}$''). Right: distribution of the
    $\phi$ resolution with an overlaid Gaussian fit.}
  \label{fig:csctf-angles}
\end{figure}

The $\phi$ angle assignment is shown in Fig.~\ref{fig:csctf-angles}. These plots
show the comparison between the $\phi$ measured by the CSCTF, with respect
to that measured by the offline muon reconstruction, and the $\phi$ resolution
with an overlaid Gaussian fit, respectively. The $\phi$ assignment between the CSCTF and the
reconstructed muons is in good agreement. In fact most of the candidates lie on
the diagonal line, as shown in the left plot in Fig.~\ref{fig:csctf-angles}.

\begin{figure}[!htb]
  \centering
  \includegraphics[angle=90, width=0.47\textwidth]{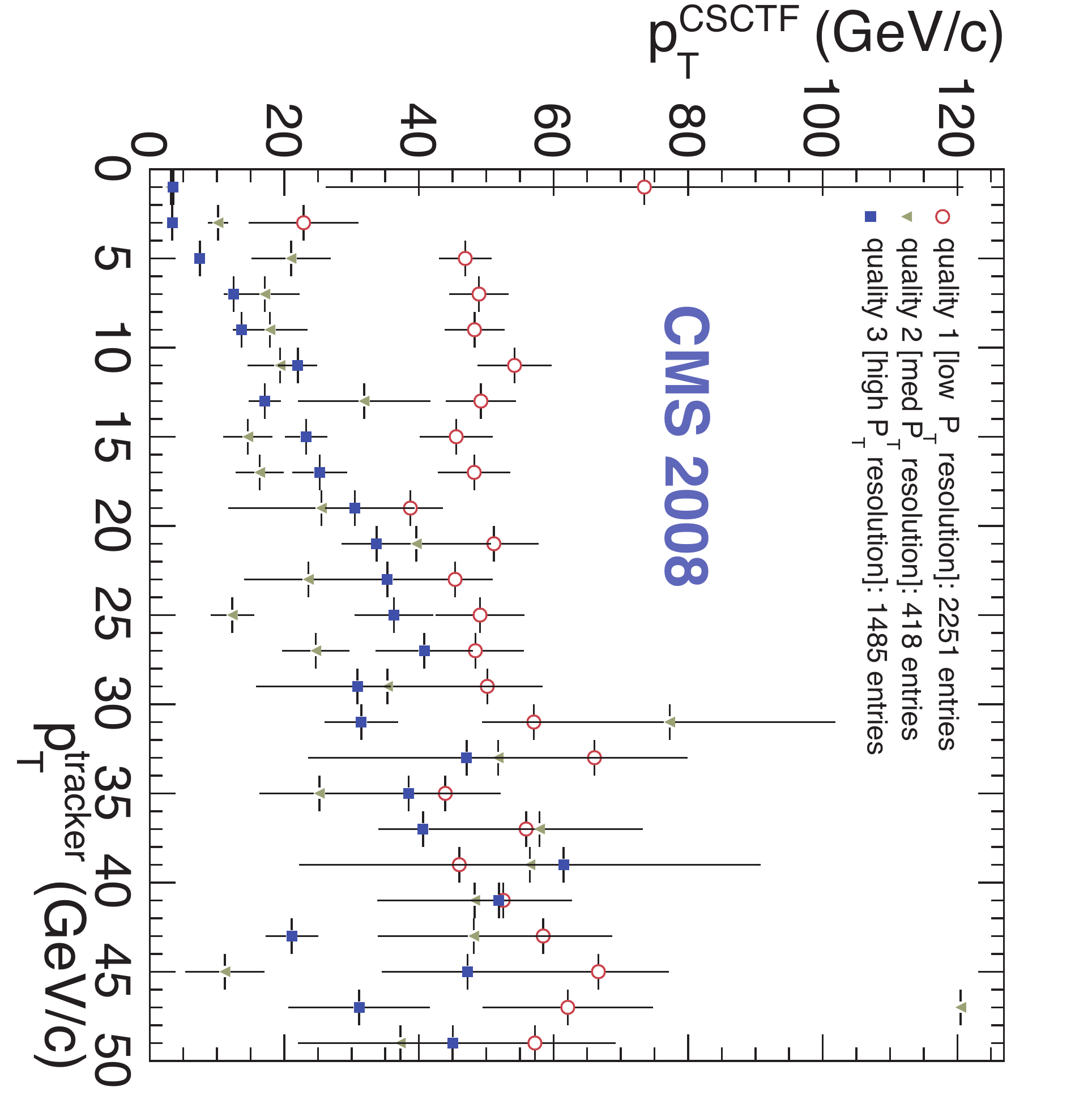}
  \caption{Mean $\pt$ assigned by the CSCTF as a function of the $\pt$ measured by
           the offline reconstruction, separated in bins of CSCTF $\pt$ resolution
           corresponding to a quality tag assigned by the CSCTF algorithms.}
  \label{fig:csctf-pt}
\end{figure}

The $\pt$ analysis requires a tighter muon selection. The CSCTF assumes that muon
tracks originate from the interaction point, and the $\pt$ assignment takes into
account loss of momentum as they traverse the detector. Therefore, only muons
following a similar trajectory, traversing the interaction region before
crossing the bottom half of the detector, are included in the analysis.  The
muon, reconstructed offline, is required to have at least one hit in the CSC detector, and
at least $10$ hits in the central tracker, to benefit from the best $\pt$
resolution for the reconstructed candidates. Fig.~\ref{fig:csctf-pt} shows the mean
$\pt$ assigned by CSCTF ($\pt^{CSCTF}$) as a function of the $\pt$ measured by
the tracker system ($\pt^{tracker}$). For a precise understanding of the
performance, the comparison is broken in bins of quality of CSCTF $\pt$
resolution. Several conclusions can be drawn. Since quality $3$ corresponds to
high $\pt$ resolution CSCTF tracks, the distribution flattens at higher $\pt$
reconstructed value, compared with the quality $2$ $\pt$ distribution (medium
$\pt$ resolution).  As expected, the CSCTF $\pt$ assignment for quality $1$ tracks 
is not well correlated with the tracker $\pt$ measurement.  It should be noted that
quality 1 tracks are only used in LHC collision trigger menus as the second leg of a
di-muon trigger, with essentially no $\pt$ requirement.

%
%

\begin{figure}[!htb]
  \centering
  \includegraphics[angle=90, width=0.47\textwidth]{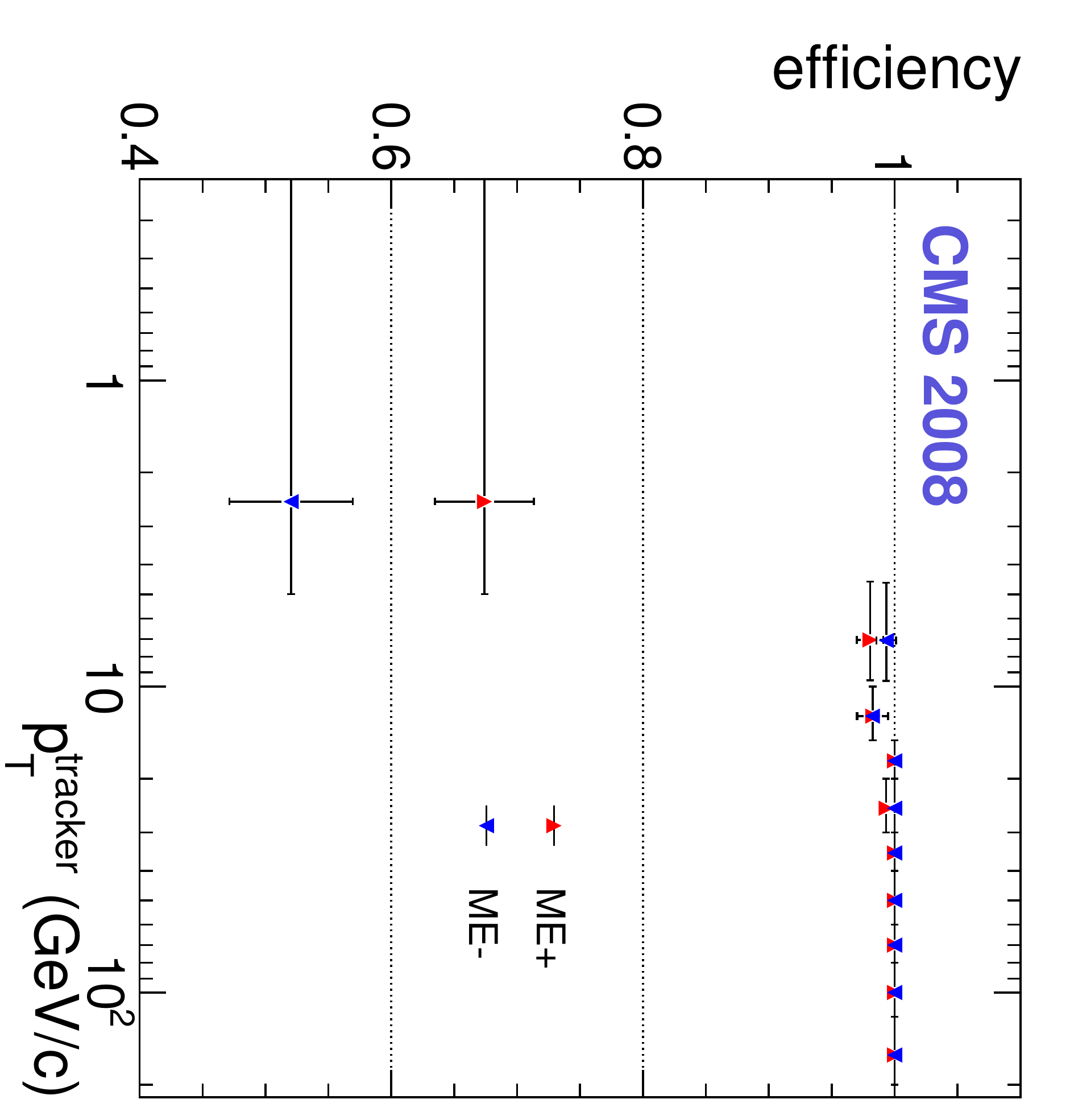}
  \includegraphics[angle=90, width=0.47\textwidth]{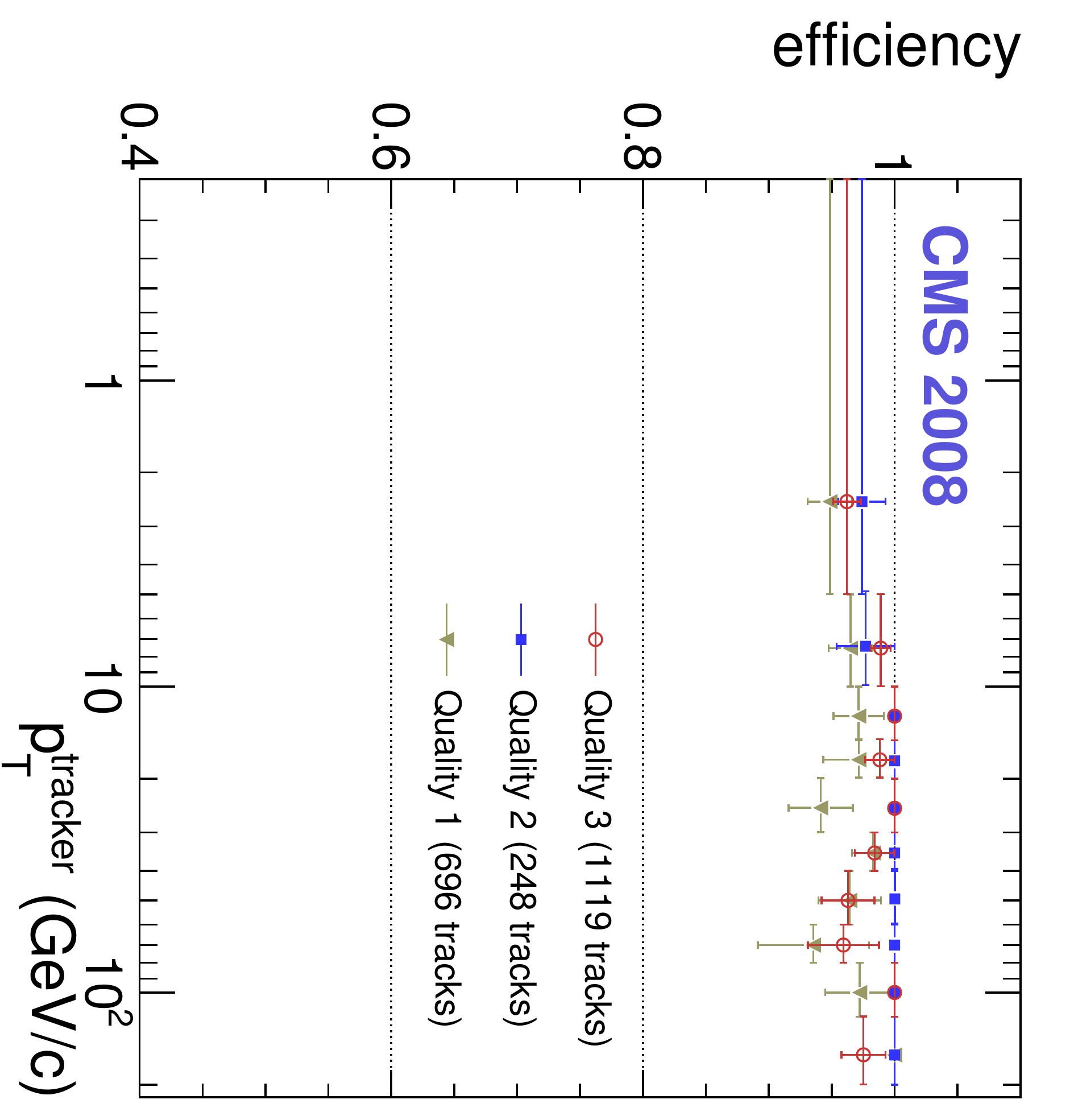}
  \caption{Left: efficiency of CSCTF in ``singles'' mode as a function of
                offline reconstructed $\pt$ in the tracker,
                $\pt^{tracker}$, for each endcap separately. Right: efficiency of CSCTF to build a muon
                candidate from several track segments separated in bins of a
                quality tag assigned by the CSCTF algorithms.}
  \label{fig:cscTFefficiency}
\end{figure}

The efficiency of CSC muon identification was studied, for both the single track
segment mode and the track-coincidence mode. For the ``singles'' mode, the
efficiency of one endcap is measured using events taken with a trigger in the
opposite endcap earlier in time. An offline reconstructed muon is required,
pointing towards the IP, with $\pt$ above 10 GeV/c. The central tracker track
is extrapolated into the CSC endcap under study, and a CSCTF candidate is
searched for within $\delta\phi < 0.3$~radians. Fig.~\ref{fig:cscTFefficiency} shows that the efficiency is greater than
$99$\% for both endcaps on $\sim2$k events with IP pointing tracks of $\pt$
above $20$ GeV/c.

In the track-segment matching mode of operation, the CSCTF builds tracks as a
coincidence of $2$, $3$, or $4$ track segments from different stations. The
CSCTF logic suppresses candidates from the ``singles'' mode if it can form a
coincidence from the received segments. The efficiency of the track coincidence mode, relative to
the singles mode, was measured by considering all CSCTF candidates (singles and
coincidence), and checking for other available segments in the other stations in
the same time bin. If segments which could form a coincidence were found, a
corresponding multi-segment CSCTF candidate was
searched for. The resulting efficiency, broken in bins of quality, is shown
in the right plot of Fig.~\ref{fig:cscTFefficiency}. The relatively large fraction of quality $1$
candidates is due to the loose geometric requirements used for cosmic ray muons.

%% file: rpc/RPCPerformance.tex
\section{Resistive Plate Chamber Trigger Performance}

The average rate of the muon candidates produced by the RPC
trigger was about 140~Hz.  However, periodic disturbances visible as
spikes in the trigger rate were observed and were the subject of
detailed studies on both the RPC chamber and the trigger electronics.  
Dedicated data were taken for the RPC noise studies.  Although
never observed before CRAFT, periodic noise
effects were found to be correlated with electric disturbances coming from
external sources and related to underground daytime activities in UXC, rather than connected with the CMS magnetic field
itself.  They were found to be sensitive to the discriminator thresholds
and completely absent in the trigger
path if no signal from the chambers was
delivered.  The noise was found to be mostly coherent and it can
only be partially eliminated by changing the trigger logic to require
more planes in coincidence. The standard configuration for cosmic rays, implemented
with the requirement of a coincidence in 3 out of 6 chamber planes,
was compared to a modified coincidence requirement of 4 out of 6
chamber planes obtained in offline emulation of the trigger. 
The modified trigger logic is found to reduce the
noise peak values from up to 1100 Hz to up to 220 Hz, but it also decreases
the overall average trigger rate, which drops from about 140 Hz to 45
Hz. It should be stressed that when the RPC readout data were processed by the 
RPC trigger emulator using LHC patterns, the rate spikes were completely eliminated.
This is due to a combination of using a coincidence of 4 out 6 planes with the fact
that LHC patterns are much narrower.  During and since CRAFT, steps were taken to identify the sources of noise.  Since the summer of 2009, the noise has been reduced to a marginal problem.





\begin{figure}[htbp]
\begin{center}
\includegraphics[angle=90, width=0.5\textwidth, height=0.5\textwidth]{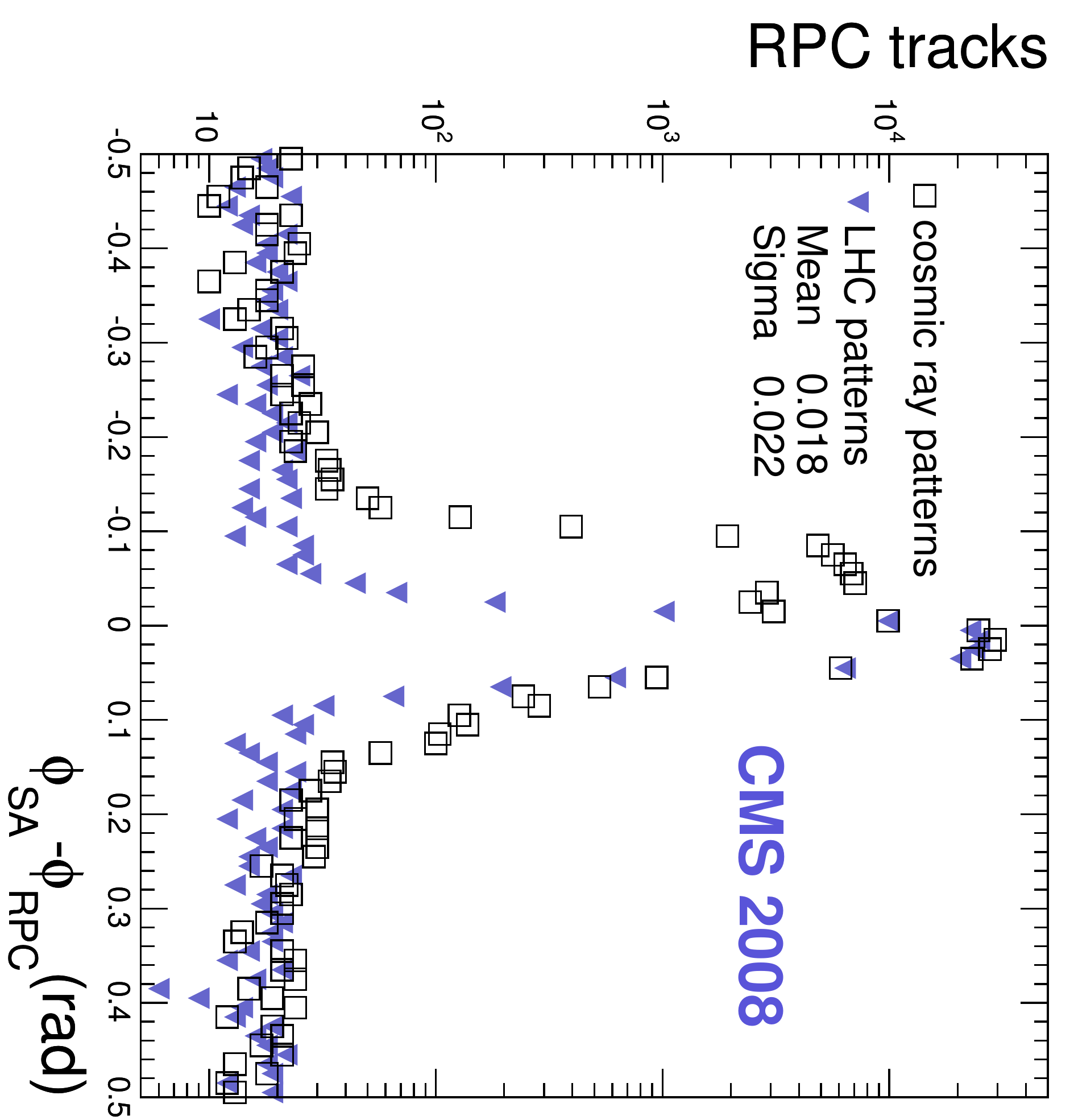}
\caption{Resolution of the RPC trigger: RPC trigger $\phi$ minus reconstructed muon
  $\phi$ for cosmic ray patterns used in CRAFT data taking (dashed line), and for beam collision patterns, produced 
by running the RPC trigger emulator over the same data sample (solid line).}
\label{fig:RPCResolution}
\end{center}
\end{figure}

The $\phi$ resolution of the RPC trigger was studied by comparing the
$\phi$ values from the RPC track finder with those from standalone
muons reconstructed using DT segments.  A typical result is shown in the open histogram of 
Fig.~\ref{fig:RPCResolution}.  A two-peak structure can be seen, which is an
artefact of the ghost removal algorithm in the case of cosmic ray
patterns.  Since a $\pt$ measurement was not made by the RPC trigger in CRAFT, the ghost removal algorithm saves only the muon candidate with higher $\phi$ value whenever two muon candidates in a given sector are found from overlapping logical cones.  Such events produce a systematic bias in the $\phi$ measurement, visible as an additional peak shifted by 6 degrees on average.  To confirm this, the resolution was obtained using patterns for LHC collisions, as shown in the solid histogram of Fig.~\ref{fig:RPCResolution}.  The additional peak is not present in this case.

\begin{figure}[htbp]
\begin{center}
\includegraphics[width=0.47\textwidth,angle=90]{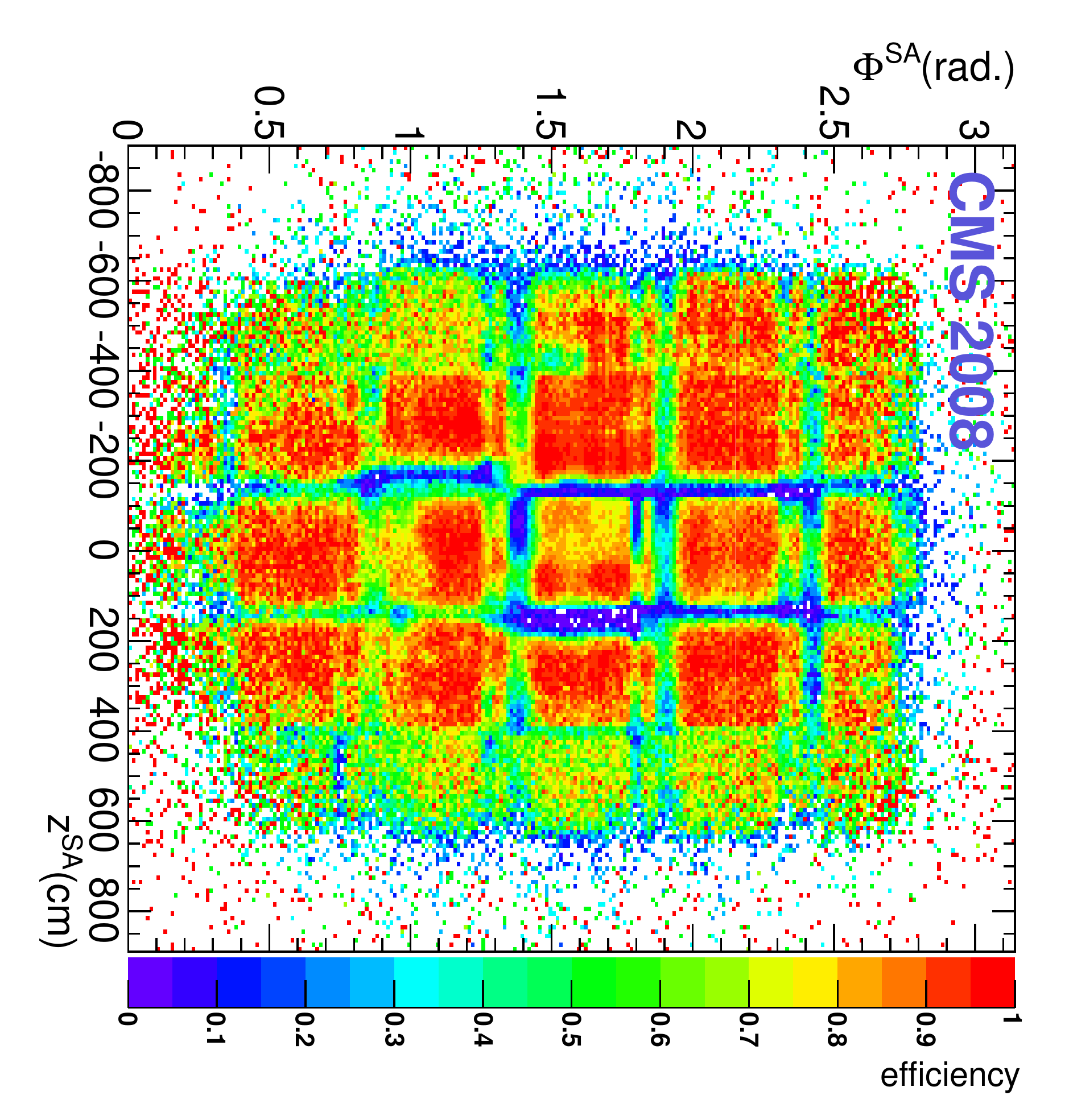}
\includegraphics[width=0.47\textwidth,angle=90]{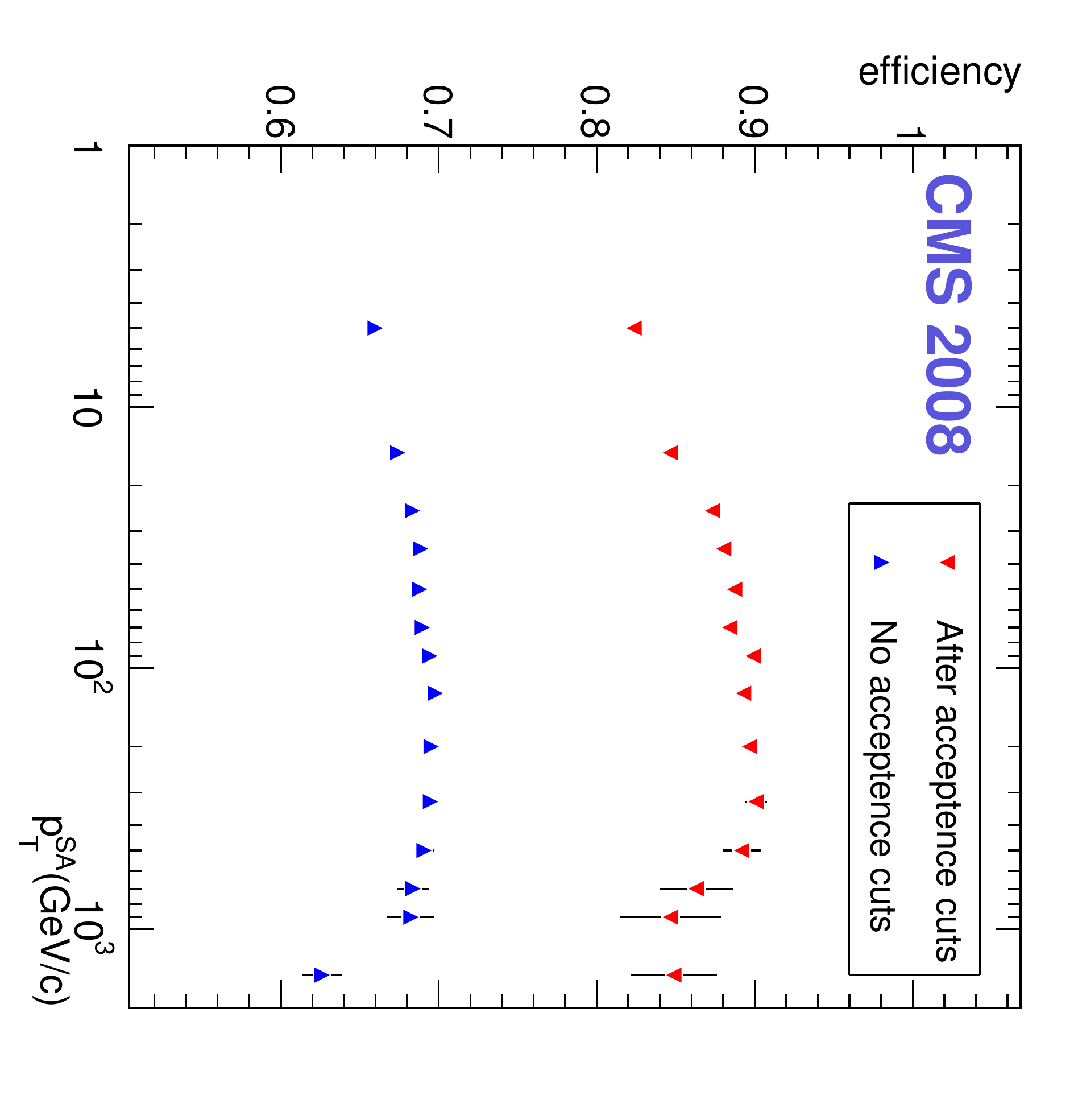}
\caption{RPC trigger efficiency as a function of ($\phi,z$) of the
  standalone reconstructed muon, computed at the position of the MB2
  station ({\em left}). RPC trigger efficiency as a function of the $\pt$ of the
  standalone muon. The two data series correspond to efficiency
  computed with and without acceptance cuts (see text).}
\label{fig:RPCEff}
\end{center}
\end{figure}

The RPC trigger efficiency is evaluated using standalone reconstructed
muon tracks, with method and selections as detailed for the DT trigger in Section \ref{sec:dtperformance}.  These data were taken at a chamber voltage of 9.2 kV.
Fig.~\ref{fig:RPCEff} ({\em left}) shows a ($\phi$,$z$) map of the efficiency
computed in this way. The main visible features are a lower
efficiency in cracks between detector sectors and very few events near $\phi = 0$ and $\phi = \pi$ due to lack of horizontal cosmic rays.
In Fig.~\ref{fig:RPCEff} ({\em right}), the efficiency versus the $\pt$ of the muon track is
shown, before and after the removal of the crack regions. The
efficiency after the removal of the cracks is between 85 and 90\% in the
central $\pt$ region, with a clear tendency to increase with $\pt$, 
while it drops to about 70\% without the acceptance cut.  Work to understand the efficiency result is ongoing.  The systematic biases due to the method are expected to play some role and are presently under evaluation.



%% file: egamma/EGPerformance.tex
\section{Electron/Photon Trigger Performance}
\label{sect-egperf}

\begin{figure}[tp]
\begin{center}
  \includegraphics[width=0.47\textwidth]{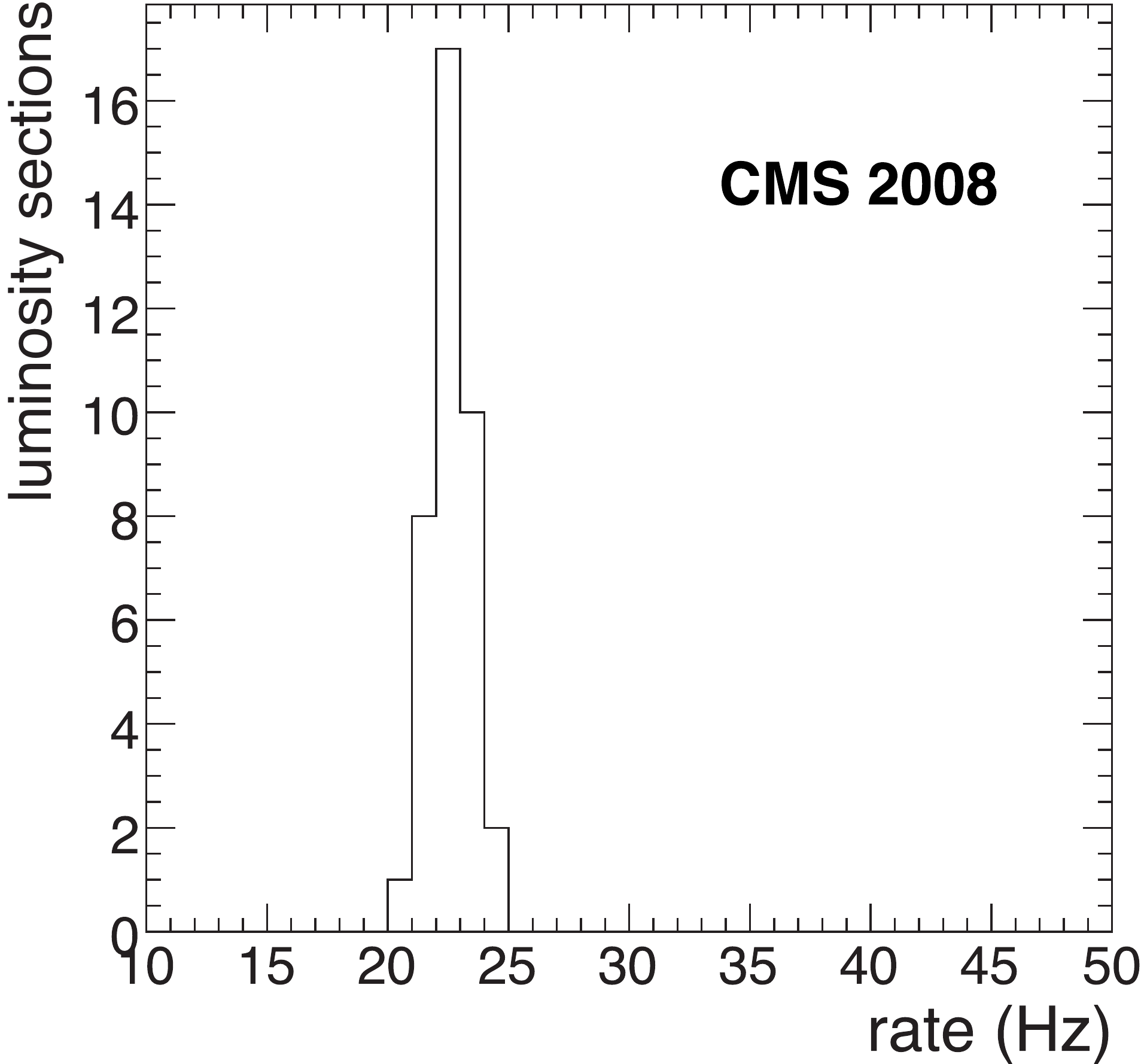}
  \caption{Rate distribution of the L1\_SingleEG1 trigger for a typical run. The average rate is compatible with that expected from the noise level measured in the ECAL.}
  \label{fig:EGRate}
\end{center}
\end{figure}

The performance of the barrel e/$\gamma$ trigger has been evaluated in terms of rate, resolution and efficiency.
The distribution of the rate of the L1\_SingleEG1 e/$\gamma$ trigger, which nominally fires whenever a single electromagnetic energy deposit above 1~GeV is detected, is shown in Fig.~\ref{fig:EGRate} for a typical CRAFT run.
The L1 decision is based on the sum of 2 towers of 25 ECAL crystals each. The average single crystal noise as measured during the entire running period was 40~MeV~\cite{CFT-09-004}, so the L1 candidate noise is expected to be around 280~MeV.
On the other hand, a rate of 22.65 Hz implies that the threshold of the L1\_SingleEG1 trigger is roughly 5$\sigma$ away from the detector noise.
As will be seen later in this section, the L1\_SingleEG1 trigger turn-on point corresponding to 50\% efficiency is measured to be 1.19~GeV, implying the noise of the particular run in Fig.~\ref{fig:EGRate} is around 240~MeV. Given that the expectation is derived from a much larger period of time, the agreement is found to be good.

\label{subsect-egtrigres}

To study the trigger resolution and efficiency, ECAL superclusters (SC) originated by muon radiation in the lead tungstate crystals and reconstructed offline as described in~\cite{CFT-09-004} are used as tags to probe for the production of L1 e/$\gamma$ trigger candidates.


\subsection{Data Selection}
The e/$\gamma$ trigger resolution and efficiency were measured using events taken with a muon trigger.  Online data reduction for the ECAL was obtained through the selective readout algorithm \cite{Almeida:2008zz}, which classifies the detector into regions of low or high interest.  Low interest regions were read out using zero-suppression on a crystal-by-crystal basis, whereas in high interest regions full readout is done, preserving the information of all crystals involved in the trigger decision.  For the full trigger chain efficiency measurement a complete configuration of the full L1 e/$\gamma$ trigger is required: namely the ECAL, the RCT and the GCT.  For the trigger primitive generator efficiency measurement, only the ECAL trigger primitive generator need be configured.  Finally, the study used only the regions of the detector that had no known hardware problems.


In selecting ECAL superclusters, at least one crystal with a reconstructed energy above 400~MeV is required, ten times the noise RMS.
This ensures accurate timing reconstruction~\cite{CFT-09-006} and by retaining events within 3.75~ns of the trigger, rejects asynchronous cosmic ray deposits.
In contrast to the cosmic ray signal reconstruction that fits an asynchronous pulse shape to the 25~ns signal samples, here the signal amplitude, and consequently $\et$, are reconstructed using a weighted sum of the signal samples.
Not only is this the standard procedure for beam collision data, it is also better suited for comparison with trigger quantities, since the trigger amplitudes determined in the detector front-ends are obtained using a similar weighted sum method.

Finally, the ECAL superclusters are required to be validated by an offline-reconstructed global muon.
Validation is based on the distance, $\Delta R$, between the ECAL supercluster position and the linear extrapolation of the muon track to the ECAL inner surface~\cite{CFT-09-005} ($\mu$), 
starting from the tracker.
Events are retained with $\Delta R\left(\text{SC},\mu\right) < 0.1$.

\subsection{Resolution and Efficiency}
\begin{figure}[tp]
\begin{minipage}[b]{0.5\textwidth}
    \centering
    \includegraphics[width=0.95\textwidth]{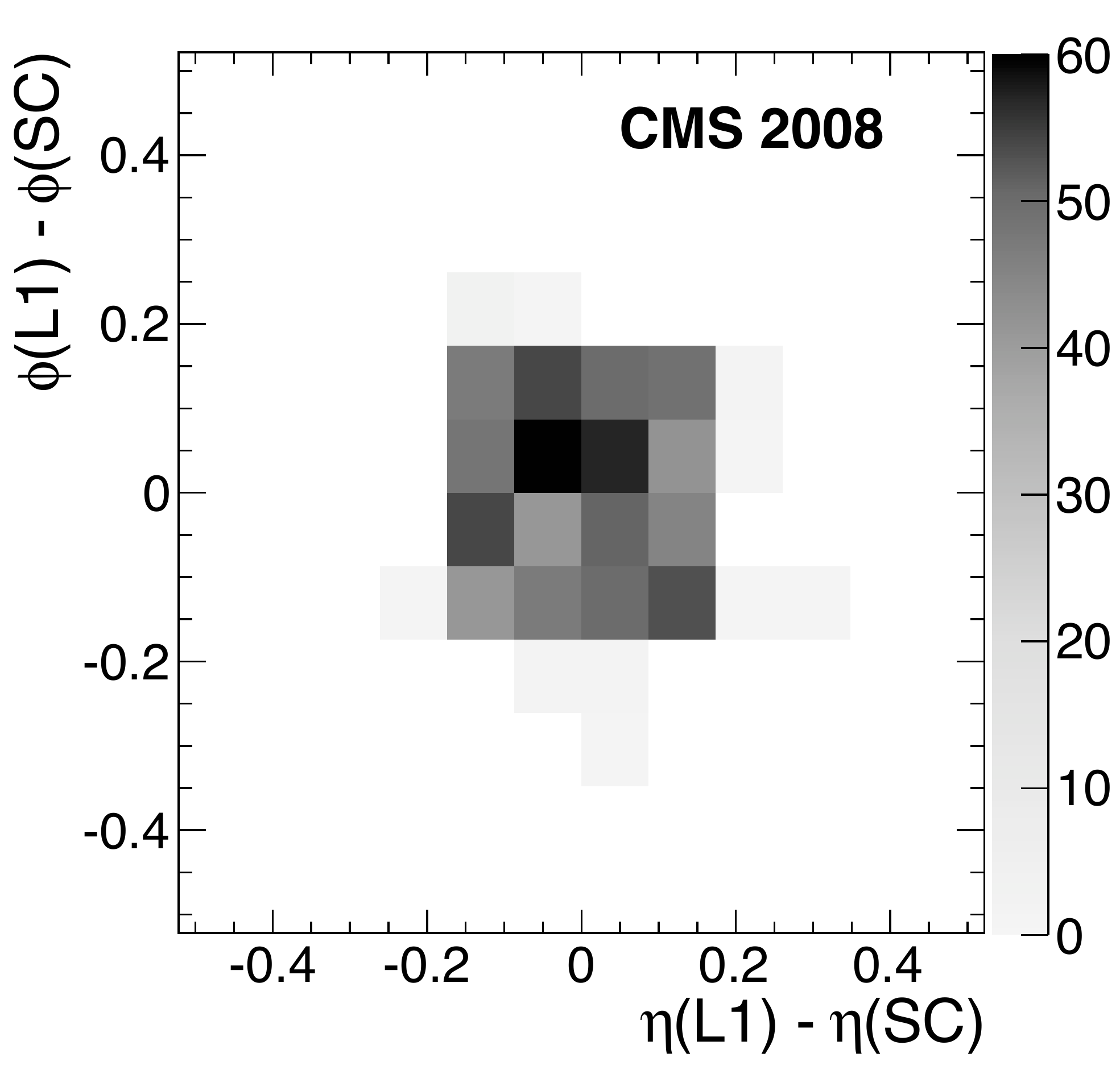}
    \caption{$\Delta\eta$ and $\Delta\phi$ of Level-1 e/$\gamma$ trigger objects with respect to reconstructed ECAL superclusters.}
    \label{fig:electron_eta_phi_delta}
\end{minipage}
\hspace{0.5cm}
\begin{minipage}[b]{0.5\textwidth}
    \centering
    \includegraphics[width=0.95\textwidth]{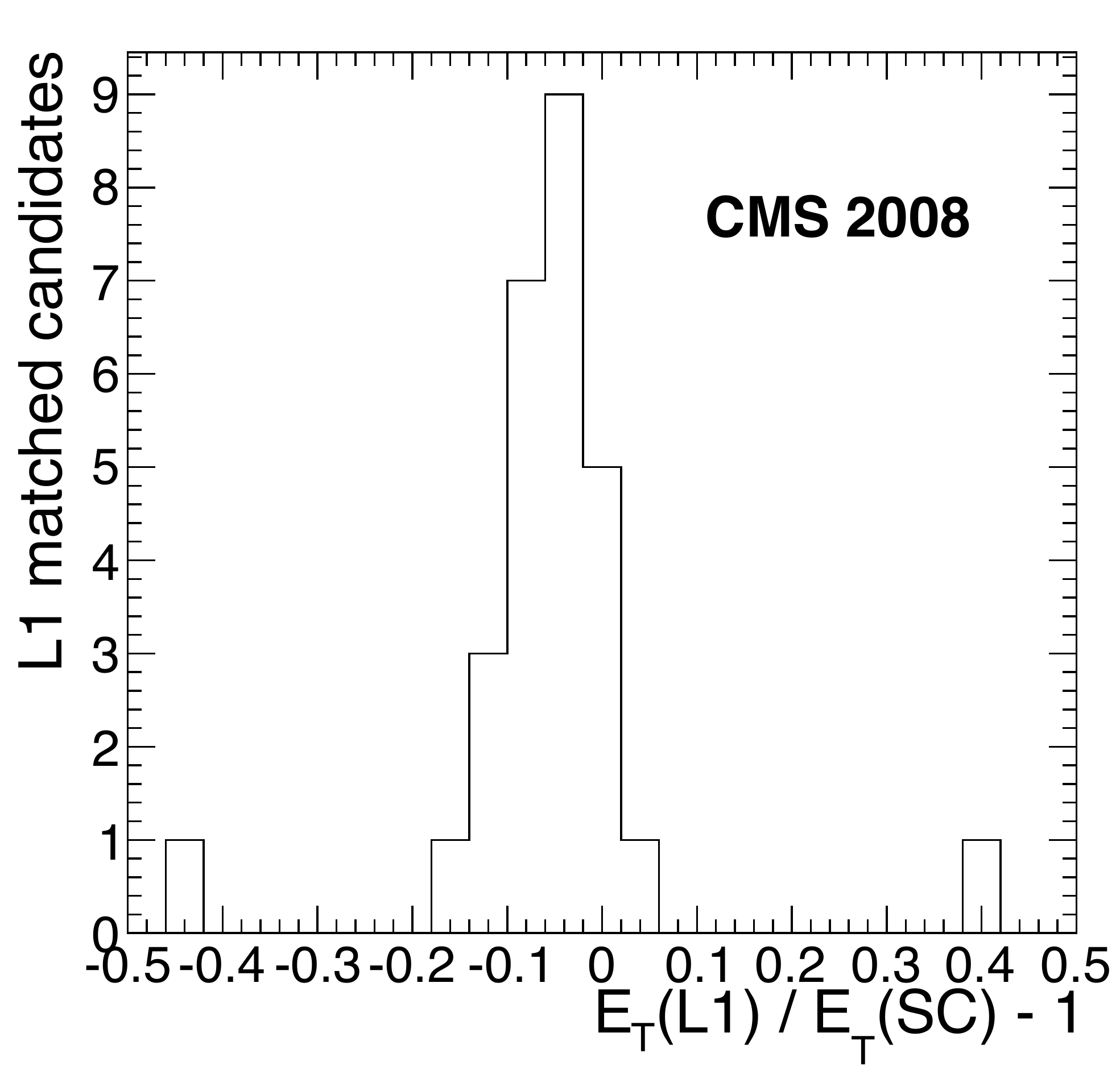}
    \caption{Level-1 e/$\gamma$ $\et$ resolution for $\et(\text{L1}) > 10$~GeV. The non-zero mean results from a combination of factors and is compatible with the LSB of $\et(\text{L1})$.}
    \label{fig:electron_ET_delta}
\end{minipage}
\end{figure}

After obtaining a pool of ECAL superclusters validated by reconstructed muons, the L1 e/$\gamma$ trigger resolution and efficiency were probed. The distributions for $\Delta\eta = \eta(\text{L1}) - \eta(\text{SC})$ and $\Delta\phi = \phi(\text{L1}) - \phi(\text{SC})$ are shown in Fig.~\ref{fig:electron_eta_phi_delta} for L1 candidates satisfying the L1\_SingleEG1 requirement of $\et(\text{L1}) > 1$~GeV.
The binning used corresponds to the dimensions of one ECAL trigger tower, 0.087 in both $\eta$ and $\phi$.
The highly populated region of $4\times 4$ ECAL trigger towers corresponds to the resolution with which the L1 e/$\gamma$ candidate position is reported.

Fig.~\ref{fig:electron_ET_delta} shows the relative difference between $\et(\text{L1})$ and $\et(\text{SC})$ for $\et(\text{L1}) > 10$~GeV.
The resolution from fitting a Gaussian to the distribution is $5.2$\% and the mean $-4.7$\%.
The non-zero mean is caused by differences in the way $\et$ is calculated in L1 and off-line reconstructions.
These are: the clustering algorithm, the signal amplitude determination, and the effect of integer truncation in the L1 $\et$ determination, which has a LSB of 500~MeV.
Each of these effects contributes to the L1 $\et$ being lower than the $\et$ which is recovered offline, the effect above 5~GeV being of the order of the LSB.  The resolution is also of the order of the LSB; due to the steeply falling energy spectrum of these events, the sample is dominated by those just above the 10~GeV threshold.

To further characterize the e/$\gamma$ trigger, two efficiencies were measured: the trigger primitive (TP) generator efficiency and the Level-1 e/$\gamma$ candidate efficiency, which correspond to the first and the last steps in the calorimeter trigger chain.
Due to the requirement of an energy deposit in the ECAL, this measurement evaluates the trigger efficiency only in the active part of the detector and is relative to the detector efficiency to detect muons and electromagnetic energy.

\begin{figure}
\begin{center}
 \includegraphics[width=0.47\textwidth]{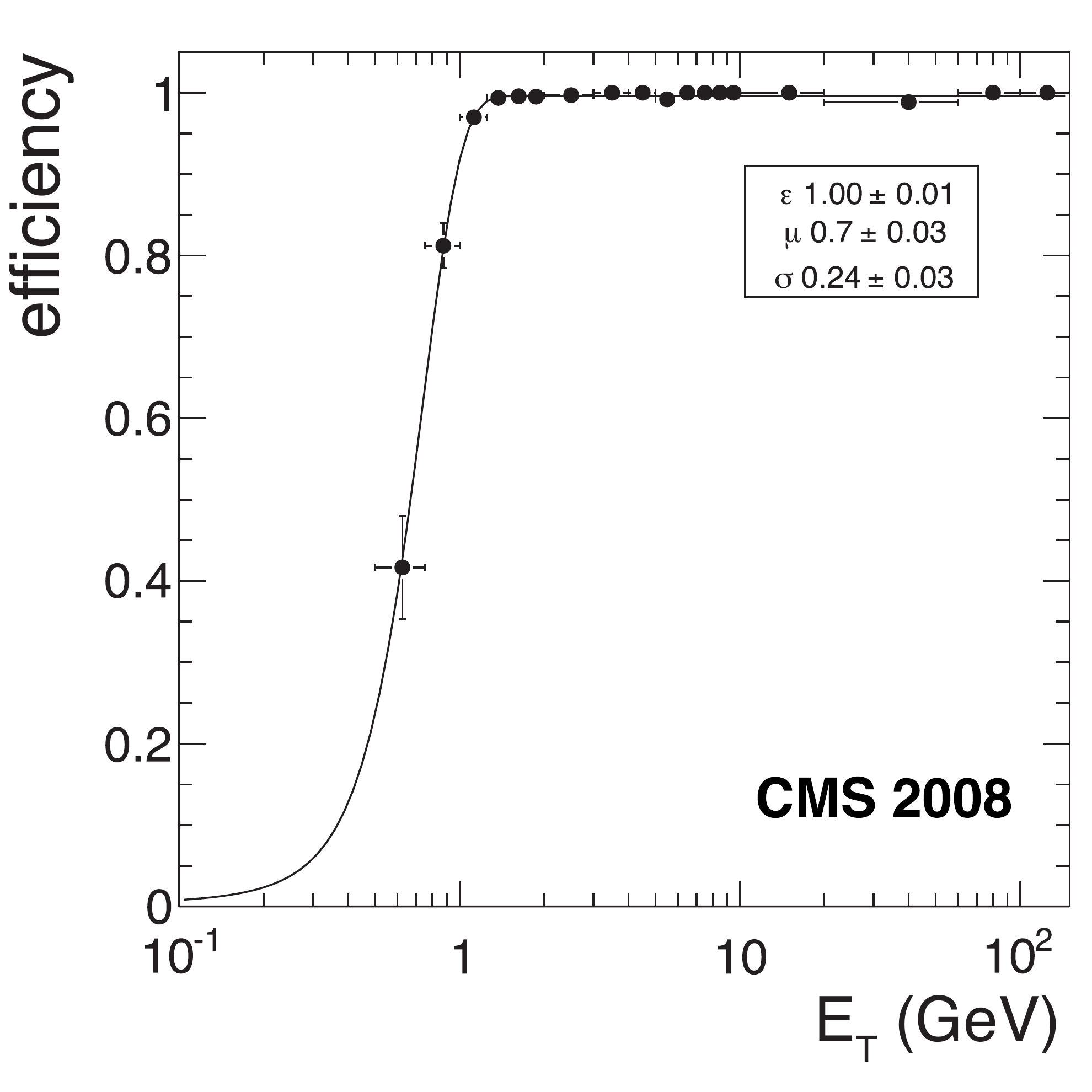}
 \includegraphics[angle=0,width=0.47\textwidth]{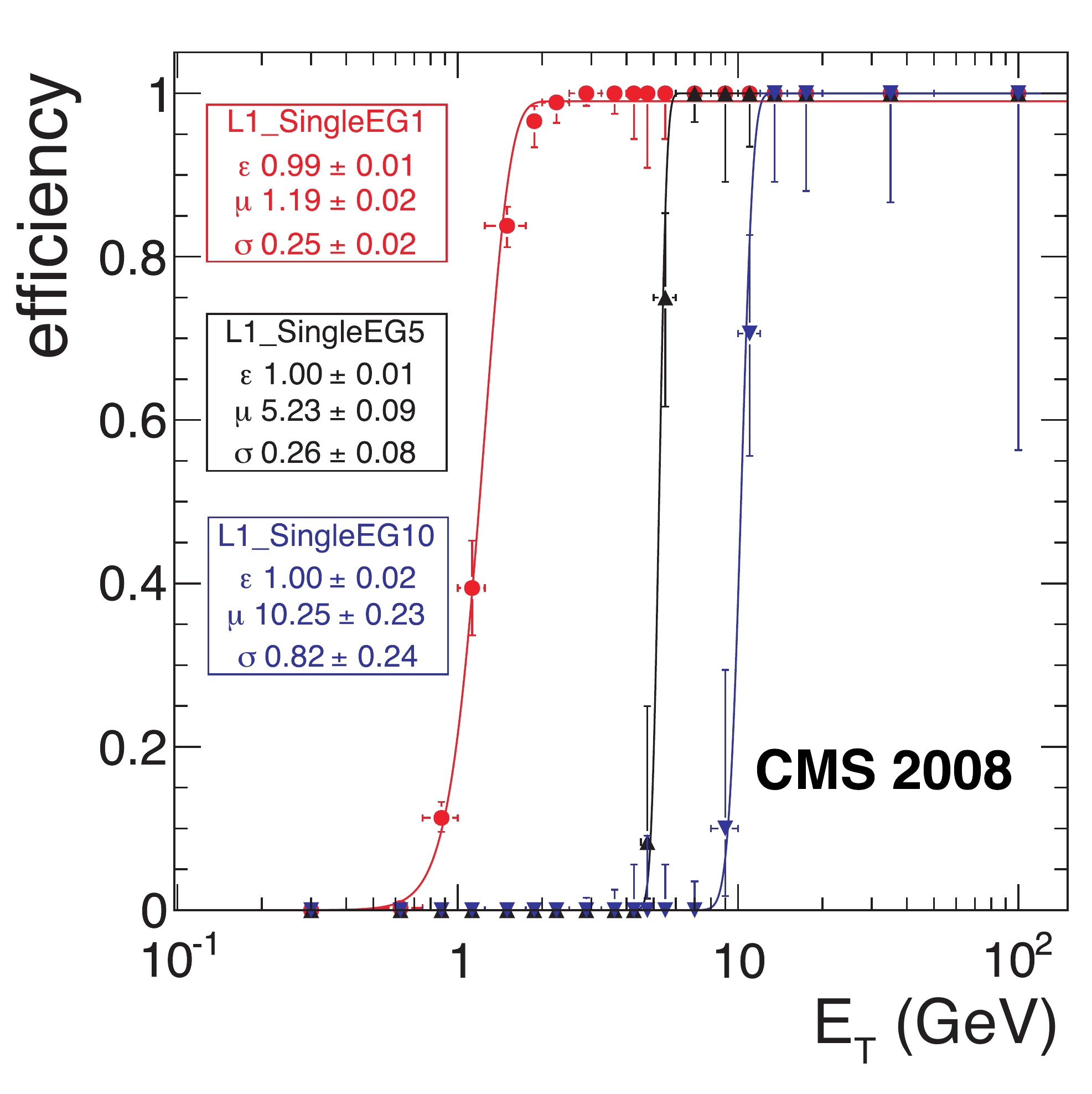}
  \caption{The ECAL trigger primitive production efficiency (left) and the full Level-1 e/$\gamma$ trigger efficiency (right) as a function of the $\et$ reconstructed offline in ECAL.  Parameters are obtained from fits of error functions to the data.  In the case of the right figure, an unbinned fit was used.}
  \label{fig:egECALEfficiency}
\end{center}
\end{figure}

The trigger primitive generator is considered efficient if a muon-tagged ECAL supercluster has an associated TP in the same ECAL trigger tower.
The TPG efficiency is shown in Fig.~\ref{fig:egECALEfficiency}~(left) as a function of $\et(\text{SC})$.
It rises with increasing $\et(\text{SC})$ before reaching a plateau of 100\%, the 50\% efficiency turn-on point being at $0.70\pm0.03 ~\text{(stat.)}\pm0.02 ~\text{(syst.)}$~GeV, compatible with the threshold set at 750~MeV (Section~\ref{sec:crafttriggers}). The systematic error is determined by varying the cuts applied to the selection.



The full e/$\gamma$ trigger chain is considered efficient if a muon-tagged ECAL supercluster has an associated L1 e/$\gamma$ candidate with energy above the L1 e/$\gamma$ threshold under study.
The measure of association used is the distance between the L1 candidates and muon-tagged ECAL superclusters, $\Delta R\left(\text{L1},\text{SC}^{\mu}\right)$.
Despite the coarse $(\eta,\phi)$ resolution of L1 e/$\gamma$ candidates (Fig.~\ref{fig:electron_eta_phi_delta}) two classes of events can be clearly distinguished: events where the L1 e/$\gamma$ candidate matches the muon-tagged ECAL supercluster around $\Delta R = 0$ and events where the objects are in opposite sides of the experiment with $\Delta R > 3$.
The latter case is expected since some muons cross the ECAL leaving 2 energy deposits, one on the top and one on the bottom.

If $\Delta R\left(\text{L1},\text{SC}^{\mu}\right) < 0.5$ and the L1 e/$\gamma$ candidate rank is above the threshold under study, the event is considered efficient.
With this selection the efficiency for the L1 e/$\gamma$ trigger chain is shown in Fig.~\ref{fig:egECALEfficiency} ({\em right}) for three different trigger algorithms: L1\_SingleEG1, L1\_SingleEG5 and L1\_SingleEG10 with nominal thresholds at 1, 5 and 10~GeV, respectively.
An unbinned maximum likelihood fit of an error-function was performed for each of the datasets.

The turn-on point of the 1~GeV threshold algorithm, L1\_SingleEG1, is found to be $1.19 \pm0.02~\text{(stat.)} \pm0.02~\text{(syst.)}$~GeV. For the corresponding 5 and 10~GeV algorithms the turn-on points are measured to be $5.23 \pm0.09~\text{(stat.)} \pm0.14~\text{(syst.)}$~GeV and $10.2 \pm0.2~\text{(stat.)} \pm0.3~\text{(syst.)}$~GeV, respectively.  Systematic errors were estimated by varying the timing selection window from $\pm1.25$~ns to $\pm5$~ns and the tagging distance $\Delta R\left(\text{SC},\mu\right)$ from 0.1 to 0.5.  The discrepancies between the measured and expected turn-on points are a reflection of the  effects already mentioned above, and mainly affect the L1\_SingleEG1, since the L1 e/$\gamma$ $\et$ LSB is 500~MeV. Above 10~GeV the effect is of no appreciable consequence.

%% file: jet/JetPerformance.tex
\section{Jet Trigger Performance}
\label{sec:jetperformance}

\begin{figure}[htbp]
  \begin{center}
   \includegraphics[width=0.47\textwidth]{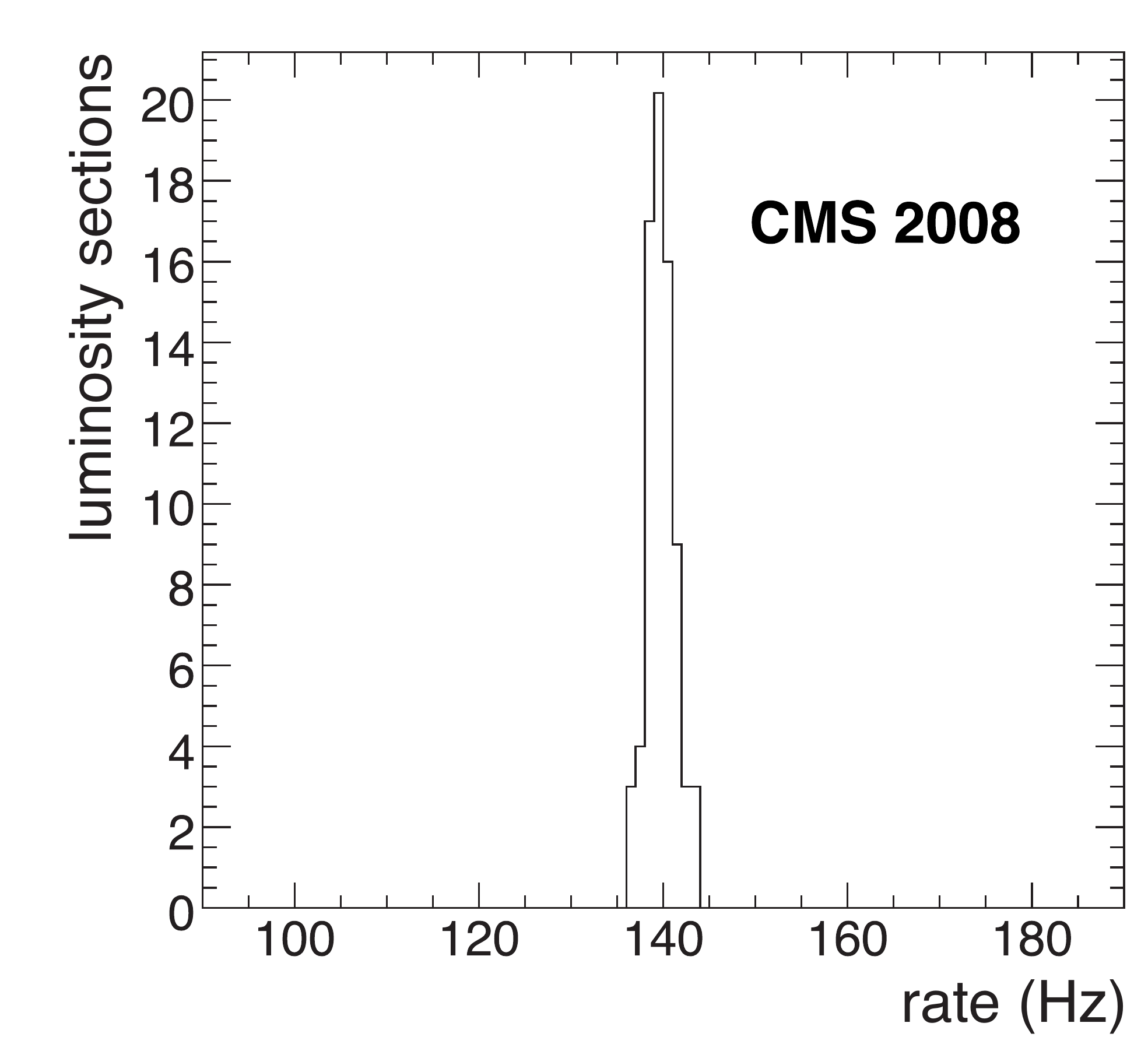}
    \caption{The distribution of the rate of the single jet trigger with an $\et$ threshold of $10 \GeV$.}
    \label{fig:jet_trig_rate}
  \end{center}
\end{figure}

Towards the end of the CRAFT data taking, a jet trigger was enabled and was active in around 20\% of the total runs. 
Due to the commissioning nature of the jet data taken, results based on only a few well-understood runs are presented.
Fig.~\ref{fig:jet_trig_rate} shows the measured rate per luminosity section in a single run for the single jet trigger with an $\et$ 
threshold of $10 \GeV$. The rate is dominated by detector noise, but is stable over the course of the run.

\begin{figure}[htbp]
\begin{center}
\includegraphics[width=0.47\textwidth,angle=90]{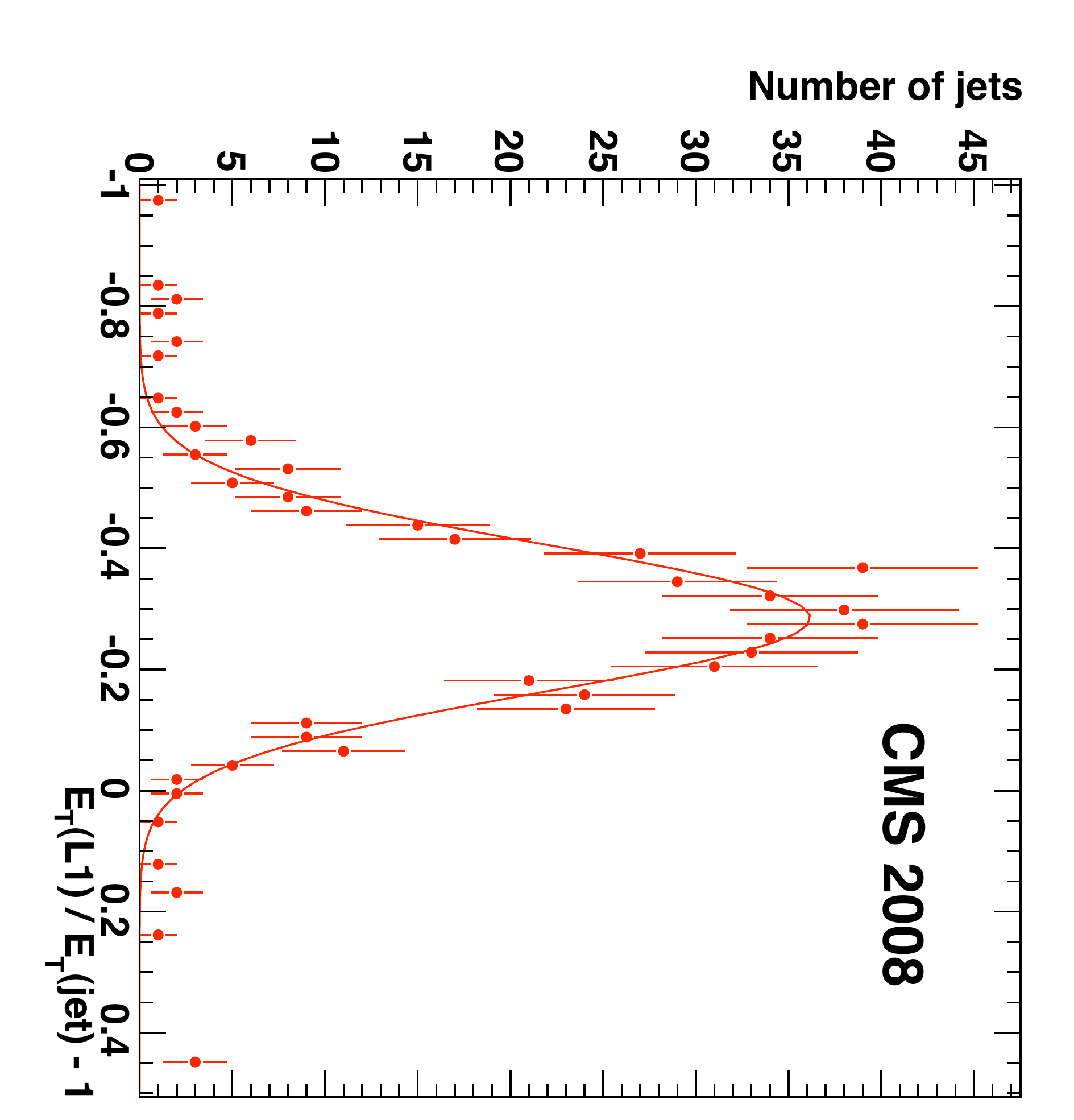}
\caption{The $\et$ resolution of L1 jets.}
\label{fig:jet_trig_et_res}
\end{center}
\end{figure}

The Level-1 jet $\et$ assignment was compared with that of offline reconstructed jets, which were found using an iterative cone algorithm with a cone size of $\Delta R=0.5$~\cite{Bayatian:2006zz}. The Level-1 jets were matched to the closest offline jet within a cone of $\Delta R=0.5$. The resulting $\et$ resolution, defined as $E_{T}(L1)/E_{T}(jet)-1$ (where $\et(L1)$ and $\et(jet)$ are the $\et$ of the matched L1 and offline jet, respectively), is shown in Fig.~\ref{fig:jet_trig_et_res}.  The RMS is $0.16$ and shows that the L1 jet $\et$ is around 70\% of the offline jet $\et$.

\begin{figure}[htbp]
  \begin{center}
    \mbox{
      \includegraphics*[width=.47\textwidth,angle=90]{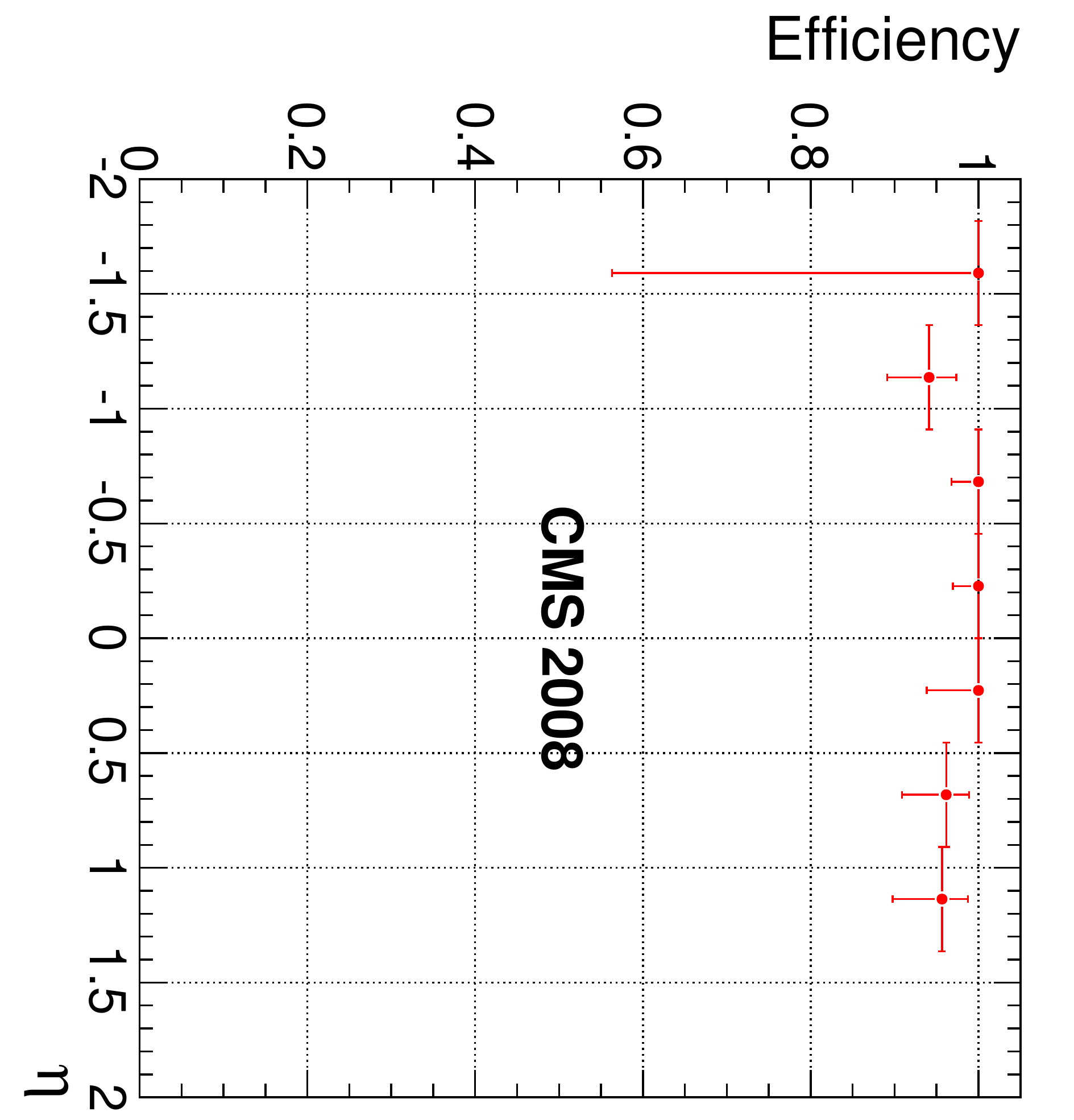}
      \includegraphics*[width=.47\textwidth,angle=90]{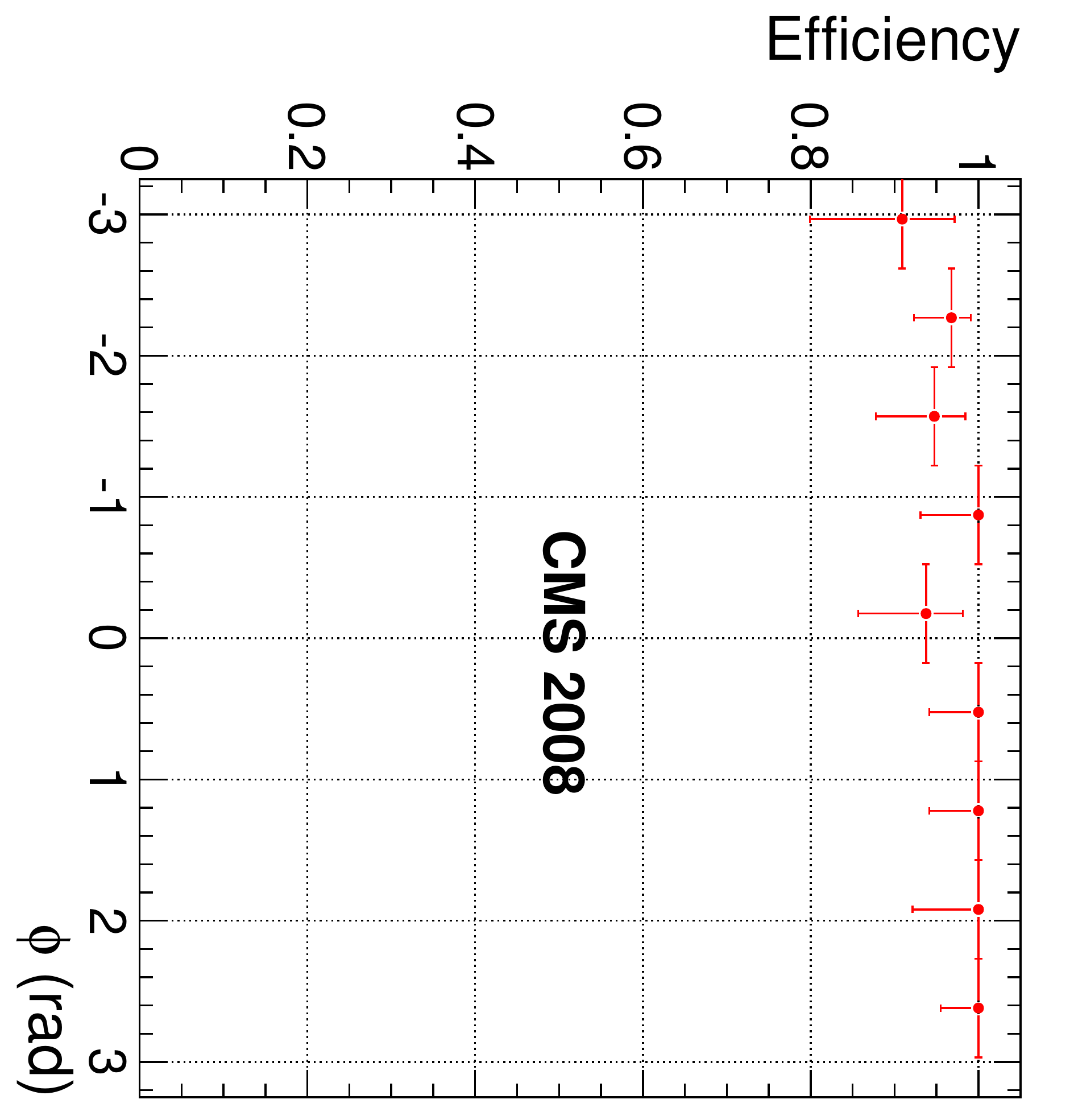}
    }
    \includegraphics*[width=.47\textwidth,angle=90]{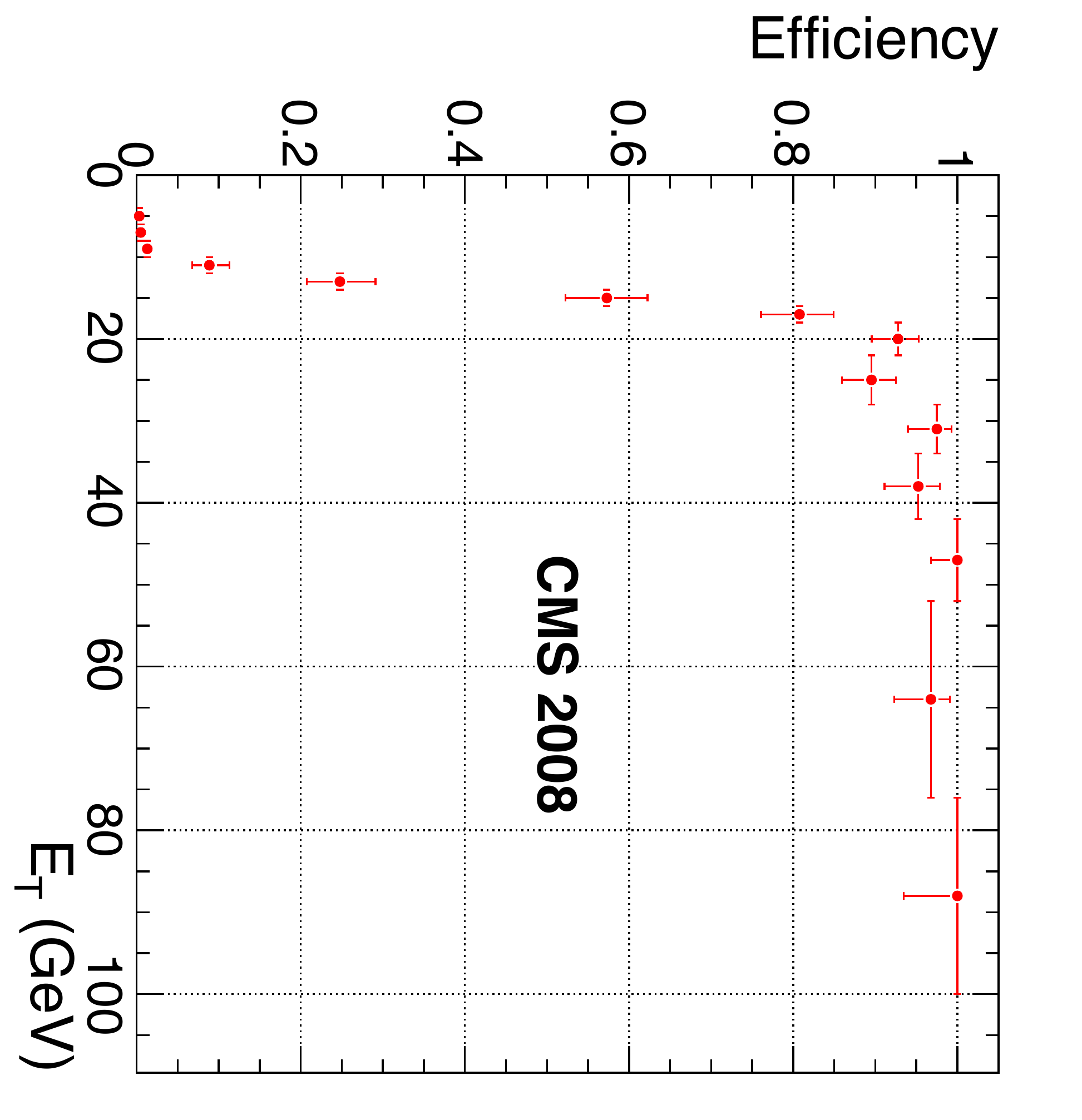}
    \caption{
      The efficiency of the single jet trigger with a L1 $\et$ threshold of $10 \GeV$, as a function of offline jet $\eta$, $\phi$, and $\et$ (with $\et > 25 GeV$ for the $\eta$ and $\phi$ plots).
    }
    \label{fig:jet_trig_eff}
  \end{center}
\end{figure}

The efficiency of the jet trigger, relative to offline reconstructed jets, was measured with a data sample triggered using the electron trigger path. The electron trigger hardware is independent of the jet trigger. An iterative cone jet-finder algorithm with a cone size of $\Delta R=0.5$ was run on the offline calorimeter data, and jets satisfying $|\eta|<3.0$ and $\et>5 \GeV$ were selected. The jet trigger efficiency was measured relative to the selected jets by demanding a L1 jet, firing the trigger, within a $\Delta R=0.5$ cone of the offline jet. A data sample containing around 5000 offline jets was used to determine the efficiency of the single jet $\et>10 \GeV$ trigger as a function of the $\et$, $\eta$, and $\phi$ of the offline jets, shown in Fig.~\ref{fig:jet_trig_eff}. The results show that the jet trigger efficiency reaches the level of 90\% at 20 GeV, and 100\% at 40 GeV. The efficiencies in $\eta$ and $\phi$ measured for jets with $\et>25 \GeV$ are uniform as expected.

%% file: CFT-09-013-authorlist.tex
\textbf{Yerevan Physics Institute,  Yerevan,  Armenia}\\*[0pt]
S.~Chatrchyan, V.~Khachatryan, A.M.~Sirunyan
\vskip\cmsinstskip
\textbf{Institut f\"{u}r Hochenergiephysik der OeAW,  Wien,  Austria}\\*[0pt]
W.~Adam, B.~Arnold, H.~Bergauer, T.~Bergauer, M.~Dragicevic, M.~Eichberger, J.~Er\"{o}
% ...
\vskip\cmsinstskip
% ...


%% file: CFT-09-013_temp.bbl
\providecommand{\href}[2]{#2}\begingroup\raggedright\begin{thebibliography}{10}

\bibitem{CMS:2008zzk}
{CMS Collaboration}, ``{The CMS experiment at the CERN LHC}'', {\em JINST} {\bf
  3} (2008)
S08004.
  \href{http://dx.doi.org/10.1088/1748-0221/3/08/S08004}{{\tt
  doi:10.1088/1748-0221/3/08/S08004}}.

\bibitem{Evans:2008zzb}
L.~Evans and P.~Bryant, ``{LHC Machine}'', {\em JINST} {\bf 3} (2008)
S08001.
  \href{http://dx.doi.org/10.1088/1748-0221/3/08/S08001}{{\tt
  doi:10.1088/1748-0221/3/08/S08001}}.

\bibitem{CRAFTGeneral}
{CMS Collaboration}, ``{The CMS CRAFT Exercise}'', {\em submitted to JINST}
  (2009).

\bibitem{Dasu:2000ge}
{CMS Collaboration}, ``{The TriDAS Project Technical Design Report: The Level-1
  Trigger}'', {\em CERN-LHCC} {\bf 2000-038} (2000).

\bibitem{Sakulin:2005qx}
H.~Sakulin and A.~Taurok, ``{Implementation and test of the first-level Global
  Muon Trigger of the CMS experiment}'', in {\em Proceedings of the 11th
  Workshop on Electronics for LHC and Future Experiments}.
\newblock 2005.

\bibitem{Guiducci:2007sc}
L.~Guiducci {et~al.}, ``{DT Sector Collector electronics design and
  construction}'', in {\em Proceedings of the Topical Workshop on Electronics
  for Particle Physics}.
\newblock 2007.

\bibitem{Ero:2008zz}
J.~Ero {et~al.}, ``{The CMS drift tube trigger track finder}'', {\em JINST}
  {\bf 3} (2008)
P08006.
  \href{http://dx.doi.org/10.1088/1748-0221/3/08/P08006}{{\tt
  doi:10.1088/1748-0221/3/08/P08006}}.

\bibitem{Acosta:2003is}
D.~Acosta {et~al.}, ``{A 3-D track-finding processor for the CMS level-1 muon
  trigger}'', in {\em Proceedings of the Conference for Computing in
  High-Energy and Nuclear Physics}.
\newblock 2003.

\bibitem{CFT-09-016}
{CMS Collaboration}, ``{Alignment of the CMS Muon System with Cosmic Ray and
  Beam Halo Muons}'', {\em submitted to JINST} (2009).

\bibitem{Andlinger:1994ve}
{RD5 Collaboration}, ``{Pattern Comparator Trigger (PACT) for the muon system
  of the CMS experiment}'', {\em Nucl. Instrum. Meth.} {\bf A370} (1996)
389--395.
  \href{http://dx.doi.org/10.1016/0168-9002(95)00861-6}{{\tt
  doi:10.1016/0168-9002(95)00861-6}}.

\bibitem{Stettler:2006zz}
M.~Stettler {et~al.}, ``{The CMS Global Calorimeter Trigger hardware design}'',
  in {\em Proceedings of the 12th Workshop on Electronics for LHC and Future
  Experiments}.
\newblock 2006.

\bibitem{Iles:2006zz}
G.~Iles {et~al.}, ``{Revised CMS global calorimeter trigger functionality and
  algorithms}'', in {\em Proceedings of the 12th Workshop on Electronics for
  LHC and Future Experiments}.
\newblock 2006.

\bibitem{Jeitler:2006zz}
M.~Jeitler {et~al.}, ``{The level-1 global trigger for the CMS experiment at
  LHC}'', in {\em Proceedings of the 12th Workshop on Electronics for LHC and
  Future Experiments}.
\newblock 2006.

\bibitem{MagransdeAbril:2005dk}
I.~Magrans~de Abril, C.~E. Wulz, and J.~Varela, ``{Concept of the CMS trigger
  supervisor}'', {\em IEEE Trans. Nucl. Sci.} {\bf 53} (2006)
474--483.
  \href{http://dx.doi.org/10.1109/TNS.2006.872631}{{\tt
  doi:10.1109/TNS.2006.872631}}.

\bibitem{Gutleber:2003cd}
J.~Gutleber, S.~Murray, and L.~Orsini, ``{Towards a homogeneous architecture
  for high-energy physics data acquisition systems}'', {\em Comput. Phys.
  Commun.} {\bf 153} (2003)
155--163.
  \href{http://dx.doi.org/10.1016/S0010-4655(03)00161-9}{{\tt
  doi:10.1016/S0010-4655(03)00161-9}}.

\bibitem{CFT-09-022}
{CMS Collaboration}, ``{Performance of the CMS Drift-Tube Local Trigger with
  Cosmic Rays}'', {\em submitted to JINST} (2009).

\bibitem{CMS-2002/022}
{CMS Collaboration}, ``{Link System and Crate Layout of the RPC Pattern
  Comparator Trigger for the CMS Detector}'', {\em CMS Note} {\bf
  \href{http://cms.cern.ch/iCMS/jsp/openfile.jsp?type=NOTE&year=2002&files=NOT%
E2002_022.pdf}{2002/022}} (2002).

\bibitem{CFT-09-012}
{CMS Collaboration}, ``{Performance of Local Muon Reconstruction in Drift Tube
  Chambers from the Analysis of Cosmic Run Data}'', {\em submitted to JINST}
  (2009).

\bibitem{Aldaya:2006cp}
M.~Aldaya {et~al.}, ``{Fine synchronization of the muon drift tubes local
  trigger}'', {\em Nucl. Instrum. Meth.} {\bf A564} (2006)
169--177.
  \href{http://dx.doi.org/10.1016/j.nima.2006.04.046}{{\tt
  doi:10.1016/j.nima.2006.04.046}}.

\bibitem{CFT-09-025}
{CMS Collaboration}, ``{Fine Synchronization of the CMS muon Drift-Tubes Local
  Trigger: from cosmic rays to LHC}'', {\em submitted to JINST} (2009).

\bibitem{CFT-09-010}
{CMS Collaboration}, ``{Measurement of RPC performance with cosmic rays}'',
  {\em submitted to JINST} (2009).

\bibitem{Ghete:2009}
{CMS Collaboration}, ``CMS Level-1 Trigger Emulator'', in {\em Proceedings of
  the International Conference on Computing in High Energy and Nuclear
  Physics}.
\newblock 2009.

\bibitem{CFT-09-004}
{CMS Collaboration}, ``{Crystal ECAL Operation and Performance}'', {\em
  submitted to JINST} (2009).

\bibitem{Almeida:2008zz}
N.~Almeida {et~al.}, ``{Data filtering in the readout of the CMS
  electromagnetic calorimeter}'', {\em JINST} {\bf 3} (2008)
P02011.
  \href{http://dx.doi.org/10.1088/1748-0221/3/02/P02011}{{\tt
  doi:10.1088/1748-0221/3/02/P02011}}.

\bibitem{CFT-09-006}
{CMS Collaboration}, ``{Time Reconstruction and Performance of Crystal ECAL}'',
  {\em submitted to JINST} (2009).

\bibitem{CFT-09-005}
{CMS Collaboration}, ``{Measurement of the muon stopping power in PWO}'', {\em
  submitted to JINST} (2009).

\bibitem{Bayatian:2006zz}
{CMS Collaboration}, ``{CMS Physics: Technical Design Report}'', {\em
  CERN-LHCC} {\bf 2006-001} (2006).

\end{thebibliography}\endgroup
